\documentclass[aps,rmp,twocolumn,showpacs,groupedaddress%
,superscriptaddress,nofootinbib]{revtex4}

\usepackage{graphicx}  
\usepackage{hyperref}  
\usepackage{dcolumn}   
\usepackage{bm}        
\usepackage{amssymb}   

\newcommand{\met}{\mbox{\ensuremath{\slash\kern-.7emE_{T}}}}
\newcommand{\pet}{\mbox{\ensuremath{\slash\kern-.7emP_{T}}}}
\newcommand{\invpb}{pb$^{-1}$}
\newcommand{\invfb}{fb$^{-1}$}
\newcommand{\et}{\mbox{\ensuremath{E_{T}}}}
\newcommand{\pt}{\mbox{\ensuremath{p_{T}}}}

\newcommand{\chn}{\tilde{\chi}^0}
\newcommand{\chc}{\tilde{\chi}^{\pm}}
\newcommand{\mweak}{m_{\text{weak}}}
\newcommand{\mmess}{M_{\text{mess}}}
\newcommand{\mgut}{M_{\text{GUT}}}

\newcommand{\mstar}{M_{*}}

\newcommand{\ev}{\text{eV}}

\newcommand{\gev}{\text{GeV}}
\newcommand{\tev}{\text{TeV}}
\newcommand{\pb}{\text{pb}}
\newcommand{\fb}{\text{fb}}

\newcommand{\cm}{\text{cm}}

\newcommand{\s}{\text{s}}

\newcommand{\eg}{{\em e.g.}}

\newcommand{\Eqref}[1]{Equation~(\ref{#1})}
\newcommand{\eqref}[1]{Eq.~(\ref{#1})}
\newcommand{\eqsref}[2]{Eqs.~(\ref{#1}) and (\ref{#2})}
\newcommand{\secref}[1]{Sec.~\ref{sec:#1}}

\newcommand{\figref}[1]{Fig.~\ref{fig:#1}}

\newcommand{\fbinv}{fb$^{-1}$}

\begin{document}


\title{Searches for Supersymmetry at High-Energy Colliders}

\author{Jonathan L.~Feng}

\affiliation{Department of Physics and Astronomy, University of
California, Irvine, CA 92697, USA}

\author{Jean-Fran\c{c}ois Grivaz}

\affiliation{Laboratoire de l'Acc\'{e}l\'{e}rateur Lin\'{e}aire, Orsay, France}

\author{Jane Nachtman}

\affiliation{University of Iowa, Iowa City, Iowa 52242, USA}

\begin{abstract}
This review summarizes the state of the art in searches for
supersymmetry at colliders on the eve of the LHC era.  Supersymmetry
is unique among extensions of the standard model in being motivated by
naturalness, dark matter, and force unification, both with and without
gravity.  At the same time, weak-scale supersymmetry encompasses a
wide range of experimental signals that are also found in many other
frameworks.  We recall the motivations for supersymmetry and review
the various models and their distinctive features.  We then
comprehensively summarize searches for neutral and charged Higgs
bosons and standard model superpartners at the high energy frontier,
considering both canonical and non-canonical supersymmetric models,
and including results from LEP, HERA, and the Tevatron.
\end{abstract}


\maketitle 
\tableofcontents

\section{Introduction}
\label{sec:introduction}

Particle physics is at a crossroads.  Behind us is the standard model
(SM), the remarkably successful theory of all known elementary
particles and their interactions.  Ahead of us is an equally
remarkable array of possibilities for new phenomena at the weak scale.
Never before has an energy scale been so widely anticipated to yield
profound insights, and never before have there been so many ideas
about exactly what these insights could be.  In this article, we
review the current state of experimental searches for supersymmetry,
the most widely studied extension of the SM.

\subsection{Motivations for New Phenomena}

There are at present many reasons to expect new physics at the weak
scale $\mweak \sim 100~\gev - 1~\tev$.  Chief among these is the Higgs
boson, an essential component of the SM that has yet to be discovered.
At the same time, there are also strong motivations for new phenomena
beyond the Higgs boson.  These motivations include naturalness, dark
matter, and unification.

\subsubsection{Naturalness}

The physical mass of the SM Higgs boson is given by
\begin{equation}
m_h^2 = m_h^{0\, 2} + \Delta m_h^2 \ ,
\end{equation}
where $m_h^{0\, 2}$ is the bare mass parameter present in the
Lagrangian, and the quantum corrections are
\begin{equation}
\Delta m_h^2 \sim \frac{\lambda^2}{16\pi^2} \int^{\Lambda} \frac{d^4
  p}{p^2} \sim \frac{\lambda^2}{16\pi^2} \Lambda^2 \ ,
\end{equation}
where $\lambda$ is a dimensionless gauge or Yukawa coupling, and
$\Lambda$ is the energy scale at which the SM is no longer a valid
description of nature.  Because $\Delta m_h^2$ is proportional to
$\Lambda^2$ (``quadratically divergent''), it is natural to expect the
Higgs mass to be pulled up to within an order of magnitude of
$\Lambda$ by quantum corrections~\cite{Weinberg:1975gm,%
Weinberg:1979bn,Susskind:1978ms,'tHooft:1980xb}.  Given that unitarity
and precision constraints require $m_h$ to be at the weak
scale~\cite{Reina:2005ae}, this implies $\Lambda \alt 1~\tev$, and new
physics should appear at the current energy frontier.  Of course, the
Higgs boson may not be a fundamental scalar, but in this case, too,
its structure requires new physics at the weak
scale~\cite{Hill:2002ap}.  For these reasons, naturalness is among the
most robust motivations for new physics at an energy scale accessible
to accelerator-based experiments.

\subsubsection{Dark Matter}

In the last decade, a wealth of cosmological observations have
constrained the energy densities of baryons, non-baryonic dark matter,
and dark energy, in units of the critical density, to
be~\cite{Komatsu:2008hk}
\begin{eqnarray}
\Omega_{\text{B}} &=& 0.0462 \pm 0.0015 \nonumber \\
\Omega_{\text{DM}} &=& 0.233 \pm 0.013 \label{omegas} \\
\Omega_{\Lambda} &=& 0.721 \pm 0.015 \ . \nonumber 
\end{eqnarray}
The non-baryonic dark matter must be stable or very long-lived and
dominantly cold or warm.  None of the particles of the SM satisfies
these conditions, and so cosmology requires new particles.

Perhaps the simplest production mechanism for dark matter is thermal
freeze
out~\cite{Zeldovich:1965,Chiu:1966kg,Steigman:1979kw,Scherrer:1985zt}.
In this scenario, a new particle is initially in thermal contact with
the SM, but as the Universe cools and expands, this particle loses
thermal contact and its energy density approaches a constant.  Under
very general assumptions, this relic energy density satisfies
\begin{equation}
\Omega_X \propto \frac{1}{\langle \sigma v \rangle} \ ,
\end{equation}
where $\langle \sigma v \rangle$ is the dark matter's
thermally-averaged annihilation cross section.  It is a tantalizing
fact that, when this cross section is typical of weak-scale particles,
that is, $\sigma v \sim \alpha^2 / \mweak^2$, where $\mweak \sim
100~\gev$, $\Omega_X$ is near the observed value of
$\Omega_{\text{DM}}$ given in \eqref{omegas}.  If thermal freeze out
is the mechanism by which dark matter is produced in the early
Universe, then, cosmological data therefore also point to the weak
scale as the natural scale for new physics.

\subsubsection{Unification}

The SM is consistent with the observed properties of all known
elementary particles.  It also elegantly explains why some phenomena,
such as proton decay and large flavor-changing neutral currents, are
not observed.  The latter fact is highly non-trivial, as evidenced by
the intellectual contortions required of model builders who try to
extend the SM.

At the same time, the SM contains many free parameters with values
constrained by experiment, but not explained.  The number of free
parameters may be reduced in unified theories, in which the symmetries
of the SM are extended to larger symmetries.  In particular, grand
unified theories, in which the SU(3) $\times$ SU(2) $\times$ U(1)
gauge structure is extended to larger groups, are significantly
motivated by the fact that the SM particle content fits perfectly into
multiplets of SU(5) and larger groups~\cite{Georgi:1974sy},
potentially explaining the seemingly random assignment of quantum
numbers, such as hypercharge.

One straightforward implication of the simplest ideas of grand
unification is that the gauge couplings of the SM must unify when
extrapolated to higher scales through renormalization group evolution.
The gauge couplings do not unify at any scale given the particle
content of the SM, but they do unify at the value $g_U \simeq 0.7$ at
$\mgut \simeq 2\times 10^{16}~\gev$ if the SM is minimally extended by
supersymmetry (SUSY) {\em and} the supersymmetric particles are at the
weak scale~\cite{Dimopoulos:1981zb,Dimopoulos:1981yj,%
Sakai:1981gr,Ibanez:1981yh,Einhorn:1981sx}.  This unification is
highly non-trivial, not only because the couplings are now so
precisely measured, but also because $g_U$ is in the perturbative
regime and $\mgut$ is in the narrow range that is both high enough to
suppress proton decay and low enough to avoid quantum gravitational
effects.  This unification is only logarithmically sensitive to the
superpartner mass scale, and the degree of its success is somewhat
model-dependent; see, \eg, the review by Raby
in Ref.~\cite{Amsler:2008zzb}.  In conjunction with the previous two
motivations, however, it provides still more evidence for new physics
at the weak scale, and selects supersymmetry as a particularly
motivated possibility.

\subsection{Experimental context}
\label{subsec:context}

There are two main areas where new phenomena could appear in particle
physics. Deviations from SM predictions could show up in measurements
performed with increasing precision.  Examples are the anomalies
observed in the forward-backward asymmetry in the production of
$b\bar{b}$ pairs in $e^+e^-$ collisions at the $Z$ peak
(see Ref.~\cite{:2005ema}, particularly section 7.3.5), or in the
anomalous magnetic moment of the muon (see, \eg, the review by
H\"ocker and Marciano in Ref.~\cite{Amsler:2008zzb}).  Even if such
anomalies receive experimental confirmation at a sufficient
significance level, their interpretation will however remain
ambiguous, because it will involve virtual contributions to the
relevant amplitudes of yet undiscovered, therefore most likely very
massive, particles. The alternative approach is to try to observe
directly the production of these new particles, which is among the
goals of the experiments at colliders operating at the highest
possible energies.

The Large Hadron Collider (LHC) at CERN will soon occupy the energy
frontier.  When it comes into operation, $pp$ collisions will take
place at a center-of-mass energy of 10 TeV, and of 14 TeV later
on. The instantaneous luminosity will be raised first to
$10^{32}~\cm^{-2}~\s^{-1}$ and progressively to
$10^{34}~\cm^{-2}~\s^{-1}$.  With the enormous data samples
accumulated, the two general purpose experiments at the LHC,
ATLAS~\cite{:1999fq,Aad:2009wy} and
CMS~\cite{:1994pu,Ball:2007zza}, will be in a position
to explore in great detail the physics at the TeV scale. Since this is
an entirely new domain, and since there are strong reasons to expect
new phenomena at that scale, as advocated in the preceding section of
this review, it may well be that ground breaking discoveries are made
at the LHC, even after a short period of operation, once the detectors
are properly aligned, calibrated, and well understood.

Until then, the most constraining results on searches for new
phenomena at high energy have been or are still being obtained at LEP,
HERA, and the Tevatron.  Providing a comprehensive account of such
searches for supersymmetry is the purpose of this review.

The large $e^+e^-$ collider (LEP) at CERN operated from 1989 to
2000. In a first phase (LEP1), the center-of-mass energy was set at or
close to 91~GeV, the peak of the $Z$ boson resonance. Four
experiments, ALEPH~\cite{Decamp:1990jra,Buskulic:1994wz},
DELPHI~\cite{Aarnio:1990vx,Abreu:1995uz}, L3~\cite{:1989kxa}, and
OPAL~\cite{Ahmet:1990eg} studied millions of $Z$ decays that allowed
them to perform stringent precision tests of the SM. From the end of
1995 on, the energy was progressively increased (LEP2) to reach
209~GeV in the center of mass during the last year of
operation. Altogether, each of the experiments collected a total of
$\sim 1$~fb$^{-1}$ of data, of which $\sim 235$~pb$^{-1}$ in 2000 at
and above 204~GeV, the data set most relevant for new particle
searches.

At DESY, the HERA collider operation was terminated in June
2007. There, $e^\pm p$ collisions were collected by two experiments,
H1~\cite{Abt:1996hi} and ZEUS~\cite{:1993ee}, at a
center-of-mass energy of $\sim 300$~GeV. This was an asymmetric
collider, with $e^\pm$ and proton beam energies of 30 and 820~GeV,
respectively. An upgrade took place in 2001 (HERA2), leading to higher
luminosities than in the previous phase (HERA1), and allowing
operation with polarized $e^\pm$ beams.  The data sets collected at
HERA1 and HERA2 with electron or positron beams altogether correspond
to an integrated luminosity of $\sim 0.5$~fb$^{-1}$ per experiment.

Until the LHC comes into operation, the highest energy collisions are
provided by the Tevatron $p\bar{p}$ collider at Fermilab. During its
first phase of operation (Run~I), the center-of-mass energy was set to
1.8~TeV, and a data sample of $\sim 110$~pb$^{-1}$ was collected by
each of the two experiments, CDF~\cite{Acosta:2004yw} and
D\O~\cite{Abazov:2005pn}.  The highlight of that period was the
discovery of the top quark in 1995. Major upgrades of the accelerator
complex and of the two detectors took place for the second phase
(Run~II), which began in 2001. The center-of-mass energy was raised to
1.96~TeV, and the instantaneous luminosity was progressively increased
to regularly approach or exceed $3\times 10^{32}~\cm^{-2}~\s^{-1}$ in
2008.  More than 5~fb$^{-1}$ of integrated luminosity had been
delivered by the Tevatron by the end of fiscal year (FY) 2008, and it
is expected that another $\sim 1.5$~fb$^{-1}$ of luminosity will be
provided per additional year of operation. At the time of writing,
running in FY 2009 is underway, running in FY 2010 is increasingly
likely, and running in FY 2011 is kept as an option.

All general purpose detectors at colliders share similar features. A
cylindrical ``barrel'' structure parallel to the beam axis surrounds
the interaction region, and is closed by ``end caps'' perpendicular to
the beam.  The first elements encountered beyond the beam pipe are
charged-particle detectors, with those closest to the interaction
point benefiting from the highest spatial precision. This tracking
system is immersed in an axial magnetic field provided by a solenoidal
magnet. Beyond the tracking system, electromagnetic calorimeters
provide electron and photon identification and energy
measurement. These are followed by hadron calorimeters for the
measurement of jet energies. Finally, track detectors are used to
identify and measure the muons which have penetrated through the
calorimeters and possibly additional absorber material.

Non-interacting particles, such as neutrinos, are detected by an
apparent non-conservation of energy and momentum. In $e^+e^-$
annihilation, the missing energy and momentum can be directly inferred
from a measurement of the final state particles, by comparison with
the center-of-mass energy of the collision.  In hadronic or $ep$
collisions, the partons participating in the hard process carry only a
fraction of the beam energy, and the beam remnants associated with the
spectator partons largely escape undetected in the beam pipe.  As a
consequence, only conservation of the momentum in the direction
transverse to the beams can be used, and the relevant quantity is the
missing transverse energy \met, rather than the total missing energy.
 
The mass reach in $p\bar p$ collisions at the Tevatron is expected to
be substantially larger than at LEP because of the higher
center-of-mass energy.  However, since the initial partons
participating in the hard process carry fractions $x_1$ and $x_2$ of
the beam energy, the effective center-of-mass energy is only
$\sqrt{\hat{s}}=\sqrt{x_1x_2s}$. Because of the rapidly falling parton
distribution functions (PDFs) as a function of those energy fractions,
increasingly large integrated luminosities are needed to probe larger
and larger $\sqrt{\hat{s}}$ values. At HERA, furthermore, the
center-of-mass energy in the $eq$ collision cannot be fully used for
new particle production, except in some very specific instances. This
is in contrast to $e^+e^-$ or $q\bar{q}$ annihilation, and to $gg$
fusion. As a consequence, the most constraining results on new
particle searches typically come from LEP and from the Tevatron.

In the following, all limits quoted are given at a confidence level of
95\%.

\section{Supersymmetric Models and Particles}
\label{sec:supersymmetry}

Supersymmetry (SUSY)~\cite{Golfand:1971iw,Volkov:1973ix,Wess:1974tw}
is an extension of Poincar\'e symmetry, which encompasses the known
spacetime symmetries of translations, rotations, and boosts.  As with
the Poincar\'e and internal symmetries, SUSY transforms particle states
to other particle states.  In contrast to these other symmetries,
however, SUSY relates states of different spin, transforming fermions
into bosons and vice versa.  None of the known particles can be
supersymmetric partners of other known particles.  As a result, SUSY
predicts many new particle states.  If SUSY were exact, these
particles would be degenerate with known particles.  Since this is
experimentally excluded, if SUSY is a symmetry of nature, it must be
broken.

SUSY is the most studied extension of the SM because it directly
addresses several of the motivations for new physics discussed in
\secref{introduction}.  In supersymmetric theories, the quadratically
divergent loop contributions to the Higgs boson mass from SM particles
are canceled by similar contributions from superpartners, ameliorating
the gauge hierarchy
problem~\cite{Davier:1979hr,Veltman:1980mj,Witten:1981nf}. Supersymmetric
theories also include excellent dark matter candidates, in the form of
neutralinos~\cite{Goldberg:1983nd,Ellis:1983ew} and
gravitinos~\cite{Pagels:1981ke,Feng:2003xh}, that may naturally have
the desired relic density.  Finally, SUSY is strongly motivated by the
hope for unifying forces, as it makes gauge coupling unification
possible in simple grand unified theories
(GUTs)~\cite{Dimopoulos:1981zb,Dimopoulos:1981yj,Sakai:1981gr,%
Ibanez:1981yh,Einhorn:1981sx}.  It is important to note that {\em all}
of these virtues are preserved only if the superpartner mass scale is
around the weak scale. The existence of SUSY in nature, although not
necessarily at the weak scale, is also motivated by string theory and
the beautiful mathematical properties of SUSY that are beyond the
scope of this review.

For these reasons, this review is devoted to searches for SUSY at
colliders.  In this Section, we present brief summaries of the
supersymmetric spectrum, parameters, and unifying frameworks to
establish our conventions and notation.  More extensive
phenomenological reviews of SUSY may be found in
Refs.~\cite{Martin:1997ns,Polonsky:2001pn,Drees:2004jm,Baer:2006rs}.

\subsection{Superpartners}
\label{sec:superpartners}

In this review, we focus our attention on the minimal supersymmetric
extension of the standard model (MSSM), the supersymmetric model with
minimal field content.  Bosonic superpartners are given names with the
prefix ``s--,'' and fermionic superpartners are denoted by the suffix
``--ino.''  Squarks and sleptons are collectively known as
``sfermions,'' and the entire group of superpartner particles are
often called ``sparticles.''

The particle content of the MSSM is in fact slightly more than a
doubling of the SM particle content.  This is because, in addition to
introducing superpartners for all known particles, the MSSM requires
two electroweak Higgs doublets.  There are two reasons for this.
First, in the SM, mass terms are generated for up- and down-type
particles by Yukawa couplings to $\varphi^*$ and $\varphi$,
respectively, where $\varphi$ is the SM Higgs field.  In SUSY, Yukawa
couplings are generalized to terms in a superpotential, a function of
superfields that contain both SM particles and their superpartners,
which generates the SM Yukawa couplings as well as all other terms
related to these by SUSY.  Complex-conjugated fields are not allowed
in the superpotential, however.  As a result, two separate Higgs
fields, denoted $H_u$ and $H_d$, are required to generate masses
through the superpotential terms
\begin{equation}
W = \lambda_u H_u Q \bar{U} + \lambda_d H_d Q \bar{D} 
+ \lambda_e H_d L \bar{E} \ ,
\end{equation}
where $Q$, $U$, $D$, $L$, and $E$ are the SU(2) quark doublet, up-type
quark singlet, down-type quark singlet, lepton doublet, and lepton
singlet superfields, respectively, and the $\lambda$ couplings are
Yukawa couplings.  Second, SUSY requires that the SM Higgs field have
fermion partners, the Higgsinos.  The introduction of these additional
fermions charged under SM gauge groups ruins anomaly cancellation,
making this theory mathematically untenable.  The introduction of an
additional Higgs doublet, with its extra Higgsinos, restores anomaly
cancellation.

The MSSM Higgs boson sector therefore consists of eight degrees of
freedom.  As in the SM, three of these are eaten to make massive $W$
and $Z$ bosons, but five remain, which form four physical particles:
\begin{equation}
\text{MSSM Higgs Bosons (Spin 0)}: \quad h, H, A, H^{\pm} \ ,
\end{equation}
where $h$ and $H$ are the CP-even neutral Higgs bosons, with $h$
lighter than $H$, $A$ is the CP-odd neutral Higgs boson, and $H^{\pm}$
is the charged Higgs boson.

The remaining supersymmetric particle content of the MSSM is
straightforward to determine and consists of the following states:
\begin{eqnarray}
\text{Neutralinos (Spin 1/2)}: && \tilde{B}, \tilde{W}^0, 
                              \tilde{H}_u^0, \tilde{H}_d^0 \nonumber \\
\text{Charginos (Spin 1/2)}: && \tilde{W}^+, \tilde{H}_u^+ \nonumber \\
                                && \tilde{W}^-, \tilde{H}_d^- \nonumber \\
\text{Sleptons (Spin 0)}: && \tilde{e}_{L,R}, \tilde{\mu}_{L,R}, 
                             \tilde{\tau}_{L,R} \nonumber \\
                     && \tilde{\nu}_{e}, \tilde{\nu}_{\mu},
                              \tilde{\nu}_{\tau} \nonumber \\
\text{Squarks (Spin 0)}: && \tilde{u}_{L,R}, \tilde{c}_{L,R},
                              \tilde{t}_{L,R} \nonumber \\
                    && \tilde{d}_{L,R}, \tilde{s}_{L,R}, 
                       \tilde{b}_{L,R} \nonumber \\
\text{Gluinos (Spin 1/2)}: && \tilde{g} \ .
\label{sparticles}
\end{eqnarray}
Each SM chiral fermion has a (complex) scalar partner, denoted by the
appropriate chirality subscript.  The dimensionless couplings of all
of these particles are fixed by SUSY to be identical to those of their
SM partners.  Note, however, that, as described in the appropriate
sections below, the states in each line of \eqref{sparticles} (except
for the last one) may mix, and mass eigenstates are in general linear
combinations of these gauge eigenstates.

Finally, most analyses of SUSY include the supersymmetric partner of
the graviton:
\begin{equation}
\text{Gravitino (Spin 3/2)}: \quad \tilde{G} \ .
\end{equation}
Although not technically required as a part of the MSSM, when SUSY is
promoted to a local symmetry, it necessarily includes gravity, and the
resulting supergravity theories include both gravitons and
gravitinos. The gravitino is therefore present if SUSY plays a role in
unifying the SM with gravity, as in string theory.

If SUSY were exact, the gravitino's properties would be determined
precisely by the graviton's, and it would be massless and have
gravitational couplings suppressed by the reduced Planck mass $\mstar
\simeq 2.4 \times 10^{18}~\gev$. However, just as Goldstone bosons
appear when conventional symmetries are spontaneously broken, a
fermion, the Goldstino $\tilde{G}_{1/2}$, appears when SUSY is broken.
The gravitino then becomes massive by eating the Goldstino.  In terms
of $F$, the mass dimension-2 order parameter of SUSY breaking, the
gravitino mass becomes
\begin{equation}
m_{\tilde{G}} \sim \frac{F}{\mstar} \ ,
\label{gravitinomass}
\end{equation}
and, very roughly, its interactions in processes probing energy scale
$E$ may be characterized by a dimensionless coupling
\begin{equation}
g_{\tilde{G}} \sim \frac{E^2}{F} \sim \frac{E^2}{m_{\tilde{G}} \mstar} \ .
\label{gravitinocoupling}
\end{equation}
Light gravitinos couple more strongly. As we will see below, in
well-motivated supersymmetric theories, these properties may take
values in the range
\begin{eqnarray}
\ev \alt &m_{\tilde{G}}& \alt 10~\tev \\
10^{-5} \agt &g_{\tilde{G}}& \agt 10^{-18} \ ,
\end{eqnarray}
where we have assumed colliders probing $E \sim \mweak$.

\subsection{Supersymmetry Parameters}
\label{sec:parameters}

As noted above, if SUSY exists in nature, it must be broken.  Although
many different Lagrangian terms could be added to break SUSY, only
some of these are allowed if SUSY is to stabilize the gauge hierarchy.
These terms, known as ``soft'' SUSY-breaking terms, include most, but
not all, Lagrangian terms with mass dimension 3 and
below~\cite{Girardello:1981wz}.  For the MSSM, they are
\begin{eqnarray}
\lefteqn{m_{\tilde{Q}}^2 |\tilde{Q}|^2 
\! + \! m_{\tilde{U}}^2 |\tilde{U}|^2 \! + \! m_{\tilde{D}}^2 |\tilde{D}|^2
\! + \! m_{\tilde{L}}^2 |\tilde{L}|^2 \! + \! m_{\tilde{E}}^2
|\tilde{E}|^2} \nonumber \\
&+& \! \! \frac{1}{2} \left\{ \left[ M_1 \tilde{B} \tilde{B} 
   + M_2\tilde{W}^j\tilde{W}^j + M_3\, \tilde{g}^k
\tilde{g}^k\right] + \text{h.c.}\right\} \nonumber \\
&+& \! \! \lambda_u A_U H_u \tilde{Q} \tilde{U} +
     \lambda_d A_D H_d \tilde{Q} \tilde{D} +
     \lambda_e A_E H_d \tilde{L} \tilde{E} \nonumber \\
&+& \! \!  m_{H_u}^2 |H_u|^2 + m_{H_d}^2 |H_d|^2 
   + \left( B H_u H_d + \text{h.c.} \right) .
\label{soft}
\end{eqnarray}
These lines are sfermion masses, gaugino masses, trilinear scalar
couplings (``$A$-terms''), and Higgs boson couplings.  In addition to
the parameters above, there are two other key parameters: the $\mu$
parameter, which enters in the Higgsino mass terms $\mu \tilde{H}_u^i
\tilde{H}_d^i$, and
\begin{equation}
\tan \beta \equiv \frac{\langle H_u^0 \rangle}{\langle H_d^0 \rangle}
\ ,
\label{beta}
\end{equation}
which parameterizes how the SM Higgs vacuum expectation value is
divided between the two neutral Higgs scalars.

The interactions of \eqref{soft} conserve
$R$-parity~\cite{Fayet:1977yc,Farrar:1978xj}.  With
$R=(-1)^{3(B-L)+2S}$, where $B$ and $L$ are the baryon and lepton
numbers, respectively, and $S$ the spin, all superpartners are odd and
all SM particles are even under $R$-parity. This implies that all
interactions involve an even number of superpartners, and so the
lightest superpartner is stable, and a potential dark matter
candidate.  $R$-parity violation generically violates both baryon and
lepton number, leading to too-rapid proton decay, which is why, for
most of this review, we limit ourselves to the $R$-parity conserving
case.

Even restricting ourselves to the $R$-parity-preserving terms of
\eqref{soft}, however, we see that SUSY introduces many new
parameters.  Note that the terms involving sfermions need not be
flavor-diagonal, and so the sfermion masses and $A$-terms are in fact
matrices of parameters in the most general case.  At the same time,
fully general flavor mixing terms violate low energy constraints on
flavor-changing neutral currents.  In addition, arbitrary complex
parameters also violate bounds on CP-violation from, for example,
$\epsilon_K$ and the electric dipole moments of the electron and
neutron.  These considerations motivate unifying frameworks, to which
we now turn.

\subsection{Unifying Frameworks}
\label{sec:unifying}

In collider searches, it is desirable to consider theories that are
both viable and simple enough to be explored fully.  For this reason,
it is common to work in simple model frameworks that reduce the number
of independent SUSY parameters.  In some cases, these model frameworks
also motivate particular collider signatures that might otherwise
appear highly unlikely or fine-tuned.

\begin{figure}
\includegraphics[width=8.5cm]{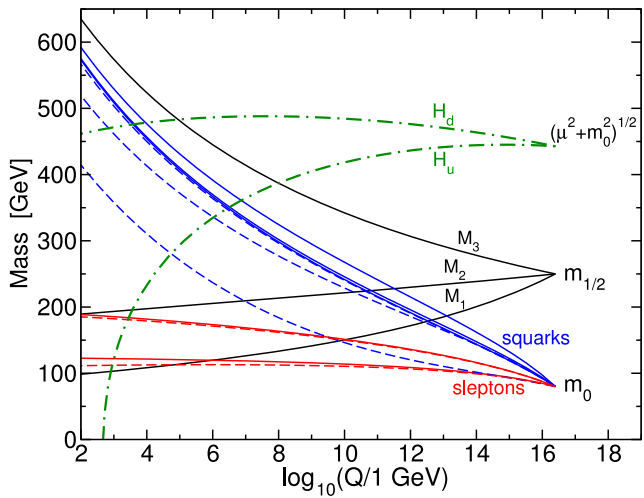}
\caption{Renormalization group evolution of scalar and gaugino mass
parameters from the GUT scale $\mgut \simeq 2\times 10^{16}~\gev$ to
the weak scale in a representative mSUGRA model.
{}From Ref.~\cite{Martin:1997ns}.
\label{fig:MSSMrun}
}
\end{figure}

In the most common unifying frameworks, SUSY is assumed to be broken
in some other sector.  SUSY breaking is then mediated to the MSSM
through a mechanism that defines the framework.  This sets
SUSY-breaking parameters at some high energy scale.  Renormalization
group evolution to the weak scale then determines the physical soft
SUSY-breaking parameters and the physical spectrum of the MSSM.  A
representative example of renormalization group evolution is shown in
\figref{MSSMrun}.  In this evolution from the high scale to the weak
scale, gauge couplings increase masses and Yukawa couplings decrease
masses.  This is central to understanding the sparticle spectrum of
many models.  In addition, it explains why $m_{H_u}^2$ becomes
negative at the weak scale --- it is the only particle to receive
large negative contributions from Yukawa couplings without
compensating large positive contributions from the strong coupling.
When $H_u$ becomes tachyonic, it breaks electroweak symmetry, and this
feature, known as ``radiative electroweak symmetry breaking,'' is a
virtue of many supersymmetric frameworks.  Note, however, that
radiative electroweak symmetry breaking makes essential use of the
large top quark mass, and so shifts the burden of understanding why
electroweak symmetry is broken to the question of why the top quark is
heavy. 

\begin{figure}
\includegraphics[width=8.5cm]{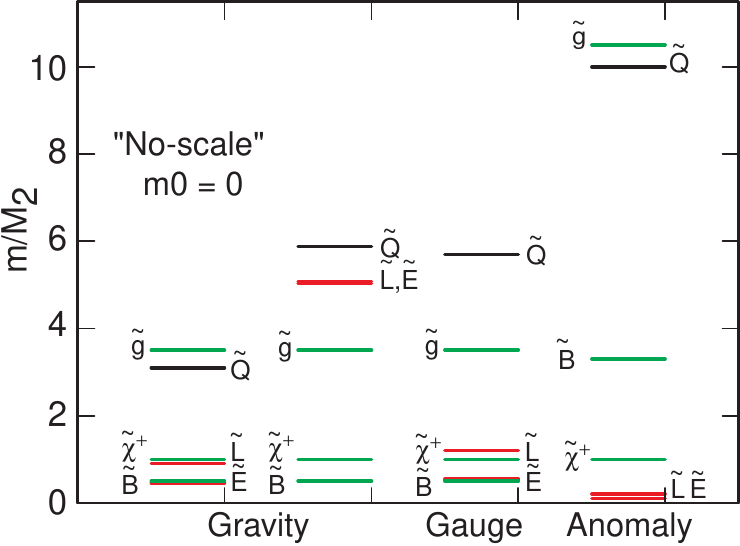}
\caption{Sparticle spectra for representative models with gravity-,
gauge-, and anomaly-mediated SUSY breaking.  The masses are normalized
to $M_2$, the Wino mass parameter at the weak scale.  In the
gravity-mediated case, two example spectra are presented: one for
``no-scale'' models with $m_0 = 0$, and another for $m_0 = 5 M_2$.  In
the anomaly-mediated case, the sleptons are tachyonic in the minimal
case --- additional effects are required to raise these to a viable
range. {}From Ref.~\cite{Peskin:2000ti}.
\label{fig:modelspectra}
}
\end{figure}

In this section, we discuss several common unifying frameworks that
have been used in collider searches, namely, models with gravity-,
gauge-, and anomaly-mediated SUSY breaking. Each of these has its
distinctive characteristics.  As a rough guide, in
\figref{modelspectra} we show representative spectra resulting from
each of these frameworks.  These spectra may be generated using
publicly available computer programs, including {\sc
isajet}~\cite{Paige:2003mg}, {\sc softsusy}~\cite{Allanach:2001kg},
{\sc spheno}~\cite{Porod:2003um}, and {\sc
suspect}~\cite{Djouadi:2002ze}.

\subsubsection{Gravity Mediation (SUGRA)}

In gravity-mediated SUSY-breaking models~\cite{Chamseddine:1982jx,%
Barbieri:1982eh,Ohta:1982wn,Hall:1983iz,AlvarezGaume:1983gj,%
Nilles:1983ge}, sometimes referred to as supergravity (SUGRA) models,
SUSY breaking in a hidden sector is mediated to the MSSM through terms
suppressed by the reduced Planck mass $\mstar$.  For example, sfermion
masses are $m_{\tilde{f}} \sim F/\mstar$.  For these to be at the weak
scale, $\sqrt{F}$ must be around $10^{11}~\gev$.  Given
\eqsref{gravitinomass}{gravitinocoupling}, the gravitino also has a
weak scale mass and couples with gravitational strength in SUGRA
models.

Without a quantum theory of gravity, the structure of gravity-mediated
SUSY parameters is unconstrained and generically violates low-energy
constraints.  To make these theories viable, {\em ad hoc} unifying
assumptions must be made.  By far the most common assumptions are
those of minimal supergravity (mSUGRA), which is specified by 4
continuous and 1 discrete parameter choice:
\begin{equation}
\text{mSUGRA:}\ m_0, m_{1/2}, A_0, \tan\beta, \text{sign}(\mu) \ ,
\end{equation}
where the first three parameters are the universal scalar mass
(including the two Higgs scalars), unified gaugino mass, and universal
$A$-parameter, which are all specified at the grand unified theory
(GUT) scale $\mgut \simeq 2\times 10^{16}~\gev$.  The remaining SUSY
parameters $|\mu|$ and the dimension-2 Higgs boson mass parameter $B$
are determined by requiring that the Higgs potential at the weak scale
give correct electroweak symmetry breaking.  At tree-level, this
requires
\begin{eqnarray}
\frac{1}{2} m_Z^2 &=& \frac{m_{H_d}^2 - m_{H_u}^2 \tan^2 \beta}
{\tan^2 \beta - 1} - |\mu|^2 \\
\sin 2 \beta &=& \frac{2B}{m_{H_d}^2 + m_{H_u}^2 + 2 |\mu|^2} \ .
\end{eqnarray}

Gaugino mass unification is motivated by the unification of gauge
couplings at $\mgut$ in the MSSM.  It leads to the prediction that the
Bino, Wino, and gluino masses are in the ratio $M_1 : M_2 : M_3 \simeq
1 : 2 : 7$ at the weak scale, as evident in \figref{modelspectra}.
Scalar mass universality is on much less solid ground.  Even in GUTs,
for example, the Higgs scalars are not necessarily in the same
multiplet as the squarks and sleptons.  This motivates a slightly less
restrictive framework, the non-universal Higgs model (NUHM) in which
$m_0$ is the universal sfermion mass, but $m_{H_u}$ and $m_{H_d}$ are
treated as independent parameters.  One may exchange these new degrees
of freedom for the more phenomenological parameters $\mu$ and $m_A$ at
the weak scale:
\begin{equation}
\text{NUHM:}\ m_0, m_{1/2}, A_0, \tan\beta, \mu, m_A \ ,
\end{equation}
The NUHM framework is employed in some MSSM Higgs boson studies
discussed in \secref{higgs}.

\subsubsection{GMSB}

In gauge-mediated SUSY-breaking (GMSB) models~\cite{Dine:1981za,%
Dimopoulos:1981au,Nappi:1982hm,AlvarezGaume:1981wy,Dine:1994vc,%
Dine:1995ag}, in addition to the gravity-mediated contributions to
soft parameters discussed above, each sparticle receives contributions
to its mass determined by its gauge quantum numbers.  These new
contributions to sfermion masses are $\sim F/\mmess$, where $\mmess$
is the mass scale of the messenger particles that transmit the SUSY
breaking.  The GMSB contributions are flavor-blind, and do not violate
low energy bounds.  For these to be dominant, one requires $\mmess
\alt 10^{14}~\gev$, and so we find that $m_{\tilde{G}} \sim F/\mstar
\ll F/\mmess \sim \mweak$ in GMSB scenarios, and the lightest
supersymmetric particle (LSP) is always the gravitino.

In GMSB models, the collider signatures are determined by the
next-to-lightest supersymmetric particle (NLSP) and its lifetime, or
equivalently, the gravitino's mass.  If the NLSP is the lightest
neutralino, the collider signature is either missing energy or prompt
photons, $Z$ or Higgs bosons from $\chn \to (\gamma, Z, h)
\tilde{G}$~\cite{Stump:1996wd,Dimopoulos:1996vz}; if the NLSP is a
slepton, the signature is typically either long-lived heavy charged
particles or multi-lepton
events~\cite{Feng:1997zr,Drees:1990yw,Goity:1993ih}.

\subsubsection{AMSB}

A third class of SUSY models are those with anomaly-mediated
SUSY-breaking (AMSB)~\cite{Randall:1998uk,Giudice:1998xp}.  These are
extra dimensional scenarios in which SUSY is broken on another
3-dimensional subspace, and transmitted to our world through the
conformal anomaly.  As with all anomalies, this effect is one-loop
suppressed.  The fundamental scale of SUSY breaking as characterized
by the gravitino mass is therefore $m_{\tilde{G}} \sim 10-100~\tev$,
with MSSM sparticle masses one-loop suppressed and at the weak scale.

The AMSB contributions to sparticle masses are completely determined
by the sparticle's gauge and Yukawa couplings.  This leads to a highly
predictive spectrum.  Unfortunately, one of these predictions is
$m_{\tilde{L}}^2, m_{\tilde{E}}^2 < 0$, but various mechanisms have
been proposed to solve this tachyonic slepton problem; see, e.g.,
Refs.~\cite{Pomarol:1999ie,Chacko:1999am,Katz:1999uw}.

The gaugino masses are determined by the corresponding gauge group
beta functions.  In particular, AMSB predicts $M_1: M_2: M_3 \simeq
2.8 : 1 : 8$; because the SU(2) coupling is nearly scale-invariant in
the MSSM, the Wino mass is the smallest.  AMSB scenarios therefore
motivate supersymmetric models with $\tilde{W}^0$ LSP and
$\tilde{W}^{\pm}$ NLSP.  This triplet may be extremely degenerate,
with the chargino traveling macroscopic distances before decaying to
soft and invisible decay products, which provides a distinctive and
challenging signature for collider searches~\cite{Feng:1999fu}.

\subsection{Supersymmetric Higgs Bosons}
\label{sec:higgs}

The MSSM Higgs potential is 
\begin{eqnarray}
V_H &=& (m_{H_u}^2 + |\mu|^2) |H_u^0|^2
+ (m_{H_d}^2 + |\mu|^2) |H_d^0|^2 \nonumber \\
&-& B (H_u^0 H_d^0 + \text{h.c.}) + \frac{1}{2} g^2 |H_u^{0*} H_d^0|^2
\nonumber \\
&+& \frac{1}{8} (g^2 + g'^2) (|H_u^0|^2 - |H_d^0|^2)^2 \ ,
\end{eqnarray}
where the parameters $\mu$, $m_{H_u}^2$, and $m_{H_d}^2$ are as
defined in \secref{parameters}, and SUSY implies that all the quartic
couplings are determined by the SU(2) and U(1) hypercharge gauge
couplings, denoted $g$ and $g'$, respectively. $V_H$ automatically
conserves CP, since any phase in the $B$ parameter can be eliminated
by a redefinition of the Higgs fields.

Assuming these parameters are such that the potential admits a stable,
symmetry-breaking minimum, the three parameter combinations $m_{H_u}^2
+ |\mu|^2$, $m_{H_d}^2 + |\mu|^2$, and $B$ may be exchanged for the
two vacuum expectation values $v_u \equiv \sqrt{2} \langle H_u^0
\rangle$, $v_d \equiv \sqrt{2} \langle H_d^0 \rangle$, and one
physical Higgs boson mass, conveniently taken to be $m_A$.  The $W$
boson mass fixes $v_u^2 + v_d^2$, leaving one additional degree of
freedom, usually taken to be $\tan\beta \equiv v_u/v_d$.  Thus, at
tree-level, the entire MSSM Higgs boson sector is determined by two
parameters, $m_A$ and $\tan\beta$.

In terms of these parameters, the physical Higgs boson masses are
\begin{eqnarray}
\label{CPevenmasses}
m^2_{\stackrel{H}{h}} \! \! &=& \! \! \! \frac{m_A^2 \! + \! m_Z^2
\! \pm \! \sqrt{(m_A^2 \! + \! m_Z^2)^2 
\! - \! 4 m_A^2 m_Z^2 c^2_{2\beta}}}{2}  \\
m_{H^{\pm}}^2 &=& m_A^2 + m_W^2 \ ,
\end{eqnarray}
where $c_{2\beta} \equiv \cos 2 \beta$.  The CP-even mass eigenstates
are related to the gauge eigenstates through
\begin{equation}
\left( \begin{array}{c} H \\ h \end{array} \right)
= \left( \begin{array}{cc} \cos\alpha & \sin\alpha \\
-\sin\alpha & \cos\alpha \end{array} \right)
\left( \begin{array}{c} \sqrt{2} \, \text{Re} H_d^0 - v_d \\ 
\sqrt{2} \, \text{Re} H_u^0 - v_u \end{array} \right) ,
\label{alpha}
\end{equation}
where the rotation angle $\alpha$ satisfies
\begin{equation}
\cos 2\alpha = - \cos 2 \beta \frac{m_A^2 - m_Z^2}{m_H^2 - m_h^2}
\end{equation}
with $-\pi/2 < \alpha < 0$.

\Eqref{CPevenmasses} implies that $m_h < m_Z |\cos 2\beta|$, a rather
disastrous relation, given that experimental bounds exclude $m_h <
m_Z$.  The results presented so far, however, are valid only at
tree-level. Large radiative corrections from top squark/quark
loops (see, \eg, Ref.~\cite{Barbieri:1991tk}),
\begin{equation}
\Delta m_h^2 \sim \frac{1}{\sin^2\beta}
\frac{3g^2 m_t^4}{8\pi^2 m_W^2}
\log\frac{m^2_{\tilde{t}}}{m^2_t} \ ,
\label{radcorr}
\end{equation}
can lift $m_h$ to values above the experimental bounds.  Note,
however, that for $\tan\beta = 1$, $m_h = 0$ at tree-level, and so
large values of $m_h$ are not possible for $\tan\beta \approx 1$.
{}From considerations of the Higgs mass alone, $\tan\beta <1$ is
possible.  However, such values imply very large top Yukawa couplings,
which become infinite well below the GUT or Planck scales. In
addition, in simple frameworks, $\tan\beta <1$ is incompatible with
radiative electroweak symmetry breaking; for a review of bounds on
$\tan\beta$, see Ref.~\cite{Haber:1993wf}.

\subsection{Neutralinos and Charginos}

The neutralinos and charginos of the MSSM are the mass eigenstates
that result from the mixing of the electroweak gauginos $\tilde{B}$
and $\tilde{W}^j$ with the Higgsinos.

The neutral mass terms are
\begin{equation}
\frac{1}{2} (\psi ^0)^T M_N \psi^0 + \text{h.c.} \ ,
\end{equation}
where $(\psi^0)^T = (-i\tilde{B},-i\tilde{W}^3, \tilde{H}^0_d,
\tilde{H}^0_u)$ and
\begin{equation}\label{neumass}
M_N =
\left( \begin{array}{cccc}
M_1             &0            &-\frac{1}{2} g' v_d & \frac{1}{2} g' v_u \\
0               &M_2          &\frac{1}{2} g v_d   & -\frac{1}{2} g v_u \\
-\frac{1}{2} g' v_d  & \frac{1}{2} g v_d &0            &-\mu           \\
 \frac{1}{2} g' v_u  &-\frac{1}{2} g v_u &-\mu         &0     \end{array}
\right) \ .
\label{neutralino}
\end{equation}
The neutralino mass eigenstates are $\chn_i = {\bf N}_{ij}\psi^0_j$,
where {\bf N} diagonalizes $M_N$.  In order of increasing mass, the
four neutralinos are labeled $\chn_1$, $\chn_2$, $\chn_3$, and
$\chn_4$.

The charged mass terms are
\begin{equation}
(\psi ^-)^T M_C \psi^+ + \text{h.c.} \ ,
\end{equation}
where $(\psi^{\pm})^T = (-i\tilde{W}^{\pm}, \tilde{H}^{\pm})$ and
\begin{equation}
\label{chamass}
M_C = \left( \begin{array}{cc}
 M_2                      &\frac{1}{\sqrt{2}} g v_u \\
\frac{1}{\sqrt{2}} g v_d  &\mu                  \end{array} \right) \ .
\label{chargino}
\end{equation}
The chargino mass eigenstates are $\tilde{\chi}^+_i = {\bf
V}_{ij}\psi^+_j$ and $\tilde{\chi}^-_i = {\bf U}_{ij}\psi^-_j$, where
the unitary matrices {\bf U} and {\bf V} are chosen to diagonalize
$M_C$, and $\chc_1$ is lighter than $\chc_2$.

\subsection{Sleptons}

Sleptons are promising targets for colliders, as they are among the
lightest sparticles in many models.  As noted in
\secref{superpartners}, sleptons include both left- and right-handed
charged sleptons and sneutrinos.  The mass matrix for the charged
sleptons is
\begin{equation}
\left( \! \! \begin{array}{cc}
m_{\tilde{L}}^2\! + \! m_{\tau}^2 \! - \! 
m_Z^2(\frac{1}{2} \! - \! s_W^2 ) c_{2\beta}
&m_{\tau} (A_{\tau} \! - \! \mu\tan\beta) \! \! \! \!  \\
m_{\tau} (A_{\tau} \! - \! \mu\tan\beta)
& \! \! \! \! m_{\tilde{E}}^2\! + \! m_{\tau}^2 \! - \! 
m_Z^2 s_W^2 c_{2\beta}
\end{array} \! \! \right) 
\label{staumass}
\end{equation}
in the basis $(\tilde{\tau}_L, \tilde{\tau}_R)$, where $s_W \equiv
\sin \theta_W$.  The sneutrino has mass
\begin{equation}
m_{\tilde{\nu}}^2 = m_{\tilde{L}}^2\! 
+ \! \frac{1}{2} m_Z^2\cos 2\beta \ .
\end{equation}
These masses are given in third-generation notation; in the presence
of flavor mixing, these generalize to full six-by-six and
three-by-three matrices.

The left-right mixing is proportional to lepton mass, and is therefore
expected to be insignificant for selectrons and smuons, but may be
important for staus, especially if $\tan\beta$ is large.  Through
level repulsion, this mixing lowers the lighter stau's mass.  As noted
in \secref{unifying}, Yukawa couplings also lower scalar masses
through renormalization group evolution.  Both of these effects imply
that in many scenarios, the lighter stau is the lightest slepton, and
often the lightest sfermion.

\subsection{Squarks}

The mass matrix for up-type squarks is
\begin{equation}
\! \! \left( \! \! \begin{array}{cc}
m_{\tilde{Q}}^2 \! + \! m_t^2 \! + \! 
m_Z^2(\frac{1}{2} \! - \! \frac{2}{3} s_W^2 )c_{2\beta} \! \! \! \! 
&m_t(A_t \! - \! \mu\cot\beta)  \\
m_t(A_t \! - \! \mu\cot\beta)
&\! \! \! \! m_{\tilde{U}}^2 \! + \! m_t^2 \! + \! 
m_Z^2 \frac{2}{3} s_W^2 c_{2\beta}
\end{array} \! \! \right) \! \!
\end{equation}
in the basis $(\tilde{t}_L, \tilde{t}_R)$, and for down-type squarks
is
\begin{equation}
\! \!  \left( \! \! \begin{array}{cc}
m_{\tilde{Q}}^2 \! + \! m_b^2 \! - \!  
m_Z^2(\frac{1}{2} \! - \! \frac{1}{3} s_W^2 )c_{2\beta} \! \! \! \! 
&m_b (A_b  \! - \!  \mu\tan\beta)  \\
m_b (A_b  \! - \!  \mu\tan\beta)
&\! \! \! \! m_{\tilde{D}}^2 \! + \! m_b^2 \! - \!
m_Z^2 \frac{1}{3} s_W^2 c_{2\beta}
\end{array} \! \! \right) \! \!
\end{equation}
in the basis $(\tilde{b}_L, \tilde{b}_R)$. Large mixing is expected in
the stop sector, and possibly also in the sbottom sector if
$\tan\beta$ is large.  Because of these mixings and the impact of
large Yukawa couplings in renormalization group evolution, the 3rd
generation squarks are the lightest squarks in many
models~\cite{Ellis:1983ed}.

\section{Searches for MSSM neutral Higgs bosons}
\label{sec:SUSYHiggs}

As already explained in Sec.~\ref{sec:supersymmetry}, two Higgs
doublets are needed in the MSSM to give mass to both up- and down-type
quarks. Under the assumption that the Higgs sector is CP conserving,
the physical states are two neutral CP-even Higgs bosons ($h$ and $H$,
ordered by increasing mass), a neutral CP-odd Higgs boson ($A$), and a
doublet of charged Higgs bosons ($H^\pm$). Further details on the
Higgs sector of the MSSM have been given in
Sec.~\ref{sec:supersymmetry}.  Here, we focus on searches for the
neutral Higgs bosons of the MSSM, while searches for charged Higgs
bosons will be discussed in Sec.~\ref{sec:charged_higgs}.

\subsection{MSSM benchmark scenarios}
\label{subsec:benchmarks}

It has been seen that two parameters are sufficient to fully determine
the MSSM Higgs sector at tree level. These are commonly taken to be
the $A$ boson mass $m_A$ and $\tan\beta$, the ratio of the vacuum
expectation values of the Higgs fields giving mass to the up- and
down-type quarks. This picture is modified significantly, however, by
large radiative corrections, arising essentially from an incomplete
cancellation of the top and stop loops.  In particular, the important
prediction $m_h<m_Z\vert\cos 2\beta\vert$ is invalidated. Among the
many parameters of the MSSM, a few have been identified as being most
relevant for the determination of Higgs boson properties. In addition
to $m_A$ and $\tan\beta$, an effective SUSY breaking scalar mass,
$M_{SUSY}$, which sets the scale of all squark masses, and a term
controlling the amount of mixing in the stop sector, $X_t$, play the
leading role.  (In \eqref{radcorr}, the stop mass is directly related
to $M_{SUSY}$, and stop mixing is neglected.)  The model is further
specified by a weak gaugino mass, $M_2$, the gluino mass, $m_{\tilde
g}$, and the SUSY Higgs mass term $\mu$. The relation $X_t=A-\mu
\cot\beta$ then allows the trilinear Higgs-squark coupling $A$
(assumed to be universal) to be calculated.  For large values of
$\tan\beta$, mixing in the sbottom sector becomes relevant too; it is
controlled by $X_b=A-\mu\tan\beta$. Finally, the top quark mass $m_t$
needs to be specified.

A few benchmark scenarios~\cite{Carena:1999xa,Carena:2002qg} were
agreed upon to interpret the searches for MSSM Higgs bosons. The most
widely considered are the so-called ``$m_h$-max'' and ``no-mixing''
ones, where $M_{SUSY}=1$~TeV, $M_2=200$~GeV, $\mu=-200$~GeV and
$m_{\tilde g}=800$~GeV. In $m_h$-max, $X_t$ is set equal to
$2M_{SUSY}$ (in the on-shell renormalization scheme), while it is set
to 0 in the no-mixing scenario. The largest value of $m_h$ is obtained
for large $m_A$ and $\tan\beta$, and is maximized (minimized) in the
$m_h$-max (no-mixing) scenario. In the $m_h$-max scenario, the maximum
value of $m_h$ is $\simeq 135$~GeV.

\subsection{Searches at LEP}
\label{subsec:LEPHiggs}

At LEP, the neutral Higgs bosons of the MSSM have been searched for in
two production processes, the Higgsstrahlung process 
$e^+e^-\to hZ$~\cite{Ellis:1975ap,Lee:1977eg,Ioffe:1976sd}
and the associated production $e^+e^-\to hA$~\cite{Pocsik:1981bg}.
Both processes are mediated by $s$-channel $Z$ boson exchange.  
With the notations of
Sec.~\ref{sec:supersymmetry}, the cross sections are
\begin{eqnarray}
\sigma_{hZ} &=& \sin^2(\beta-\alpha)\sigma^{SM}_{hZ} \\
\sigma_{hA} &=& \cos^2(\beta-\alpha)\bar\lambda\sigma^{SM}_{hZ} \ ,
\end{eqnarray}
where $\beta$ and $\alpha$ are defined in \eqsref{beta}{alpha},
\begin{equation}
\sigma^{SM}_{hZ}={{G^2_F m^4_Z}\over{96\pi s}}(v_e^2+a_e^2)\lambda_{hZ}^{1/2}
{{\lambda_{hZ}+12m^2_Z/s}\over{(1-m^2_Z/s)^2}}
\end{equation} 
is the SM Higgs boson production cross section,
$s$ is the square of the center-of-mass energy, and
\begin{eqnarray}
\lambda_{ij} &=& [1-(m_i+m_j)^2/s][1-(m_i-m_j)^2/s] \ , \\
\bar\lambda &=& \lambda^{3/2}_{hA}[\lambda^{1/2}_{hZ}(\lambda_{hZ}
+12m^2_Z/s)] \ .
\end{eqnarray}

It is apparent from the above formulae that the two processes are
complementary. In practice, the Higgsstrahlung process dominates for
values of $\tan\beta$ close to unity, while associated production
dominates for large values of $\tan\beta$, if kinematically allowed.
In large regions of the MSSM parameter space, the $h$ decay branching
fractions are similar to those of the SM Higgs boson. For a mass of
115~GeV, these are 74\% into $b\bar b$, 7\% into both $\tau^+\tau^-$
and $gg$, 8\% into $WW^\ast$, and 4\% into $c\bar
c$~\cite{Djouadi:1997yw}.  The $A$ boson couples only to fermions, so
that its decay branching fraction into $b\bar b$ is always close to
90\%, with most of the rest going into $\tau^+\tau^-$.  These same
branching fractions also hold for the $h$ boson for large values of
$\tan\beta$~\cite{Djouadi:1997yw}.

Searches for Higgs bosons were performed at LEP first in $Z$ boson
decays during the LEP1 era, and subsequently at increasing
center-of-mass energies at LEP2, up to 209~GeV in 2000. In the
following, only the searches performed at the highest energies are
described.

The four LEP experiments carried out searches for the SM Higgs boson
produced via Higgsstrahlung, $e^+e^-\to HZ$, and the results were
combined to maximize the sensitivity.\footnote{Production by vector
boson fusion, $e^+e^-\to He^+e^-$ or
$H\nu\bar\nu$~\cite{Jones:1979bq}, was also considered, but its
contribution was found to be negligible in practice.}  Four final
state topologies were analyzed to cope with the various decay modes of
the Higgs and $Z$ bosons: a four-jet topology with two $b$-tagged
jets, for $(H\to b\bar b)(Z\to q\bar q)$; a two $b$-tagged jets and
two-lepton topology, for $(H\to b\bar b)(Z\to \ell^+\ell^-)$, with
$\ell = e$ or $\mu$; a two $b$-tagged jets and missing energy
topology, for $(H\to b\bar b)(Z\to \nu\bar\nu)$; and a two-jet and
two-$\tau$ topology for $(H\to b\bar b)(Z\to \tau^+\tau^-)$ and
$(H\to\tau^+\tau^-)(Z\to q\bar q)$. A few candidate events were
observed at the edge of the sensitivity domain, but the overall
significance was only at the level of 1.7~$\sigma$. A lower mass limit
was therefore derived, excluding a SM Higgs boson with mass smaller
than 114.4~GeV~\cite{Barate:2003sz}.

\begin{figure}
\includegraphics[width=8.5cm]{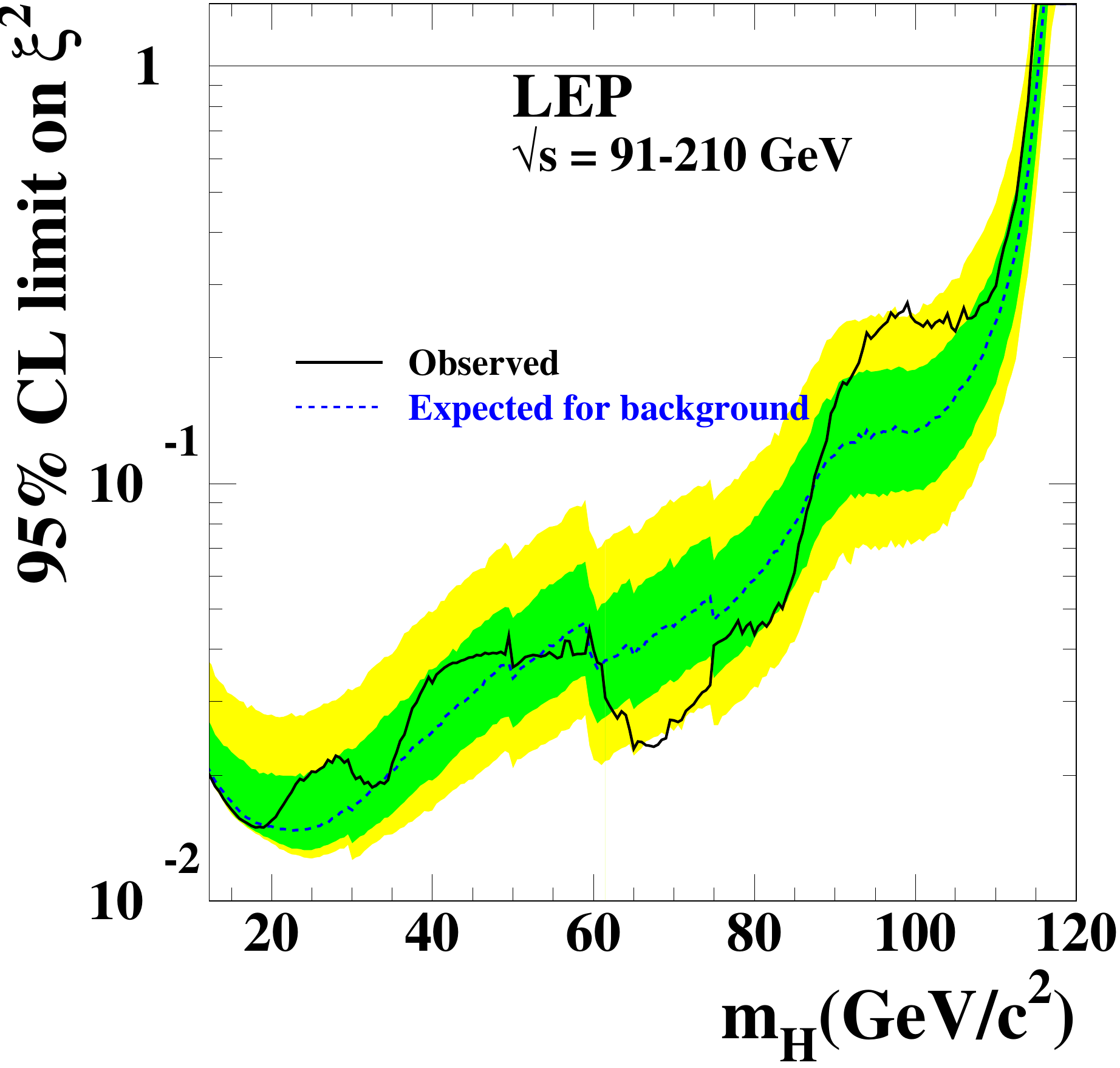}
\caption{\label{ksi} The upper bound on the factor $\xi^2$ by which
the square of the SM $HZZ$ coupling is multiplied, as provided by the
LEP experiments~\cite{Barate:2003sz}. The full curve is the observed
limit, the dashed curve the median expected limit in the absence of
signal, and the green and yellow bands are the 68\% and 95\%
probability regions around the expected limit.}
\end{figure}

The Higgs boson mass lower limit depends on the strength of the $HZZ$
coupling, and the LEP collaborations also provided, as a function of
the mass of a SM-like Higgs boson, an upper limit on $\xi^2$, where
$\xi$ is a multiplicative factor by which the SM $HZZ$ coupling is
reduced~\cite{Barate:2003sz}.  By SM-like, it is meant that the decay
branching fractions are similar to those expected from a SM Higgs
boson. This result is shown in Fig.~\ref{ksi}.  Constraints on the
MSSM parameter space can be deduced from this, since in that case $\xi
= \sin(\beta-\alpha)$.

For Higgs boson masses accessible at LEP, the structure of the MSSM
Higgs sector is such that the $h$ and $A$ masses are similar whenever
associated production is relevant, i.e., for large values of
$\tan\beta$.  Searches for $hA$ associated production were performed
in the four $b$-jet final state for $(h\to b\bar b)(A\to b\bar b)$ and
in the two $b$-jet and two-$\tau$ topology for $(h/A\to b\bar
b)(A/h\to\tau^+\tau^-)$. The constraint that the $h$ and $A$ boson
candidate masses should be similar was imposed.  The backgrounds from
multijet and $WW$ production were largely reduced by the $b$-jet
identification requirements, leaving $ZZ$ as an irreducible
background.

\begin{figure}
\includegraphics[width=8.5cm]{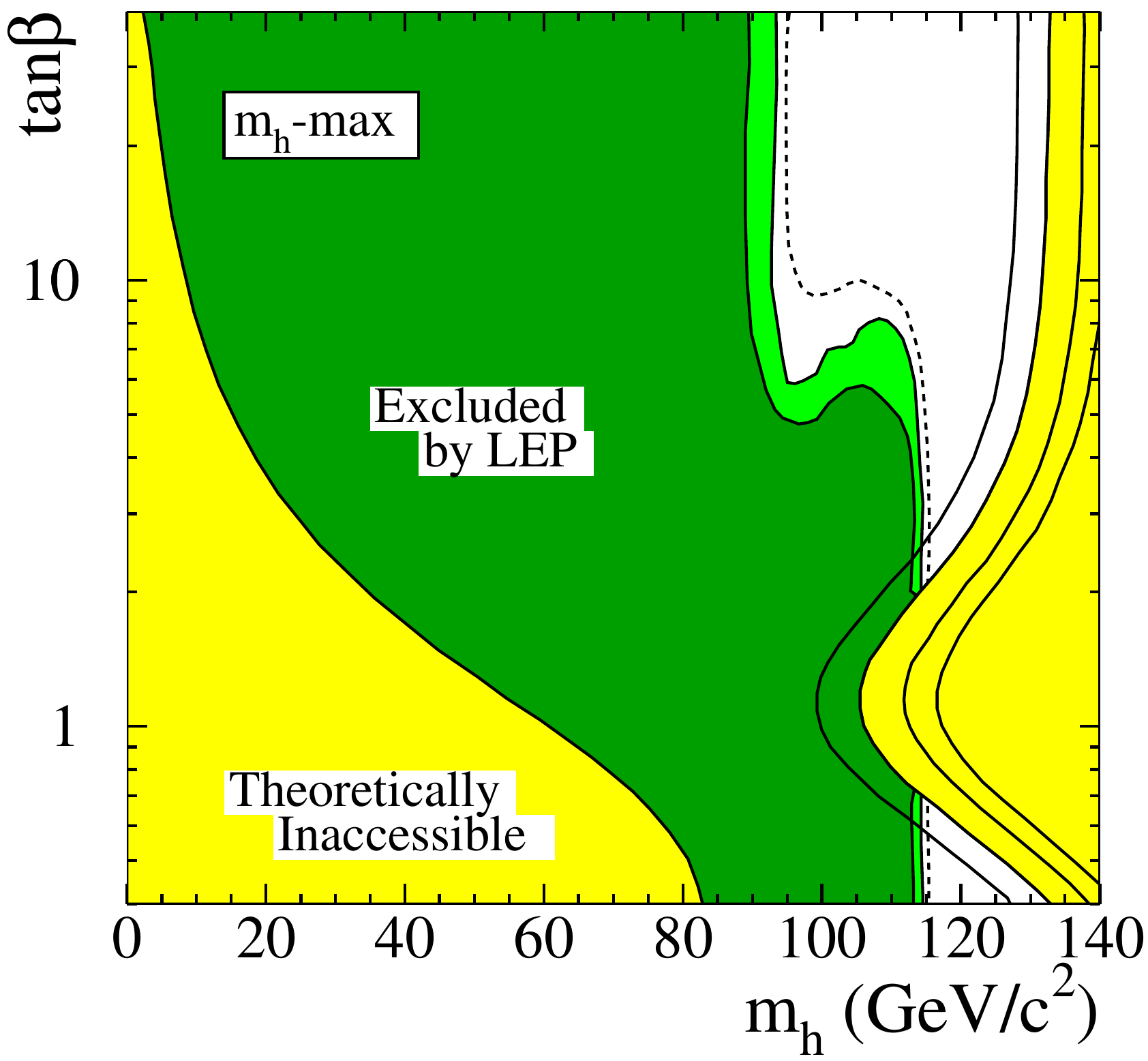}\\
\includegraphics[width=8.5cm]{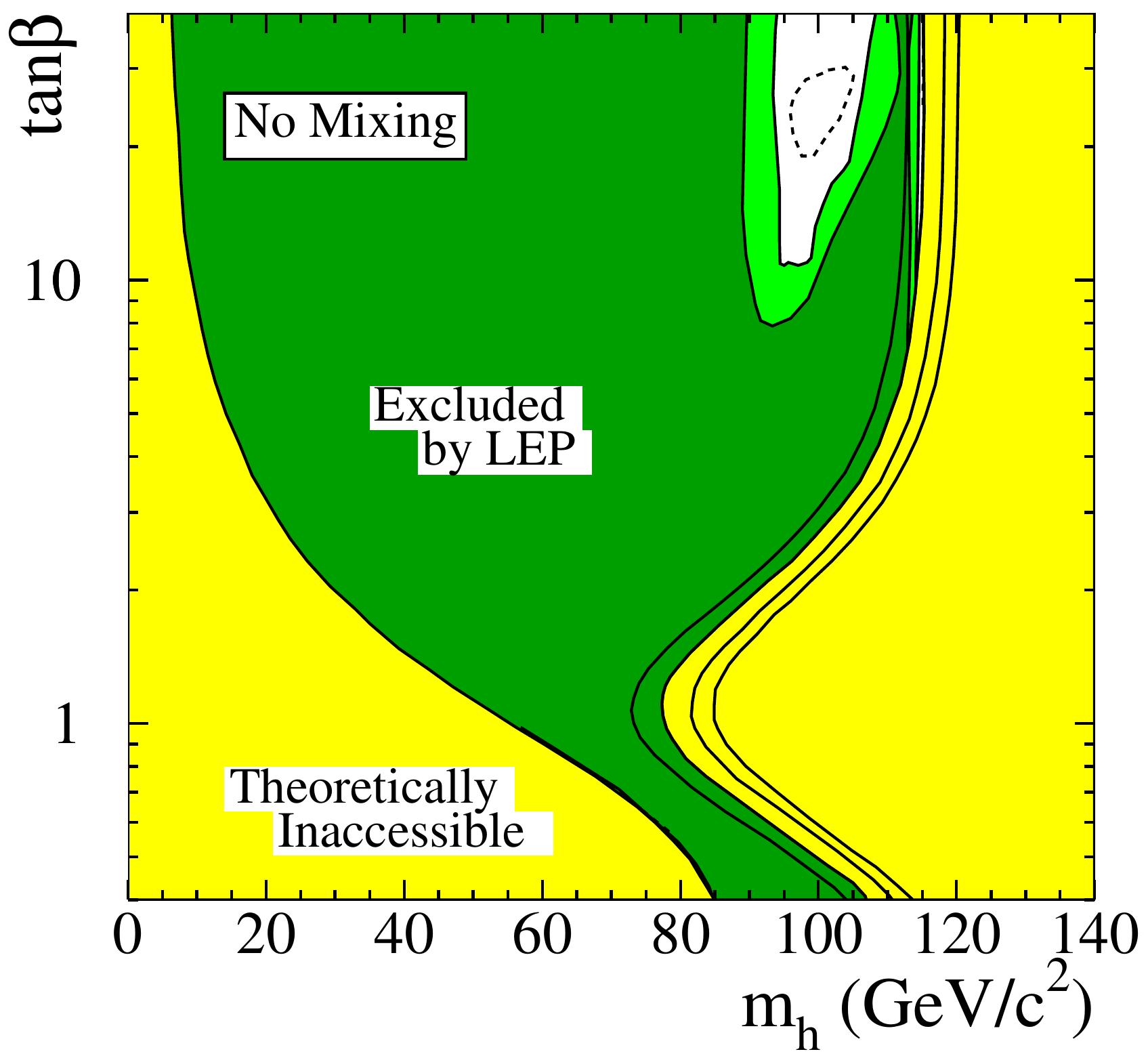}
\vspace{.08in}
\caption{\label{mhtb} Domains excluded at 95\% CL (light green) and
99.7\% CL (dark green) by the four LEP
experiments~\cite{Schael:2006cr} in the $(m_h,\tan\beta)$ plane in the
$m_h$-max (top) and no-mixing (bottom) benchmark scenarios, with
$m_t=174.3$~GeV. The yellow regions are not accessible
theoretically. The dashed lines represent the boundaries of the
domains expected to be excluded at 95\% CL in the absence of
signal. The upper boundaries of the physical regions are indicated for
four top quark masses: 169.3, 174.3, 179.3, and 183~GeV, from left to
right.}
\end{figure}

No significant excess over the SM background expectation was observed,
and production cross section upper limits were derived as a function
of $m_h \simeq m_A$. For each benchmark scenario, a scan was performed
as a function of $m_A$ and $\tan\beta$, and in each point of the scan
the cross section upper limit was compared to the corresponding
prediction, taking into account the slight modifications expected for
the values of the $h$ and $A$ branching fractions into $b\bar b$ and
$\tau^+\tau^-$, as well as the non-negligible difference between $m_h$
and $m_A$ which develops at lower values of $\tan\beta$. If the cross
section upper limit was found to be smaller than the prediction, the
$(m_A,\tan\beta$) set was declared excluded. The result of the
combination of the searches in the $hZ$ and $hA$ channels by the four
LEP experiments~\cite{Schael:2006cr} is shown in Fig.~\ref{mhtb},
projected onto the $(m_h,\tan\beta)$ plane in the $m_h$-max and
no-mixing scenarios. In the derivation of those results, contributions
of the $e^+e^-\to HZ$ and $HA$ processes were also taken into account
whenever relevant, where $H$ is the heavier CP-even Higgs boson.

In the most conservative scenario, i.e., $m_h$-max, it can be seen in
Fig.~\ref{mhtb} that the lower limit on the mass of the SM Higgs boson
holds also for $m_h$ as long as $\tan\beta$ is smaller than about 5,
and that values of $\tan\beta$ between $\simeq 0.7$ and 2 are excluded
for the current average value of the top quark mass, $173.1 \pm
1.3$~GeV~\cite{:2009ec}. A lower mass limit of 93~GeV is obtained for
$m_h \simeq m_A$ for large values of $\tan\beta$.

The benchmark scenarios were chosen such that the Higgs bosons do not
decay into SUSY particles. An interesting possibility is that the
$h\to\tilde\chi^0_1\tilde\chi^0_1$ decay mode is kinematically
allowed, where $\tilde\chi^0_1$ is the LSP. If $R$-parity is
conserved, the LSP is stable and, since it is weakly interacting, the
Higgs boson decay final state is invisible. Searches for such an
``invisible'' Higgs boson were performed by the LEP experiments, and
the combination~\cite{:2001xz} yields a mass lower limit identical to
that set on the SM Higgs boson if the production cross section is the
SM one, as is the case for low values of $\tan\beta$.

To cope with fine-tuned choices of MSSM parameters, the LEP
collaborations considered yet other possibilities, e.g., that the
$h\to AA$ decay mode is kinematically allowed, or that the $h\to b\bar
b$ decay is suppressed.  For example, dedicated searches for $hA\to
AAA\to b\bar b b\bar b b\bar b$ and for $hZ$, with $h\to q\bar q$ in a
flavor-independent way, have been performed~\cite{:2001yb}.  In the
end, the sensitivity of the standard searches is only slightly
reduced, except for rather extreme parameter choices leading, for
instance, to $m_h \simeq 100$~GeV, while at the same time $m_A <
2m_b$. This last possibility is however less unnatural in extensions
of the MSSM, such as the NMSSM where an additional Higgs singlet field
is introduced~\cite{Dermisek:2005gg}.

Finally, the possibility that CP is violated in the Higgs sector has
also been considered. While CP is conserved at tree level, radiative
corrections may introduce such a CP violation if the relative phase of
$\mu$ and $A$ is not vanishing. In such a case, the three mass
eigenstates all share properties of $h$, $H$ and $A$, so that the
signatures of Higgs boson production are less distinct. The
constraints are accordingly weaker. A dedicated ``CPX''
scenario~\cite{Carena:2000yi,Carena:2000ks} was set up to perform
quantitative studies. As an example, a region around $m_h=45$~GeV and
$\tan\beta=5$ is not excluded for $M_{SUSY}=500$~GeV, $M_2=200$~GeV,
$\mu=2$~TeV, and $m_{\tilde g}=1$~TeV, when $\vert A\vert=1$~TeV and
arg$(A)=90^\circ$. Further details can be found in
Ref.~\cite{Schael:2006cr}.

\subsection{Searches at the Tevatron}
\label{subsec:TevHiggs}

At the Tevatron, i.e., in $p\bar p$ collisions at 1.96~TeV, the
dominant production mechanism for the SM Higgs boson is via gluon
fusion, $gg\to H$~\cite{Wilczek:1977zn,Georgi:1977gs}.  In the mass
range that is of interest for a SM-like Higgs boson of the MSSM,
namely $m_h < 135$~GeV, the dominant decay mode is $H\to b\bar
b$. Such a two-jet final state is totally overwhelmed by standard jet
production via the strong interaction, even after $b$-jet
identification. This is why the SM Higgs boson searches at the
Tevatron have been performed in the associated production processes
$q\bar q\to (W/Z)H$~\cite{Glashow:1978ab}, which proceed via
$s$-channel $W$ or $Z$ exchanges in a similar way to the
Higgsstrahlung in $e^+e^-$ collisions. In spite of cross sections an
order of magnitude smaller than that of gluon fusion, these processes
offer better discrimination against the multijet background, by making
use of the leptonic decays of the $W$ and $Z$ ($W\to\ell\nu$,
$Z\to\ell^+\ell^-$ and $Z\to\nu\bar\nu$).  These searches for the SM
Higgs boson apply equally well for the $h$ boson of the MSSM in the
low $\tan\beta$ regime.  Their sensitivity is, however, still not
sufficient to provide any significant constraint.

The situation is much more favorable for large values of
$\tan\beta$. In this regime, the $A$ boson is almost mass degenerate
with either the $h$ or $H$ boson, depending on whether $m_A$ is less
than or greater than $m_h^{\text{max}}$, where $m_h^{\text{max}}$ is
the maximum value that $m_h$ can take, e.g., 135~GeV in the $m_h$-max
scenario.  In the following, the two nearly degenerate Higgs bosons
are collectively denoted $\phi$.  Their couplings to $b$ quarks and
$\tau$ leptons are enhanced by a factor $\tan\beta$ with respect to
the SM couplings. As a result, the contribution of the $b$ quark loop
to their production via gluon fusion is enhanced by a factor
$2\tan^2\beta$. Although this is not sufficient to render feasible a
detection in the $\phi\to b\bar b$ decay mode, this is not the case
for the $\phi\to\tau^+\tau^-$ decay mode, which has a branching
fraction of $\simeq 10\%$.

\begin{figure}
\includegraphics[width=8.5cm]{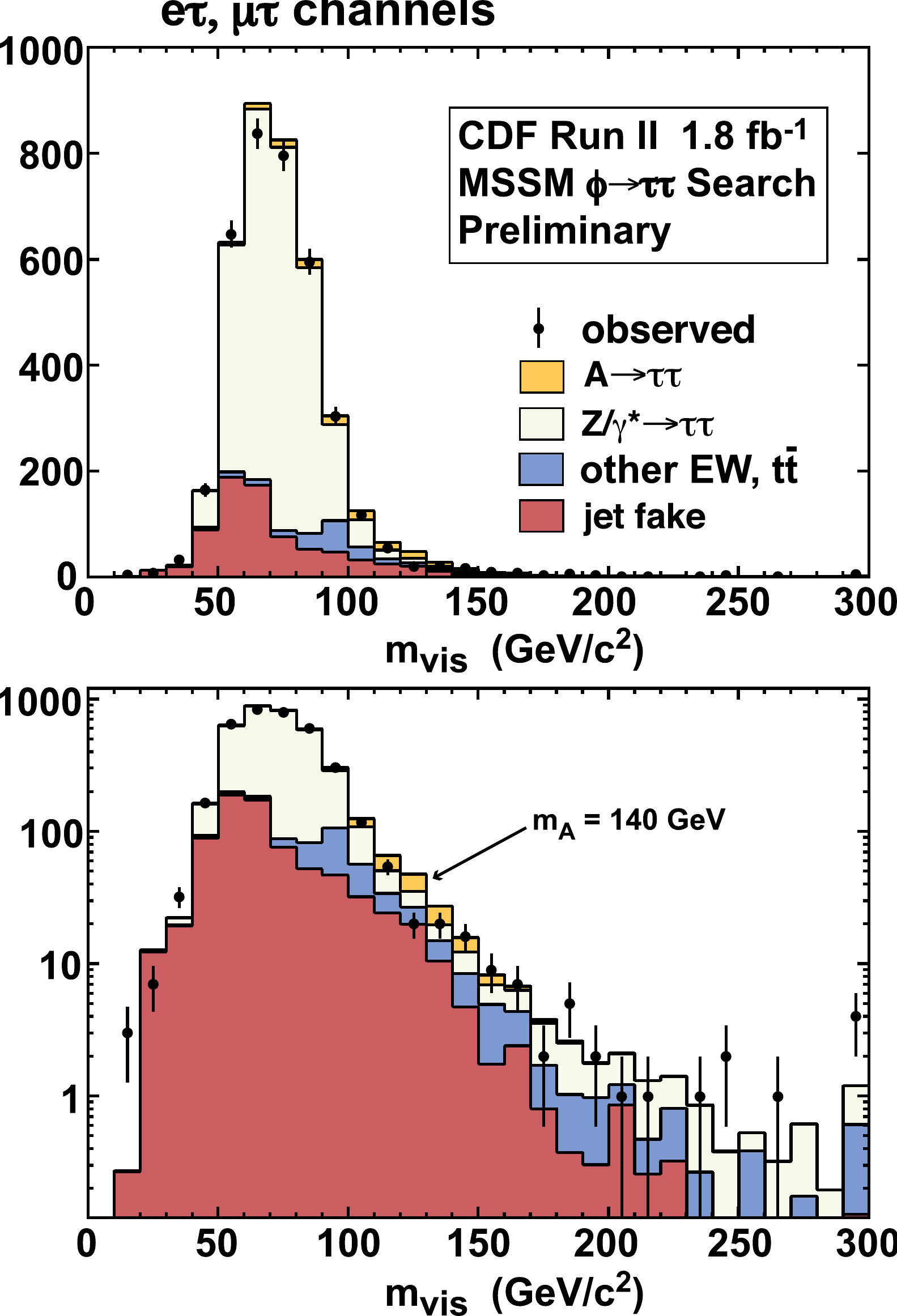}
\caption{\label{CDFttmvis} Visible mass distribution in the
$(e/\mu)\tau_{\text{had}}$ channels from the CDF search for
$\phi\to\tau\tau$~\cite{CDFtautau}.  The signal contribution indicated
corresponds to the cross section upper limit set with this data.}
\end{figure}

\begin{figure}
\includegraphics[width=8.5cm]{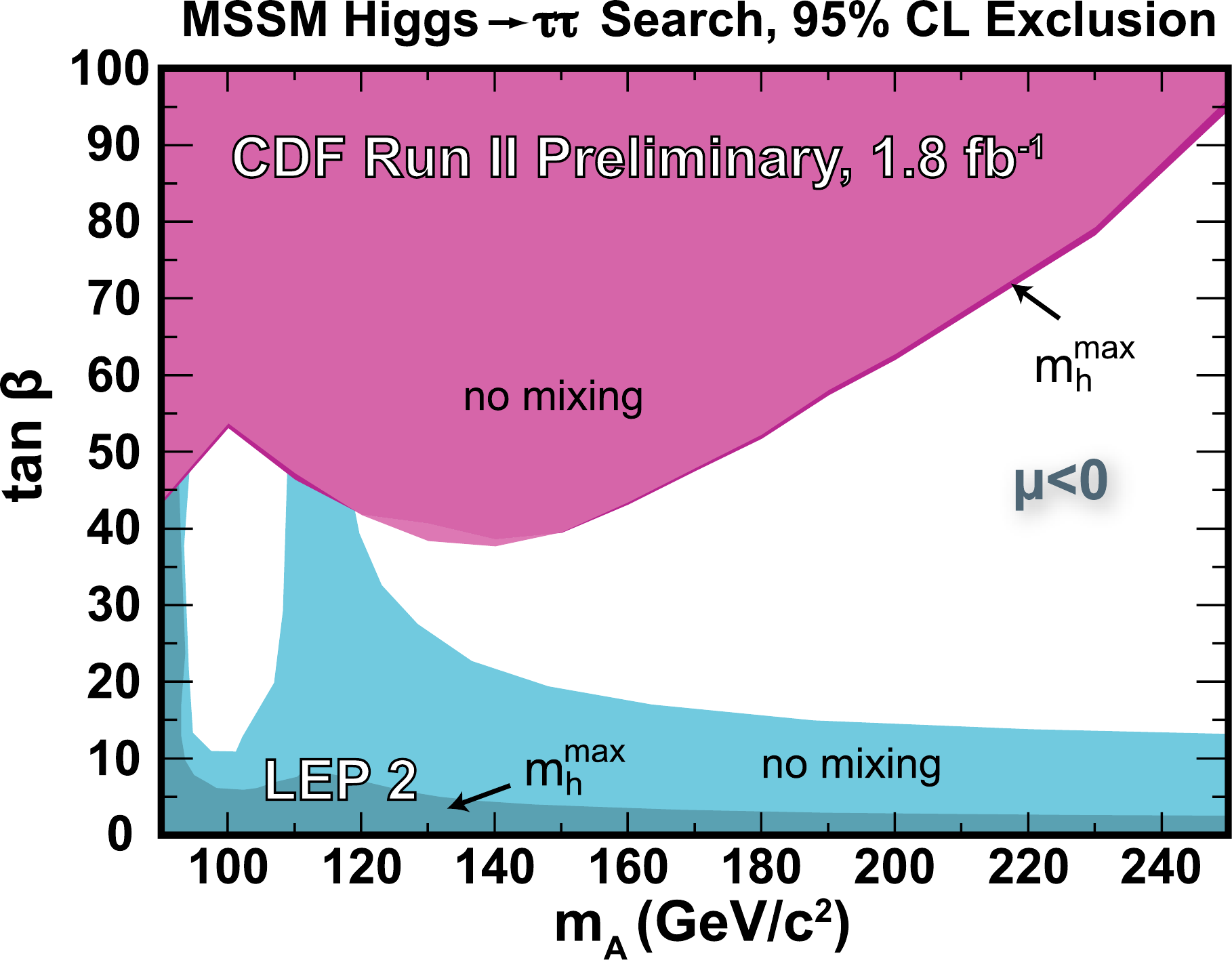}
\caption{\label{CDFttmAtb} Domains in the $(m_A,\tan\beta)$ plane
excluded by the CDF search for $\phi\to\tau\tau$~\cite{CDFtautau}. The
domains excluded at LEP are also indicated.}
\end{figure}

Both CDF and D\O\ required one of the two $\tau$ leptons to decay
leptonically ($\tau\to (e/\mu)\nu\nu$) to ensure proper
triggering. Three final state topologies were considered:
$e\tau_{\text{had}}$, $\mu\tau_{\text{had}}$, and $e\mu$, all with
missing transverse energy \met\ from the $\tau$ decay neutrinos. Here
$\tau_{\text{had}}$ denotes a $\tau$ lepton decaying into hadrons and
a neutrino.  The dominant, irreducible background comes from $Z$
production with $Z\to\tau^+\tau^-$, but there also remains a
substantial component from $(W\to\ell\nu)$+jet, where the jet is
misidentified as a $\tau$ lepton. This background was reduced, for
instance, by requiring a low transverse mass of the lepton and the
\met. The final discriminating variable was chosen to be the visible
mass $m_{\text{vis}}= \sqrt{(P_{\tau_1}+P_{\tau_2}+\pet)^2}$,
constructed from the $\tau$ visible products and from the \met. The
distribution of $m_{\text{vis}}$ obtained by CDF~\cite{CDFtautau} in a
1.8~\invfb\ data sample is shown in Fig.~\ref{CDFttmvis}. From this
distribution, as well as from a similar one in the $e\mu$ channel, a
cross section upper limit on $\phi$ production was derived, which in
turn was translated into exclusion domains in the $(m_A,\tan\beta)$
plane within benchmark scenarios. The result obtained in the $m_h$-max
and no-mixing scenarios is shown in Fig.~\ref{CDFttmAtb}.  Similar
results have been obtained by D\O~\cite{DZtautau}.  The calculations
of Ref.~\cite{Hahn:2006my} were used to derive these results as well
as those reported in the rest of this section.

\begin{figure}
\includegraphics[width=8.5cm]{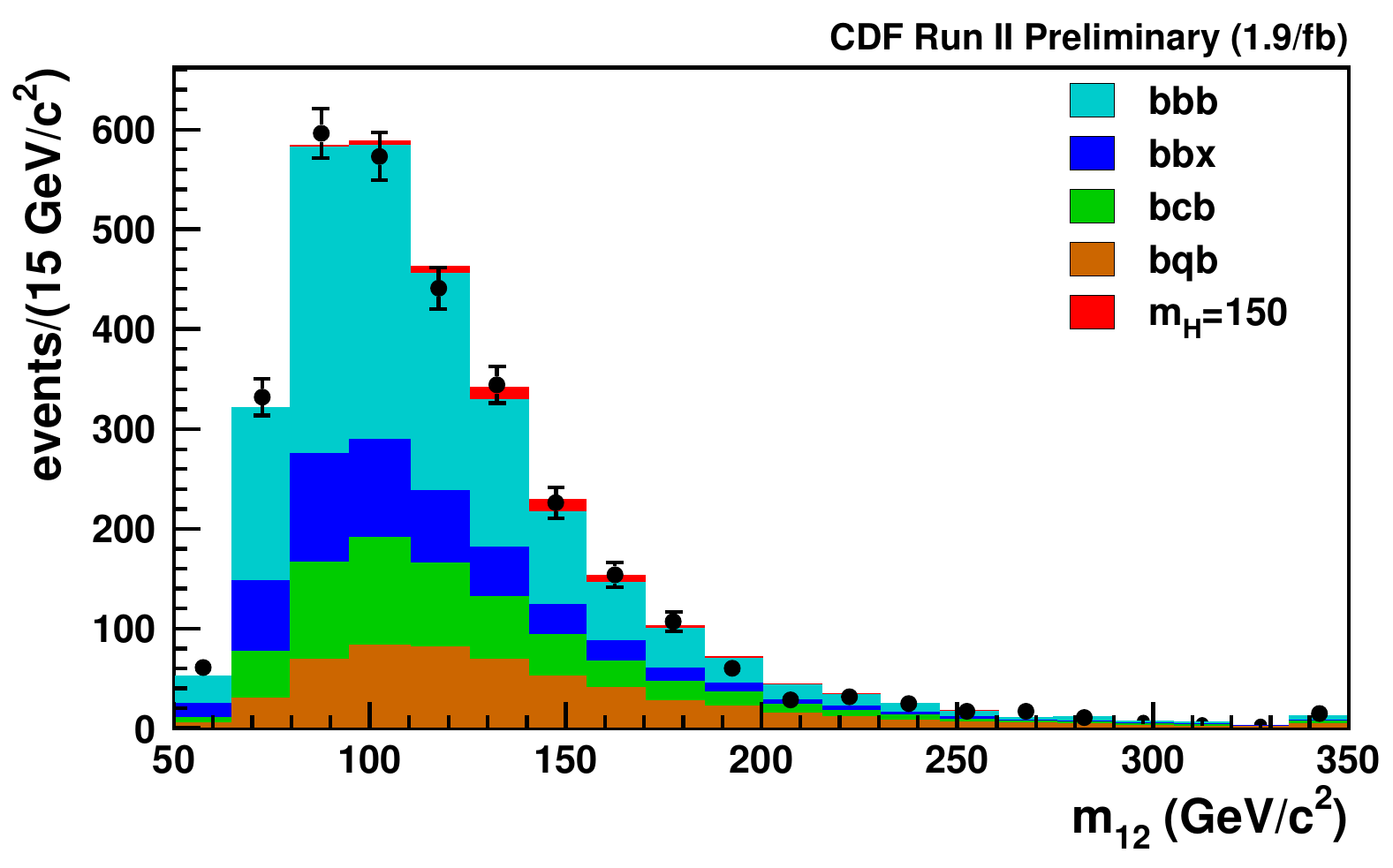}
\caption{\label{CDFbbbmbb} 
Fit to the mass of the two jets with highest transverse momenta in the
CDF sample of events with three $b$-tagged jets~\cite{CDFbbb}.  The
contributions of the various multijet backgrounds and of a signal with
a mass of 150~GeV are indicated.}
\end{figure}

\begin{figure}
\includegraphics[width=8.5cm]{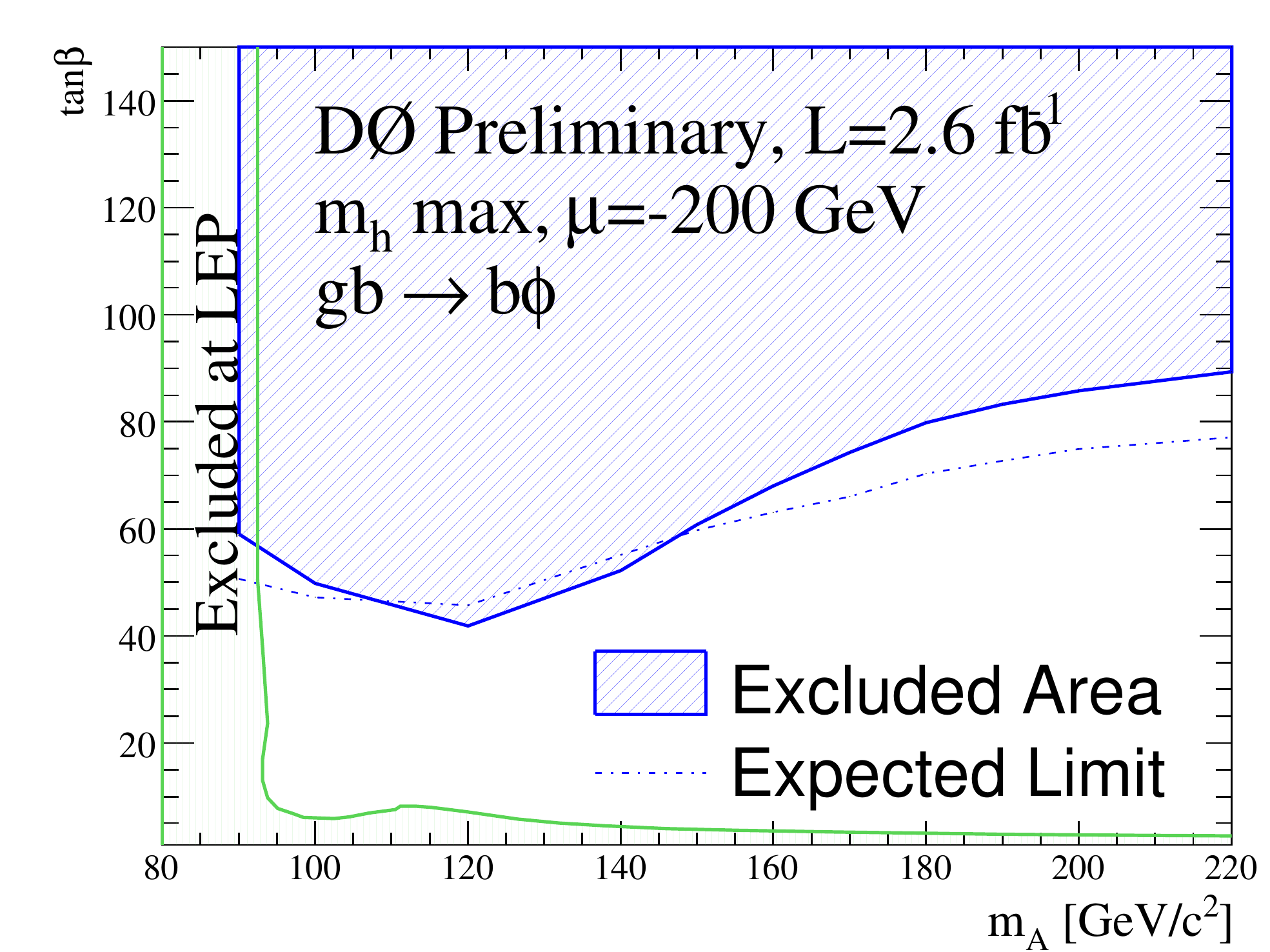}
\caption{\label{DZbbbmAtb}
Domain in the $(m_A,\tan\beta)$ plane excluded by the D\O\ search for
$\phi\to b\bar b$ in events with three $b$-tagged jets in the
$m_h$-max scenario~\cite{DZbbb}.}
\end{figure}

Because of the enhanced coupling of $\phi$ to $b$ quarks at high
$\tan\beta$, the production of Higgs bosons radiated off a $b$ quark
may be detectable in the $\phi\to b\bar b$ decay mode in spite of the
large background from multijet events produced via the strong
interaction (``QCD background'').  This process can be described in
the so-called four-flavor or five-flavor schemes, and it has been
shown that the two approaches yield very similar
results~\cite{Campbell:2004pu}.  In the four-flavor scheme, the main
contribution comes from gluon fusion, $gg\to b\bar b\phi$, while the
main one in the five-flavor scheme comes from $gb\to b\phi$. Because
one of the final state $b$ quarks (a spectator $b$ quark in the
five-flavor scheme) tends to be emitted with a low transverse
momentum, the searches required only three $b$ jets to be identified.
The signal was searched for by inspecting the mass distribution of the
two jets with highest transverse momenta in the sample of events with
three $b$-tagged jets.  Further discrimination against the QCD
background was provided by the mass of the charged particles in the
tagged jets (at CDF~\cite{CDFbbb}) or by the inclusion of additional
kinematic variables in a likelihood discriminant (at
D\O~\cite{DZbbb}). The QCD background was modeled using a combination
of information from control samples in the data, where one of the jets
is not $b$-tagged, and from Monte Carlo simulations of the various
processes contributing to the background ($bbb$, $bbc$, $bbq$, $ccc$,
$ccq$, etc., where $q$ represents a light quark, $u$, $d$, $s$, or a
gluon).  The mass distribution obtained by CDF in a 1.9~\invfb\ data
sample is shown in Fig.~\ref{CDFbbbmbb}, with the individual
background contributions displayed. No signal was observed, and
production cross section upper limits were derived, from which
exclusion domains in the $(m_A,\tan\beta)$ plane were determined in
various benchmark scenarios. The D\O\ result obtained with 2.6~\invfb\
of data in the $m_h$-max scenario is shown in Fig.\ref{DZbbbmAtb}.  In
the derivation of the cross section upper limits and exclusion
domains, special attention was given to a proper handling of the Higgs
boson width, which is enhanced by a factor $\tan^2\beta$ at tree level
and therefore becomes large with respect to the mass resolution. (This
is not the case for $\phi\to\tau^+\tau^-$ because of the degradation
of the mass resolution due to the missing neutrinos.). It should also
be noted that the exclusion domain is quite sensitive to the model
parameters. It is smaller in the no-mixing scenario, and also if $\mu$
is positive.  These effects due to potentially large SUSY loop
corrections to the production cross sections and decay widths tend to
cancel in the search for $\phi\to\tau^+\tau^-$ described
above~\cite{Carena:1998gk,Carena:1999bh}.
 
\begin{figure}
\includegraphics[width=8.5cm]{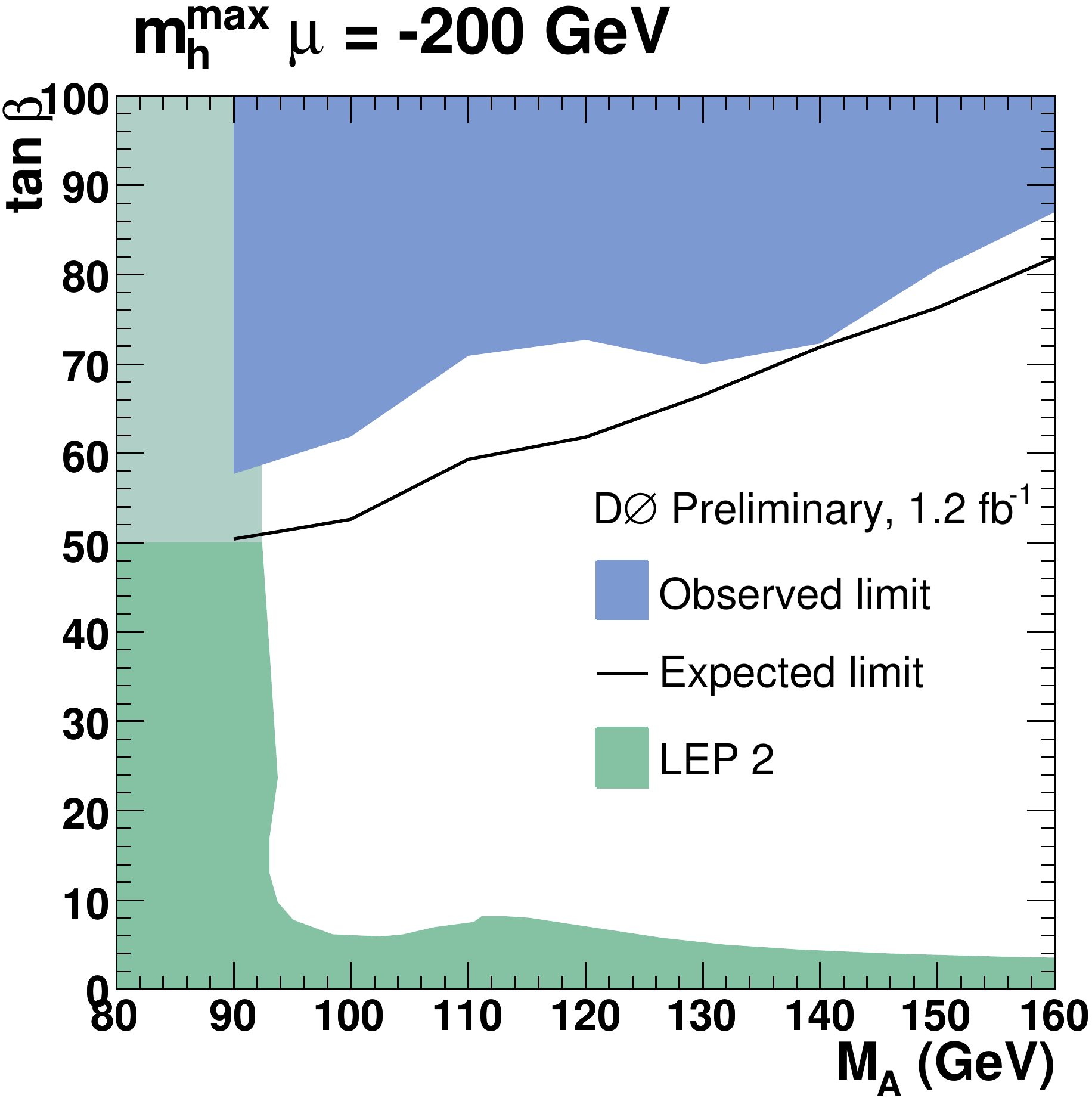}
\caption{\label{DZbttmAtb}
Domain in the $(m_A,\tan\beta)$ plane excluded by the D\O\ search for
$\phi\to\tau^+\tau^-$ produced in association with a $b$ quark in the
$m_h$-max scenario~\cite{DZbtt}. The light green shaded region is the
extension of the LEP exclusion region to $\tan \beta > 50$.}
\end{figure}

Finally, Higgs bosons produced in association with $b$ quarks can also
be searched for in the $\phi\to\tau^+\tau^-$ decay mode. Although the
branching fraction is an order of magnitude smaller than the one of
$\phi\to b\bar b$, the signal is much easier to disentangle from the
background.  A D\O\ analysis~\cite{DZbtt} was performed where one of
the $\tau$ leptons decays into a muon and neutrinos, while the other
decays into hadrons and a neutrino. Furthermore, a $b$-tagged jet was
required, at which point the main background comes from top quark pair
production, $t\bar t\to \mu\nu b\tau\nu \bar b$. A neural network was
used to discriminate signal and $t\bar t$ background, taking advantage
of the large differences in their kinematic properties. The result,
based on 1.2~\invfb\ of data, is shown in Fig.~\ref{DZbttmAtb}. Given
the limited amount of integrated luminosity used up to now, this
channel appears to be quite promising.

\section{Searches for charged Higgs bosons}
\label{sec:charged_higgs}

Many extensions of the SM involve more than one complex doublet of
Higgs fields.  Two-Higgs doublet models (2HDMs) fall into three main
categories.  In Type I models, all quarks and leptons couple to the
same Higgs doublet.  In Type II models, down-type fermions couple to
the first Higgs doublet, and up-type fermions couple to the second
Higgs doublet.  Flavor-changing neutral currents are naturally avoided
in Type I and Type II 2HDMs.  In Type III models, fermions couple to
both doublets, and flavor-changing neutral currents must be avoided
using other strategies.  In addition to the three neutral Higgs bosons
discussed in the previous section, 2HDMs involve a pair of charged
Higgs bosons, $H^\pm$.  Most of the experimental results on charged
Higgs bosons have been obtained within the context of Type II 2HDMs,
of which the MSSM is a specific instance.  Further details on extended
Higgs boson sectors may be found in
Ref.~\cite{Gunion:1989we,Gunion:1992hs}.

In Type II 2HDMs, the charged Higgs boson decay width into a fermion
pair $\bar{f}_u f_d$ is
\begin{eqnarray}
\Gamma ( H^- \to \bar{f}_u f_d ) &=& 
\frac{N_c g^2 m_{H^{\pm}}}{32 \pi m_W^2} 
 \left(1-\frac{m_{f_u}^2} {m_{H^{\pm}}^2} \right)^2
\nonumber \\
&\times & \left( 
m_{f_d}^2 \tan^2\beta + m_{f_u}^2 \cot^2 \beta \right) , 
\end{eqnarray}
where $N_c$ is the number of colors, and we have approximated $m_{f_d}
\ll m_{H^{\pm}}$ in the phase space factor.  Charged Higgs bosons
therefore decay into the heaviest kinematically-allowed fermions:
$\tau^-\bar\nu_{\tau}$ at large $\tan\beta$ and $\bar{c}s$ at low
$\tan\beta$ for charged Higgs boson masses to which current
accelerators are sensitive.

\subsection{Searches at LEP}

At LEP, charged Higgs bosons are produced in pairs through $e^+e^-\to
H^+H^-$~\cite{Chang:1978ke}.  The production cross section depends
only on SM parameters and on the mass of the charged Higgs boson.  The
process $e^+ e^-\to H^+W^-$ has a significantly lower cross section.

The charged Higgs boson can decay into $c\bar{s}$ or
$\tau\nu_{\tau}$. In searches for Type I 2HDM Higgs bosons, the decay
$H^\pm\to A W^{\pm*}$~\cite{Akeroyd:1998dt} was also considered, as in
Refs.~\cite{Abdallah:2003wd,:2008be}.  The interpretation of the
search results generally assumed that Br($H^\pm\to\tau \nu_{\tau}$) +
Br($H^\pm\to q\bar{q}'$) = 1, where the dominant $q\bar{q}'$ flavors
are $c\bar{s}$, due to the Cabbibo suppression of $c\bar{b}$.  This
assumption leads to the consideration of three topologies for
pair-produced charged Higgs bosons: four jets from $H^+H^-\to c\bar{s}
\bar{c}s$, two jets, a $\tau$ lepton and missing energy from
$H^+H^-\to c\bar{s}\tau\bar\nu_\tau$ and two charge conjugate,
acoplanar\footnote{The acoplanarity angle is the angle between the
projections of the $\tau$ momenta on a plane transverse to the beam
axis. If this angle is less than $180^\circ$, the $\tau$ leptons are
said to be acoplanar.} $\tau$ leptons from
$H^+H^-\to\tau^+\nu_\tau\tau^-\bar\nu_\tau$.

Direct searches for pair production of charged Higgs bosons have been
published by all four LEP
experiments~\cite{Heister:2002ev,Abdallah:2003wd,Achard:2003gt,:2008be}.
Each topological analysis began with a general selection for the
expected number of jets and $\tau$ leptons, followed by more
sophisticated techniques.  The main difficulty in these analyses was
separating the signal from the nearly identical signature of $W^+W^-$
production; selection criteria usually included a mass-dependent
optimization.  Techniques such as linear discriminants, likelihood
estimators, and jet-flavor tagging were used in these analyses.  The
$H^+H^-\to\tau^+\nu_\tau \tau^-\bar{\nu}_\tau$ channel had additional
complexity due to the missing neutrinos, which removed the possibility
of reconstructing the $H^\pm$ candidate masses and of improved
discrimination from the equal-mass constraint.  However, final states
with $\tau$ leptons can benefit from extracting information about
their polarization; the $\tau^+$ lepton from a $H^+$ boson (a scalar)
is produced in a helicity state opposite to that of a $\tau^+$ lepton
from $W^+$ decay.

\begin{figure}
\includegraphics[width=8.5cm]{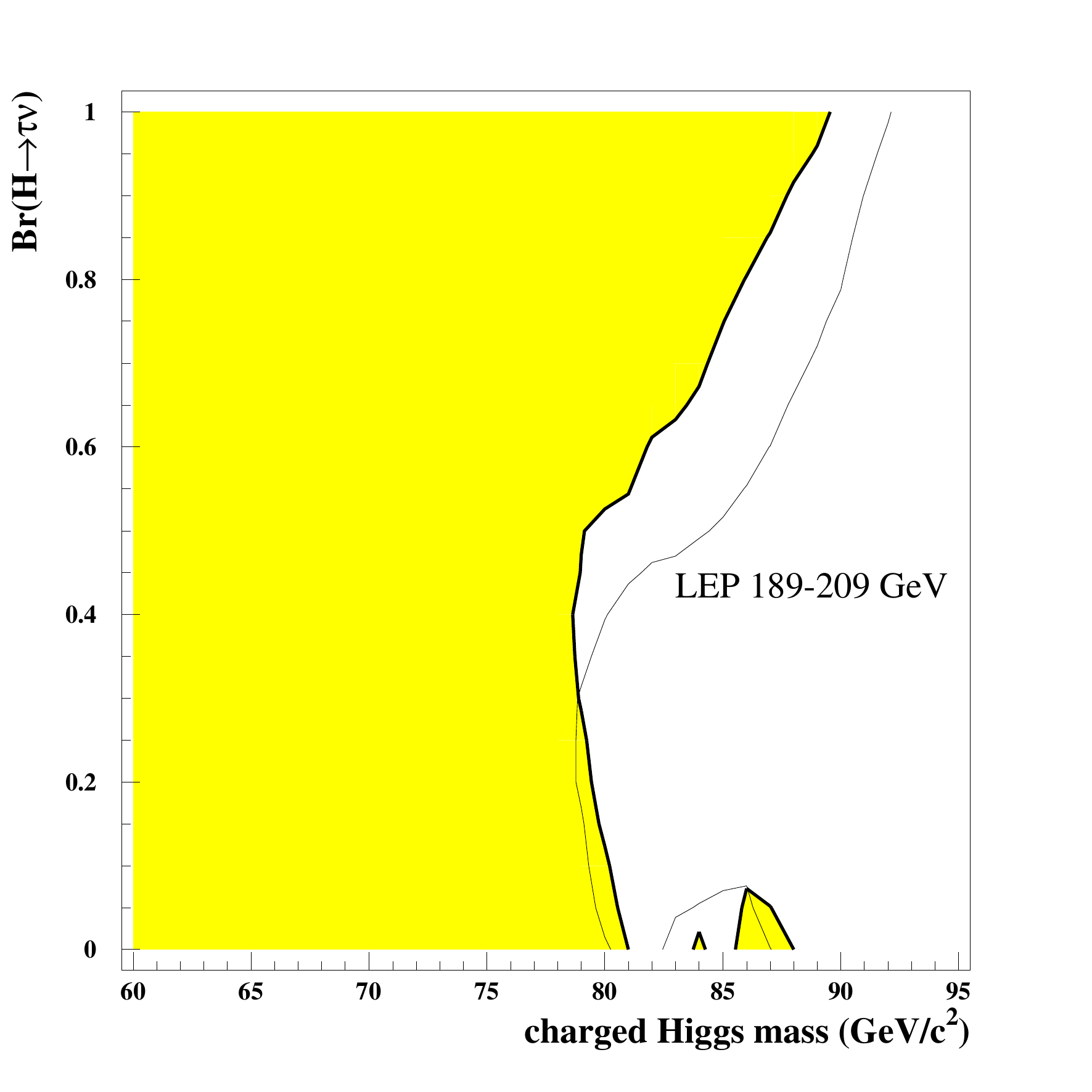}
\caption{\label{fig:lep-hpm-combo}
Limit on the charged Higgs boson mass as a function of
Br($H^+\to\tau^+\nu_\tau$), from the combined data of the four LEP
experiments at center-of-mass energies from 189 to 209 GeV.  The
expected exclusion limit is shown as a thin solid line and the
observed limit as a thick solid line; the shaded region is
excluded~\cite{:2001xy}.}
\end{figure}

The LEP experiments have combined the results of their searches for
charged Higgs bosons into one result~\cite{:2001xy} based on
common assumptions. The total dataset has an integrated luminosity of
2.5~\fbinv, ~collected at center-of-mass energies between 189 and
209~GeV.  The possible decays were restricted to $H^+\to c\bar{s}$ and
$\tau^+\nu_\tau$ in a general 2HDM framework.  The combined mass limit
is shown in Fig.~\ref{fig:lep-hpm-combo} as a function of
Br($H^+\to\tau^+\nu_\tau$).  A lower bound of 78.6~GeV holds for any
value of the branching ratio.

\subsection{Searches at the Tevatron}

At the Tevatron, pair production of charged Higgs bosons is expected
to occur at a very low rate.  However, in contrast to searches at LEP,
advantage can be taken of the large mass of the top quark, which opens
new ways to search for evidence of charged Higgs bosons. Two
approaches have been considered, depending on whether the charged
Higgs boson is lighter or heavier than the top quark.  In the first
case, the top quark can decay into a $H^+$ boson and a $b$
quark~\cite{Chang:1978ke}. For heavier charged Higgs bosons, resonant
production of a single $H^+$ boson followed by the decay $H^+\to
t\bar{b}$ is the most promising process~\cite{Gunion:1986pe}.
  
\begin{figure}
\includegraphics[width=8.5cm]{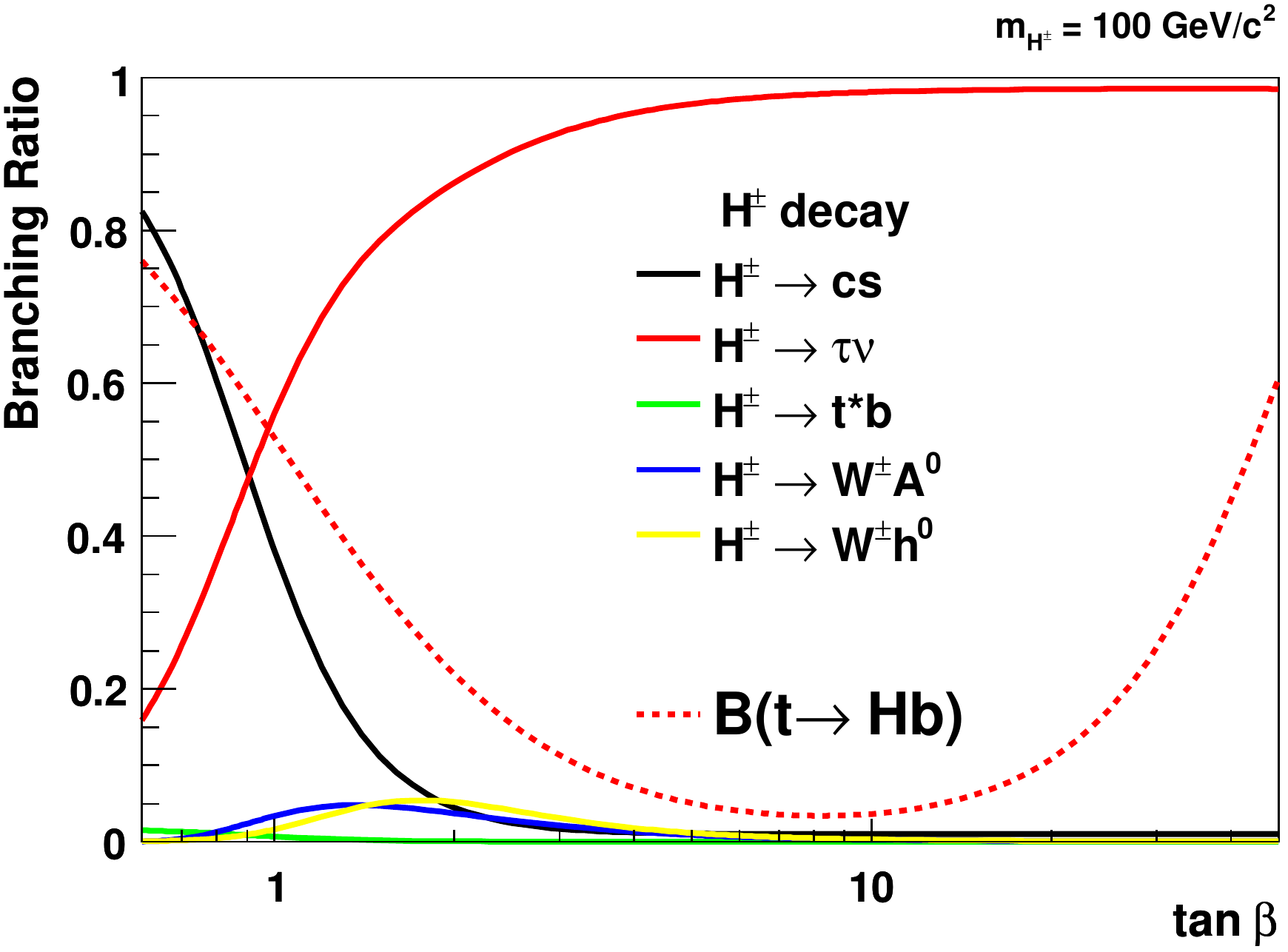}
\caption{\label{fig:tev-br}
For a charged Higgs boson mass of 100~GeV, the branching ratios for
the top quark decay into $H^+b$, under the assumption that Br($t\to
W^+b$)+Br($t\to H^+b$)=1, and for the various $H^+$ decay channels, as
a function of tan$\beta$.  From Ref.~\cite{ref:d0-h-from-t}.}
\end{figure}

In the SM, the top quark decays almost exclusively into a $W$ boson
and a $b$ quark, and the possible signatures of $t\bar{t}$ pair
production are associated with the various combinations of $W$-boson
decay channels. If the charged Higgs boson is lighter than the top
quark, the decay $t\to H^+b$ will compete with the standard $t\to
W^+b$ mode. The decay of the charged Higgs boson, with branching
ratios different from those of the $W$ boson, will modify the
fractions of events observed in the various topologies, compared to
the SM expectations.\footnote{The branching ratios for topologies
arising from SM $t\bar{t}$ pair production are roughly 50\% in six
jets; 14\% in each of $e$, $\mu$, and $\tau + \text{4 jets} +\met$;
1\% in each of $ee$, $\mu\mu$, and $\tau\tau +\text{2 jets} +\met$;
and 2\% in each of $e\mu$, $e\tau$, and $\mu\tau+ \text{2 jets}
+\met$.  In each of these channels, there are two $b$ jets.}  The
qualitative aspects and magnitude of these modifications depend on the
model parameters.  The dependence on $\tan\beta$ of the top quark
decay ratio to $H^+b$ and of the various charged-Higgs boson decay
channels is shown in Fig.~\ref{fig:tev-br} for $m_{H^+}=100$~GeV and a
typical set of MSSM parameters, with QCD, SUSY-QCD and electroweak
radiative corrections to the top and bottom quark Yukawa couplings
calculated with the {\sc CPsuperH} code~\cite{Lee:2003nta}.  The
dominant $H^+$ boson decay channels are $c\bar{s}$ at low values of
$\tan\beta$ and $\tau^+\nu_\tau$ at high values; with this set of
parameters, $H^+$ boson decays to $W^+A/h$ are also allowed, although
always at a small rate.  The $H^+\to t^\ast\bar{b}\to W^+b\bar{b}$
decay mode becomes relevant for charged Higgs boson masses closer to
the top quark mass. It can be seen that charged Higgs bosons will be
most prominent at high and low values of $\tan\beta$. Two simplified
models address each of these regions: the tauonic model, with
Br($H^+\to\tau^+\nu_\tau$)=1, and the leptophobic model, with
Br($H^+\to c\bar{s}$)=1.  The tauonic model is a very good
approximation to the MSSM with $\tan\beta \agt 15$, while purely
leptophobic charged Higgs bosons can be found in some
multi-Higgs-doublet models~\cite{Grossman:1994jb}.
 
\begin{figure}
\includegraphics[width=8.5cm]{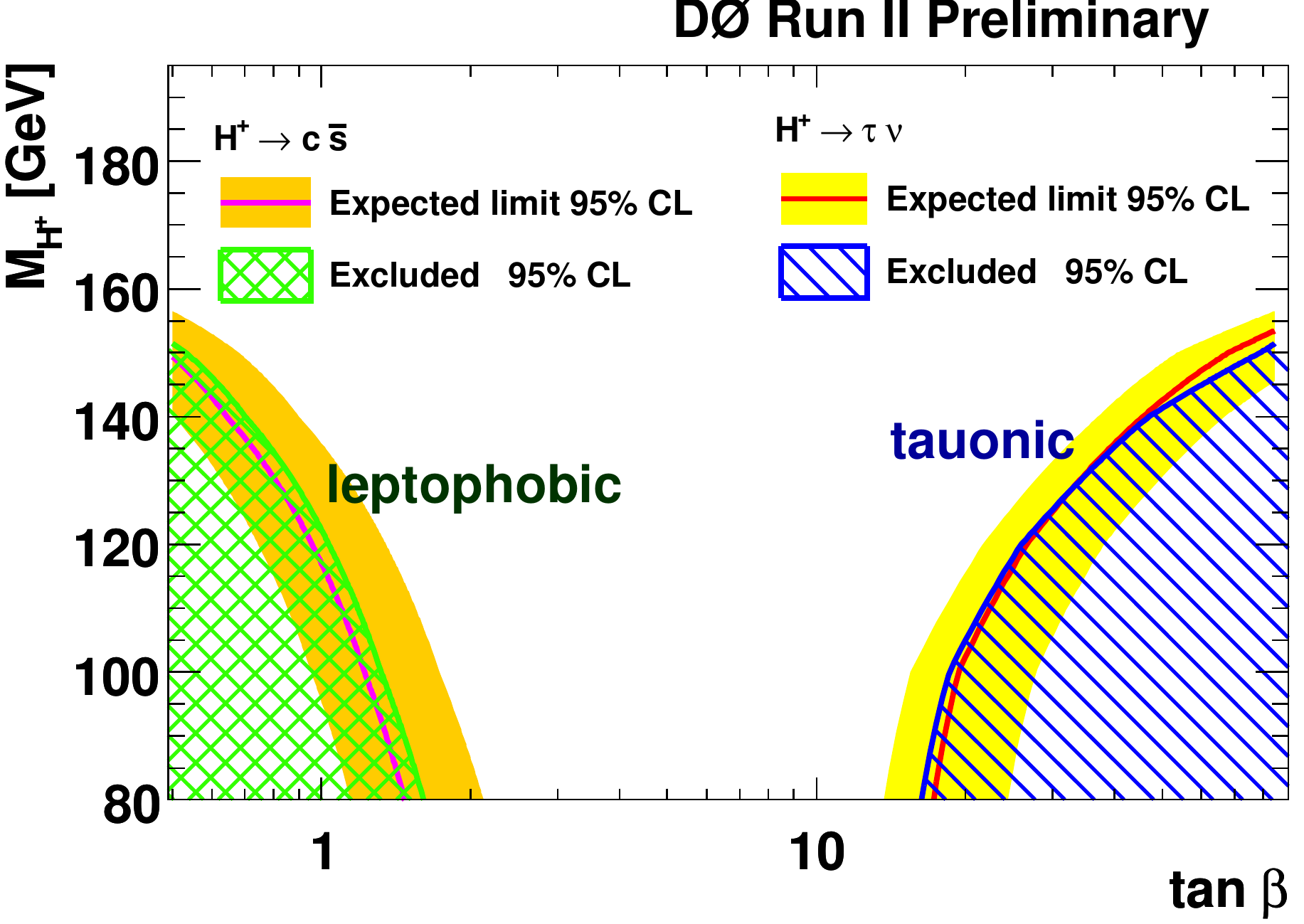}
\caption{\label{fig:d0-h-from-t}
Limit on the mass of the charged Higgs boson as a function of
$\tan\beta$ from the D\O\ search in top quark
decays~\cite{ref:d0-h-from-t}.}
\end{figure}

Analyses based on measurements of $t\bar{t}$ final states include an
earlier CDF search in $200~\pb^{-1}$ of data~\cite{Abulencia:2005jd}
and a recent D\O\ analysis of $1~\fb^{-1}$ of
data~\cite{ref:d0-h-from-t}.  The yields observed in the various
topologies were compared to what would be expected in models with
charged Higgs bosons, taking into account the $t\to H^+b$ and $H^\pm$
decay branching ratios predicted as a function of the Higgs boson mass
and $\tan\beta$.  In particular, no excess of final states involving
$\tau$ leptons was observed, nor was any disappearance of final states
with one or two leptons, jets and \met, as would be expected at large
and small $\tan\beta$, respectively.  Figure~\ref{fig:d0-h-from-t}
displays the exclusion domain in the plane of the charged Higgs boson
mass and $\tan\beta$ from the D\O\ analysis~\cite{ref:d0-h-from-t},
for leptophobic and tauonic models.  The CDF analysis excludes
Br($t\to H^+b$)~$>$~0.4 for a tauonic $H^\pm$
boson~\cite{Abulencia:2005jd}.

\begin{figure}
\includegraphics[width=8.5cm]{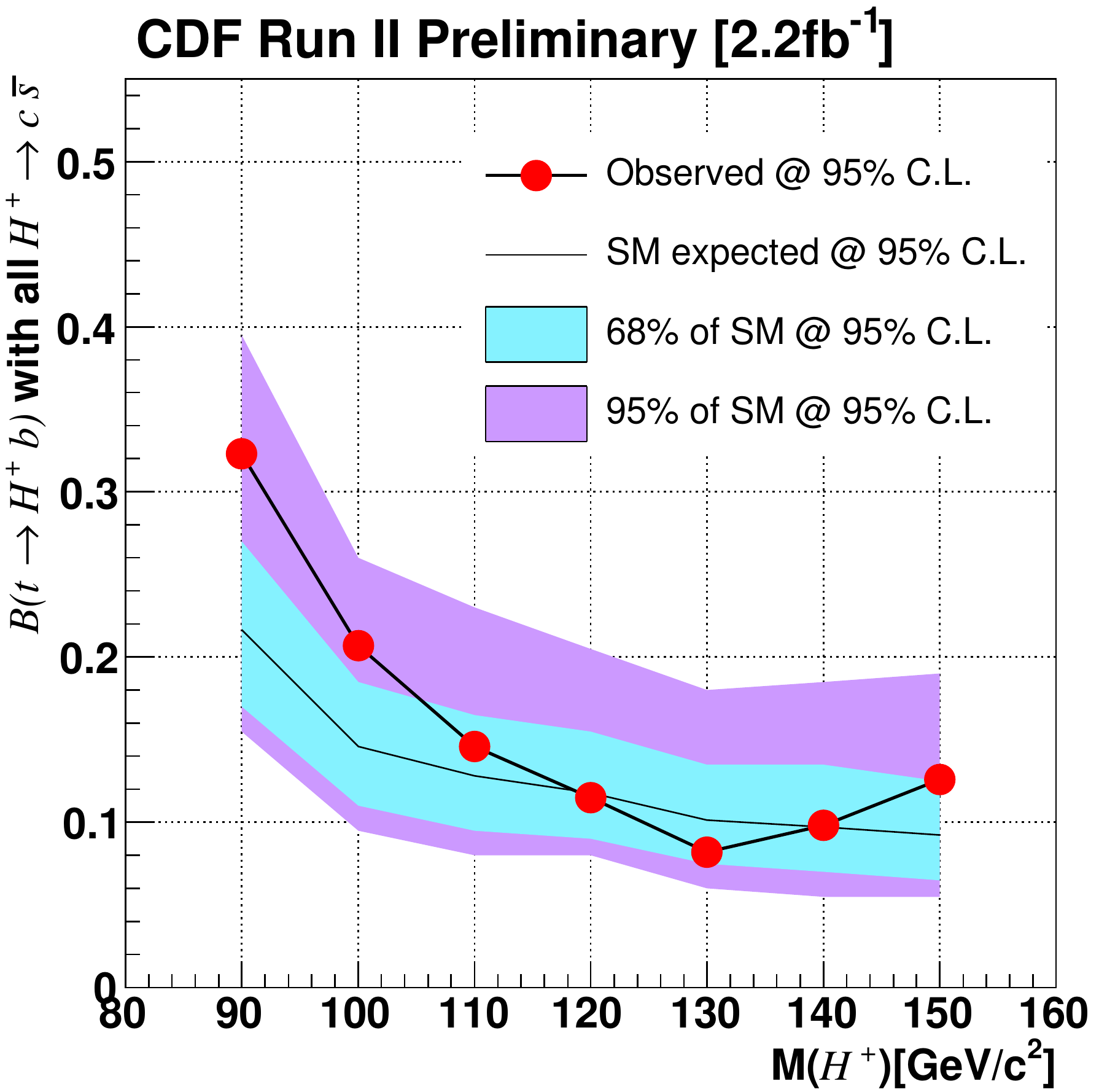}
\caption{\label{fig:cdf-h-from-t-2}
For a leptophobic charged Higgs boson, upper limit on the branching
ratio Br($t\to H^+ b$) as a function of the Higgs boson mass from a
CDF search in top quark decays~\cite{ref:cdf-h-from-t-2}.}
\end{figure}

In a recent analysis based on a data sample of 2.2~\fbinv, the CDF
collaboration used a different approach to search for a leptophobic
charged Higgs boson in top quark decays~\cite{ref:cdf-h-from-t-2}. The
search was performed in the $\text{lepton}+ \text{jets} +\met$ final
states with two $b$-tagged jets, where the lepton (electron or muon),
the neutrino (responsible for the missing \et), and a $b$ jet were the
signature of a $t\to Wb\to \ell\nu b$ decay, while the other top quark
of the $t\bar{t}$ pair was assumed to decay to either $Wb\to q\bar{q}'
b$ or $Hb\to c\bar{s}b$.  The $t\bar{t}$ events were fully
reconstructed, taking the masses of the $W$ boson and of the top quark
into account as constraints to assign correctly each of the $b$ jets
to its parent $t$ or $\bar{t}$.  Templates of the mass of the dijet
system reconstructed from the non-$b$ jets were used to extract limits
on the branching ratio of $t\to H^+ b$, as shown in
Fig.~\ref{fig:cdf-h-from-t-2}.

If the charged Higgs boson is heavier than the top quark, it will
decay dominantly into $t\bar{b}$.  The resonant production of such a
charged Higgs boson leads to a final state similar to the one
resulting from single top $s$-channel production, $q\bar{q}\to
W^\ast\to t\bar{b}$. Therefore the analyses developed for the search
for single top production can be applied to the search for a charged
Higgs boson.  Such an analysis was performed by the D\O\
collaboration~\cite{Abazov:2008rn}, in the topology arising from a
subsequent $t\to Wb\to\ell\nu b$ decay.  The large $H^\pm$ mass,
reconstructed from the decay products imposing the $W$ boson and top
quark mass constraints, was used as discriminating variable. No excess
was observed over SM background predictions, and upper limits were set
on the production of a charged Higgs boson. The results are, however,
not sensitive to Type II 2HDMs, but provide some exclusion in Type I
2HDMs.

\section{Searches for supersymmetric particles}
\label{sec:SUSYparticles}

\subsection{General features of SUSY models}
\label{subsec:modelsSUSY}

As explained in Sec.~\ref{sec:supersymmetry}, the main features of
SUSY models for phenomenology are related to the type of mediation
mechanism for SUSY breaking, to the choice of soft breaking terms, and
to whether or not $R$-parity is assumed to be conserved.

The most widely studied models involve gravity-mediation of SUSY
breaking.  In the minimal form of such models, mSUGRA, $R$-parity is
conserved, and only five parameters are needed beyond those already
present in the standard model: a universal gaugino mass $m_{1/2}$, a
universal scalar mass $m_0$, and a universal trilinear coupling $A_0$,
all defined at the scale of grand unification, and $\tan\beta$ and the
sign of $\mu$. The low energy parameters, including $|\mu|$, are
determined by the renormalization group equations and by the condition
of electroweak symmetry breaking. In addition, it is commonly assumed
that the LSP is the lightest neutralino $\tilde\chi^0_1$. A somewhat
less constrained model keeps $\mu$ and $m_A$ as independent low energy
parameters, which is in effect equivalent to decoupling the Higgs
scalar masses from the masses of the other scalars. Such a model was
largely used at LEP.

Many studies have been performed where the assumption of $R$-parity
conservation is dropped, while keeping unchanged the other features of
those mSUGRA inspired models. If $R$-parity is violated, the
superpotential is allowed to contain lepton or baryon number-violating
terms~\cite{Barger:1989rk}
\begin{equation}
W_{R_p} = 
\lambda_{ijk}L_iL_j\bar{E}_k + \lambda'_{ijk}L_iQ_j\bar{D}_k +
\lambda''_{ijk}\bar{U}_i\bar{D}_j\bar{D}_k \ , 
\end{equation}
where $L$ and $Q$ are lepton and quark doublet superfields, $E$ and
$D$ are lepton and down-type quark singlet superfields, and $i$, $j$
and $k$ are generation indices. These terms are responsible for new
couplings through which the LSP decays to SM particles. The
simultaneous occurrence of different coupling types is however
strongly constrained, e.g., by the bounds on the proton lifetime,
which is why it is commonly assumed that only one of the $R$-parity
violating terms is present in the superpotential.

In models with gauge-mediated SUSY breaking (GMSB), the LSP is a very
light gravitino $\tilde{G}$, and the phenomenology is governed by the
nature of the NLSP. In the minimal such model, mGMSB, all SUSY
particle masses derive from a universal scale $\Lambda$, and in most
of the parameter space the NLSP is either the lightest neutralino
$\tilde\chi^0_1$ or the lighter stau $\tilde\tau_R$, the latter
occurring preferentially at large $\tan\beta$. The couplings of the
gravitino depend on yet another parameter, the SUSY-breaking scale
$\sqrt{F}$, which can be traded for the lifetime of the NLSP.

Anomaly-mediation of SUSY breaking (AMSB) generically leads to a
neutralino LSP which is almost a pure wino $\tilde{W}^0$, and has a
small mass splitting with the lighter chargino. As a consequence, this
chargino may acquire a phenomenologically relevant lifetime, possibly
such that it behaves like a stable particle.

\subsection{Signatures and strategies}
\label{subsec:strategies}

Most of the searches for SUSY particles were performed within a
``canonical scenario,'' the main features of which are borrowed from
mSUGRA: $R$-parity conservation, universal gaugino mass terms, a
universal sfermion mass term, and a neutralino LSP. Because of
$R$-parity conservation, SUSY particles are produced in pairs, and
each of the produced SUSY particles decays into SM particles
accompanied by an LSP. Since the LSP is neutral and weakly
interacting, it appears as missing energy, which is the celebrated
signature of SUSY particle production.

Alternatively, if $R$-parity is not conserved, the LSP decays to SM
particles, so that no missing energy is expected beyond that possibly
arising from neutrinos. The signature of SUSY particle production is
therefore to be sought in an anomalously large multiplicity of jets or
leptons. The $R$-parity violating couplings can also make it possible
that SUSY particles are produced singly, rather than in pairs.

In $R$-parity conserving scenarios other than the canonical one,
additional or different features are expected. In GMSB, each of the
pair-produced SUSY particles decays into SM particles and an NLSP. The
NLSP further decays into its SM partner and a gravitino. With a
neutralino NLSP in the mass range explored up to now, the dominant
decay is $\tilde\chi^0_1\to\gamma\tilde{G}$, so that the final state
contains photons, with missing energy due to the escaping gravitinos.
With a stau NLSP, the decay is $\tilde\tau_R\to\tau\tilde{G}$. If the
stau lifetime is so long that it escapes the detector before decaying,
the final state from stau pair production does not exhibit any missing
energy, but rather appears as a pair of massive stable particles. A
similar final state may also arise from chargino pair production in
AMSB. Long-lived gluinos can lead to spectacular signatures if they
are brought to rest by energy loss in the detector material.

Except for the gluino, all SUSY particles are produced in a democratic
way in $e^+e^-$ collisions via electroweak interactions. It is
therefore natural that the searches at LEP were targeted toward the
lightest ones. The results of these searches could further be combined
within a given model, thus providing constraints on the model
parameters. In contrast, it is expected that the most copiously
produced SUSY particles in hadron collisions, such as $p\bar p$ at the
Tevatron, will be colored particles, namely squarks and gluinos. Their
detailed signature however depends on the mass pattern of the other
SUSY particles, which may be present in their decay chains. This is
why a specific model, usually mSUGRA, is needed to express the search
results in terms of mass constraints. Thanks to lower masses and more
manageable backgrounds, the search for gauginos produced via
electroweak interactions can be competitive at hadron colliders for
model parameter configurations where their leptonic decays are
enhanced.

In $e^+e^-$ collisions, the production cross sections of SUSY
particles are similar to those of their SM partners, except for the
phase space reduction due to their larger masses. The data collected
at the highest LEP energies, up to 209~GeV, are therefore the most
relevant for SUSY particle searches. Mixing effects may however reduce
these cross sections, as is the case for instance for neutralinos with
a small Higgsino component, in which case the large integrated
luminosity accumulated by the LEP experiments at lower energies also
contributes to the search sensitivity.

Although the center-of-mass energy of 1.96~TeV in $p\bar p$ collisions
at the Tevatron allows higher new particle masses to be probed, large
integrated luminosities are needed because of the rapid PDF fall off
at high $x$, as explained in Sec.~\ref{subsec:context}.  The search
for SUSY particles at the Tevatron is also rendered more challenging
than at LEP because of the large cross sections of the background
processes. In the searches for squarks and gluinos, signal production
cross sections of the order of 0.1~pb at the edge of the sensitivity
domain are to be compared to the total inelastic cross section of
80~mb. In the searches for gauginos, with similar signal production
cross sections in the mass range probed, the main backgrounds are
$W\to\ell\nu$ and $Z\to\ell\ell$, with cross sections at the 2.7~nb
and 250~pb level per lepton flavor.

In $ep$ collisions at HERA, the most promising SUSY
particle production process is single squark resonant production via
an $R$-parity violating $\lambda'_{1j1}$ or $\lambda'_{11k}$ coupling,
with a cross section depending not only on the squark mass, but also
on the value of the coupling involved. The decay of the squark
produced could be either direct, via the same $\lambda'$ coupling as
for its production, or indirect through a cascade leading to the LSP,
which in turn decays to two quarks and a neutrino or an electron. The
mass reach at HERA is the full center-of-mass energy of 320~GeV, but
the production of squarks with masses close to this bound involves
quarks at large $x$ values, so that the effective reach is
substantially smaller, even for large values of the $\lambda'$
coupling.
    
\subsection{Searches in the canonical scenario}
\label{subsec:canon}

As mentioned above, the characteristic signature of SUSY particle
production in the canonical scenario is missing energy carried away
from the detector by the LSPs at the end of the decay chains.

\subsubsection{Searches at LEP}
\label{subsubsec:cononLEP}

The main channels for SUSY particle searches in $e^+e^-$ collisions
are slepton~\cite{Farrar:1980uy,Bartl:1987zg},
chargino~\cite{Barger:1983wc,Bartl:1985fk} and
neutralino~\cite{Ellis:1983er,Dicus:1983fe,Bartl:1986hp}
production. Squark pair production~\cite{Almeida:1983df} can also be
relevant in some specific cases~\cite{Bartl:1996wt}.

\begin{figure}
\includegraphics[width=8.5cm]{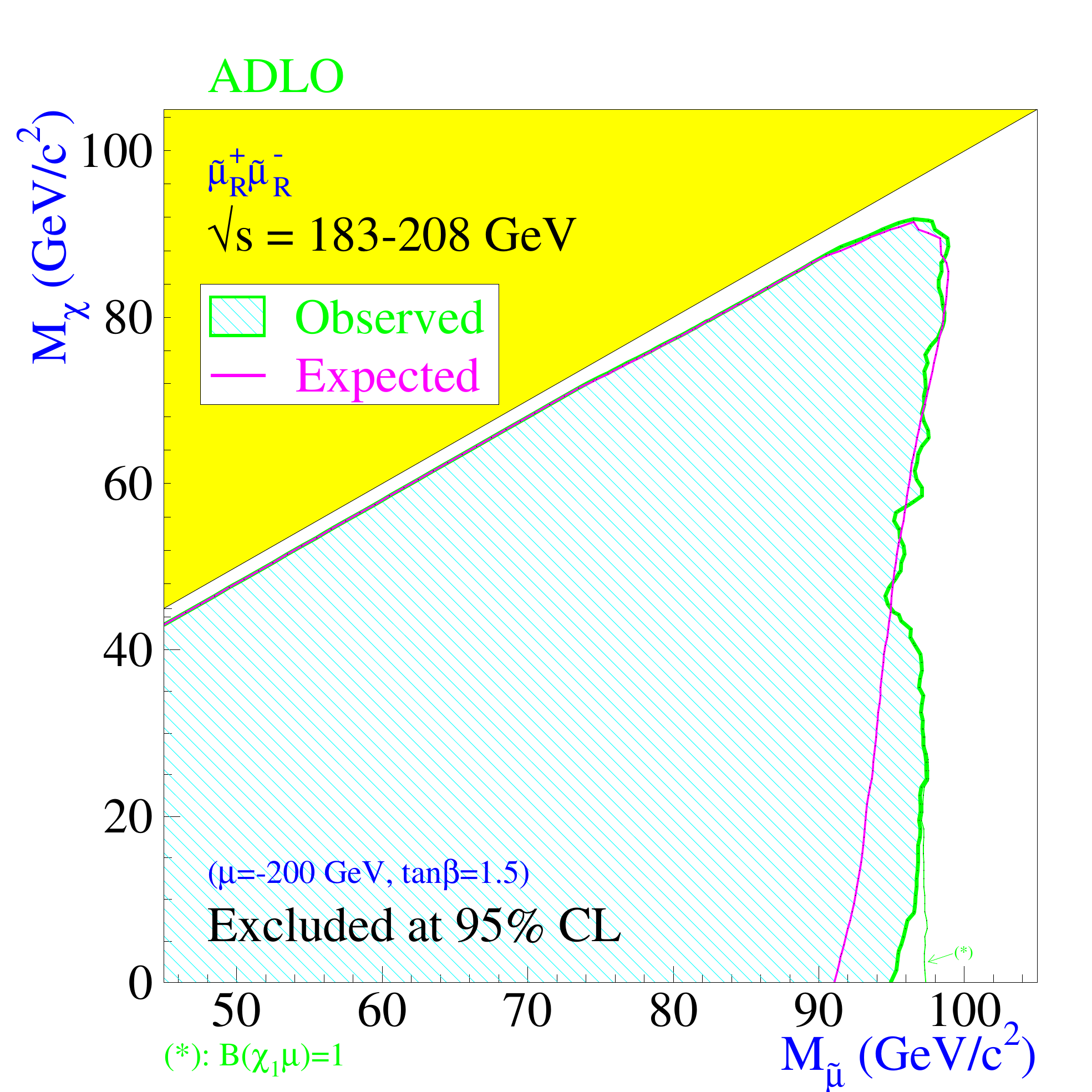}
\caption{\label{smuonLEP}
Region in the ($m_{\tilde\mu_R},m_{\tilde\chi^0_1}$) plane excluded by
the searches for smuons at LEP~\cite{lepslep}.  The dotted contour is
drawn under the assumption that the smuon decay branching ratio into
$\mu\tilde\chi^0_1$ is 100\%.
}
\end{figure}

{\bf Sleptons:} In $e^+e^-$ annihilation, the search for SUSY
particles that involves the least set of hypotheses for its
interpretation is the search for smuons. Pair production proceeds via
$Z/\gamma^\ast$ exchange in the $s$-channel. Because of the small mass
of the muon, the smuon mass eigenstates can be identified with the
interaction eigenstates, of which $\tilde\mu_R$ is the lighter one in
models with slepton and gaugino mass unification. The search results
were interpreted under this assumption, which is furthermore
conservative, as the coupling of the $\tilde\mu_R$ to the $Z$ boson is
smaller than that of the $\tilde\mu_L$. Only one parameter is needed
to calculate the smuon pair production cross section, the smuon mass
$m_{\tilde\mu_R}$. The sole decay mode of a $\tilde\mu_R$ NLSP is
$\tilde\mu_R\to\mu\tilde\chi_1^0$, so that smuon pair production leads
to a final state consisting of two acoplanar muons with missing energy
and momentum. The topology of this final state also depends on the
mass of the LSP. If $m_{\tilde\chi_1^0}$ is small, the final state is
very similar to that arising from $W$ pair production, with both $W$
bosons decaying to a muon and a neutrino. If the
$\tilde\mu_R-\tilde\chi_1^0$ mass difference is small, the final state
muons carry little momentum, so that the selection efficiency is
reduced. In that configuration, the main background comes from
``$\gamma\gamma$ interactions,''
$e^+e^-\to(e^+)\gamma^\ast\gamma^\ast(e^-)\to(e^+)\mu^+\mu^-(e^-)$,
where the spectator electrons $(e^\pm)$ escape undetected in the beam
pipe.  The LSP mass $m_{\tilde\chi_1^0}$ is therefore needed, in
addition to the smuon mass, to interpret the search results. The
constraints obtained in the $(m_{\tilde\mu_R},m_{\tilde\chi_1^0})$
plane by the four LEP experiments~\cite{lepslep} are shown in
Fig.~\ref{smuonLEP}.  If the assumption that the smuon is the NLSP is
dropped, further specification of the model is needed to turn the
search results into mass constraints. An example is shown in
Fig.~\ref{smuonLEP} in the case of gaugino mass unification, for the
specified values of $\mu$ and $\tan\beta$. A slight reduction of the
excluded domain is observed for low values of $m_{\tilde\chi_1^0}$,
due to the competition of the $\tilde\mu_R\to\mu\tilde\chi_2^0$ decay
mode, with $\tilde\chi_2^0\to\gamma\tilde\chi_1^0$. Depending on
$m_{\tilde\chi_1^0}$, smuon masses smaller than 95 to 99~GeV are
excluded, except for $\tilde\mu_R-\tilde\chi_1^0$ mass differences
below 5~GeV.

Because of the larger $\tau$ mass, compared to the muon mass, the
hypothesis that the stau mass eigenstates can be identified with the
interaction eigenstates may not hold, especially for large values of
$\tan\beta$ that enhance the off-diagonal elements of the mass matrix
in \eqref{staumass}. The coupling to the $Z$ boson of the lighter stau
mass eigenstate $\tilde\tau_1$ may therefore be reduced with respect
to the smuon coupling, and even vanish.  Moreover, because there is at
least one neutrino in each $\tau$ decay, the visible energy of the
final state arising from stau pair production is smaller than in the
case of smuons, so that the selection efficiency is reduced. The mass
lower limits obtained at LEP are therefore lower for staus than for
smuons, from 86 to 95~GeV, depending on $m_{\tilde\chi_1^0}$, provided
the $\tilde\tau_1 - \tilde\chi_1^0$ mass difference is larger than 7
GeV~\cite{lepslep}.

As for smuons, the selectron mass eigenstates can be identified with
the interaction eigenstates. But because of the contribution of
$t$-channel neutralino exchange to selectron pair production, the
gaugino sector of the model, mass spectrum and field contents, has to
be specified to interpret the results of the searches for acoplanar
electrons. With gaugino mass unification and for $\tan\beta=1.5$ and
$\mu=-200$~GeV, a selectron mass lower limit of 100~GeV was obtained
for $m_{\tilde\chi_1^0}<85$~GeV~\cite{lepslep}.  Neutralino
$t$-channel exchange can furthermore mediate associated $\tilde e_L
\tilde e_R$ production. This process is useful if the $\tilde e_R -
\tilde\chi_1^0$ mass difference is small, because the electron from
the $\tilde e_L\to e\tilde\chi_1^0$ decay can be energetic enough to
lead to an apparent single electron final state. Both gaugino and
slepton mass unifications have to be assumed for the masses of the two
selectron species to be related. Under these assumptions, a lower
limit of 73~GeV was set on $m_{\tilde e_R}$, independent of the
$\tilde e_R - \tilde\chi_1^0$ mass
difference~\cite{Heister:2002jca,Achard:2003ge}.

From the measurement of the invisible width of the $Z$
boson~\cite{:2005ema}, a general mass lower limit of 45~GeV can be
deduced for a sneutrino LSP or NLSP.

{\bf Charginos and neutralinos:} As evident from \eqref{chargino},
three parameters are sufficient to fully specify the masses and field
contents in the chargino sector.  These may be taken to be $M_2$,
$\mu$, and $\tan\beta$. The lighter of the two charginos will simply
be denoted ``chargino'' in the following.  To specify the neutralino
mass matrix of \eqref{neutralino}, one more parameter, $M_1$, is
needed.  If gaugino mass unification is assumed, the two gaugino
masses are related by $M_1=(5/3)\tan^2\theta_W M_2\simeq 0.5
M_2$. Unless otherwise specified, this relation is assumed to hold in
the following.  Charginos are pair produced via $s$-channel
$Z/\gamma^\ast$ and $t$-channel $\tilde\nu_e$ exchanges, the two
processes interfering destructively. The three-body final states
$f\bar f'\tilde\chi_1^0$ are reached in chargino decays via virtual
$W$ or sfermion exchange. If kinematically allowed, two-body decays
such as $\tilde\chi^\pm\to\ell^\pm\tilde\nu$ are dominant. Similarly,
neutralino pair or associated production proceed via $s$-channel $Z$
and $t$-channel selectron exchanges, and $\tilde\chi_2^0$ three-body
decays to $f\bar f\tilde\chi_1^0$ via virtual $Z$ or sfermion
exchange; whenever kinematically allowed, two-body decays such as
$\tilde\chi_2^0\to\nu\tilde\nu$ are dominant.

If sfermions are heavy, chargino decays are mediated by virtual $W$
exchange, so that the final states arising from chargino pair
production are the same as for $W$ pairs, with additional missing
energy from the two neutralino LSPs: all hadronic ($q\bar
q'\tilde\chi_1^0 q\bar q'\tilde\chi_1^0$), mixed ($q\bar
q'\tilde\chi_1^0\ell\nu\tilde\chi_1^0$), and fully leptonic
($\ell\nu\tilde\chi_1^0\ell\nu\tilde\chi_1^0$). Selections were
designed for these three topologies and for various
$m_{\tilde\chi^\pm}-m_{\tilde\chi_1^0}$ regimes, with no excess
observed over SM backgrounds. From a scan over $M_2$, $\mu$, and
$\tan\beta$, a chargino mass lower limit of 103~GeV was derived for
$m_{\tilde\nu}>200$~GeV~\cite{lepchar}.  For smaller sneutrino masses,
the limit is reduced by the destructive interference in the
production.  This limit holds for $M_2<\simeq 1$~TeV. For larger $M_2$
values, the selection efficiency decreases rapidly as the
$\tilde\chi^\pm - \tilde\chi_1^0$ mass difference becomes smaller. If
this mass difference becomes so small that even the
$\tilde\chi^\pm\to\pi^\pm\tilde\chi_1^0$ decay mode is closed, the
chargino becomes long lived. Searches for charged massive stable
particles, in which advantage is taken of their larger ionization
power, were designed to cope with this configuration. For slightly
larger mass differences, the visible final state is so soft that even
triggering becomes problematic. Chargino pair production can however
be ``tagged'' by an energetic photon from initial state radiation,
$e^+e^-\to\gamma\tilde\chi^+\tilde\chi^-$, providing access to those
almost invisible charginos, although at a reduced effective center of
mass energy. The combination of these analysis techniques allowed
chargino masses smaller than 92~GeV to be excluded, irrespective of
the $\tilde\chi^\pm - \tilde\chi_1^0$ mass
difference~\cite{lepchardegen}.

\begin{figure}
\includegraphics[width=8.5cm]{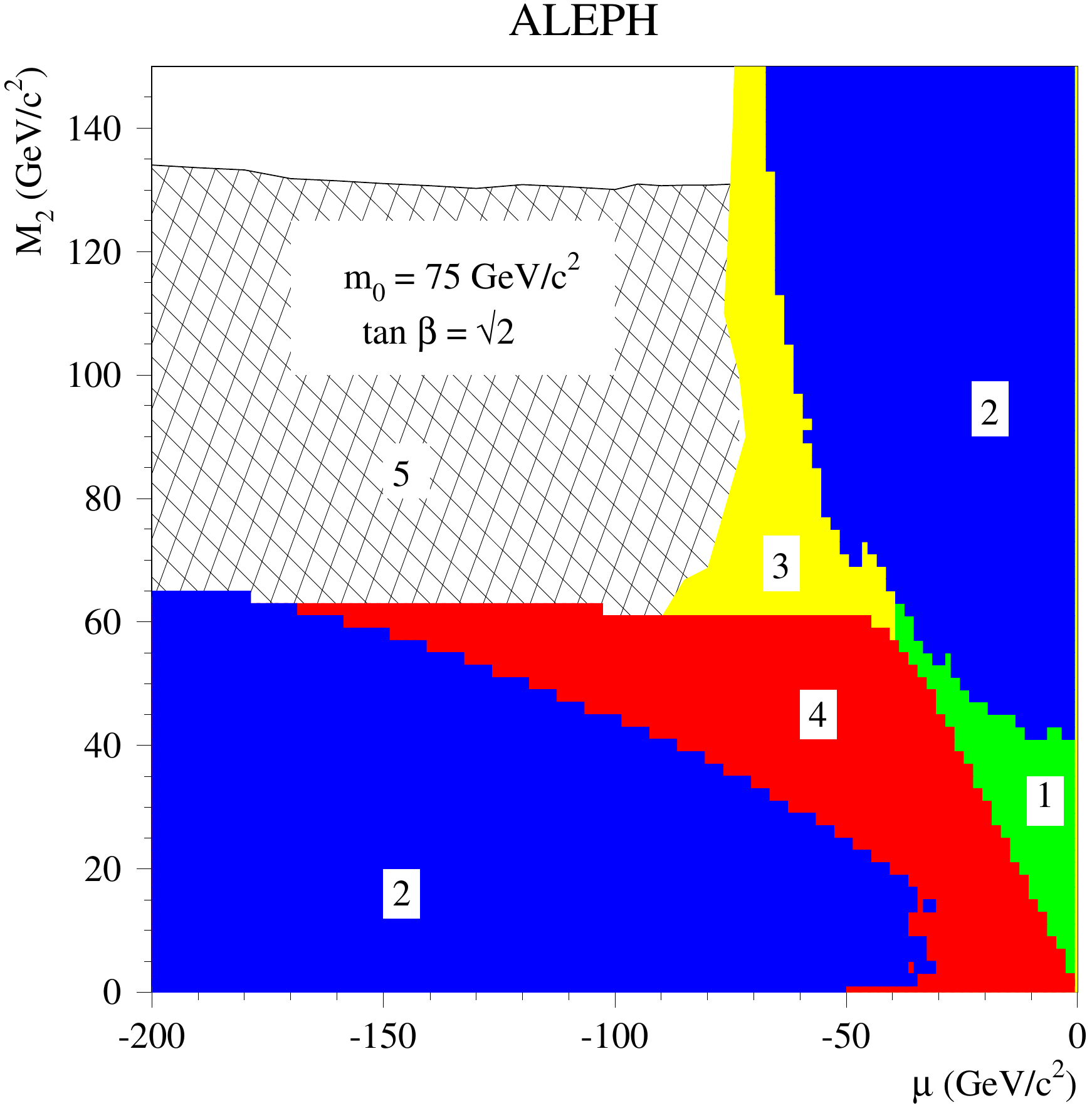}
\caption{\label{m2mucombi}
Regions in the $(\mu,M_2)$ plane excluded by the LEP~I constraints
(1), and by the searches for charginos (2), neutralinos (3) and
sleptons (4) at LEP II, for $\tan\beta=\sqrt{2}$ and $m_0=75$~GeV. The
region (5) is excluded by the Higgs boson searches at LEP II. This
figure is from Ref.~\cite{Barate:1999fs}.}
\end{figure}

For lower sfermion masses, the sensitivity of the former analyses is
reduced first because of the destructive interference between the
$s$-channel $Z/\gamma^\ast$ and $t$-channel sneutrino exchanges, and
second because of the opening of two-body decays. The latter effect is
specifically detrimental in the ``corridor'' of small $\tilde\chi^\pm
- \tilde\nu$ mass differences, where the final state from the
$\tilde\chi^\pm\to\ell\tilde\nu$ decays becomes invisible in
practice. Gaugino mass unification allows indirect limits on charginos
to be obtained, based on constraints on the parameter space resulting
from searches for pair or associated neutralino production, e.g.,
$e^+e^-\to\tilde\chi_2^0\tilde\chi_2^0$ or
$\tilde\chi_1^0\tilde\chi_2^0$. In order to relate all production
cross sections and decay branching fractions, it is however necessary
to fully specify the sfermion spectrum, which is done with the
assumption of sfermion mass unification. The results of the chargino
and neutralino searches are then expressed as exclusion domains in the
$(\mu,M_2)$ plane for selected values of $\tan\beta$ and $m_0$. The
invisible two-body decay $\tilde\chi_2^0\to\nu\tilde\nu$ can however
cause a large sensitivity reduction. Since this configuration occurs
for low $m_0$ values, constraints arising from the slepton searches
can be used to mitigate this effect. With gaugino and sfermion mass
unification, the slepton masses are related to the model parameters by
$m^2_{\tilde{\ell_R}}\simeq m_0^2+0.22M_2^2-\sin^2\theta_W m^2_Z \cos
2\beta$, so that a limit on $m_{\tilde\ell_R}$ can be turned into a
limit on $M_2$ for given values of $\tan\beta$ and $m_0$. After a
proper combination of the searches for charginos, neutralinos and
sleptons, an example of which is shown in Fig.~\ref{m2mucombi}, it
turns out that the chargino mass limit obtained in the case of heavy
sfermions is only moderately degraded.

\begin{figure}
\includegraphics[width=8.5cm]{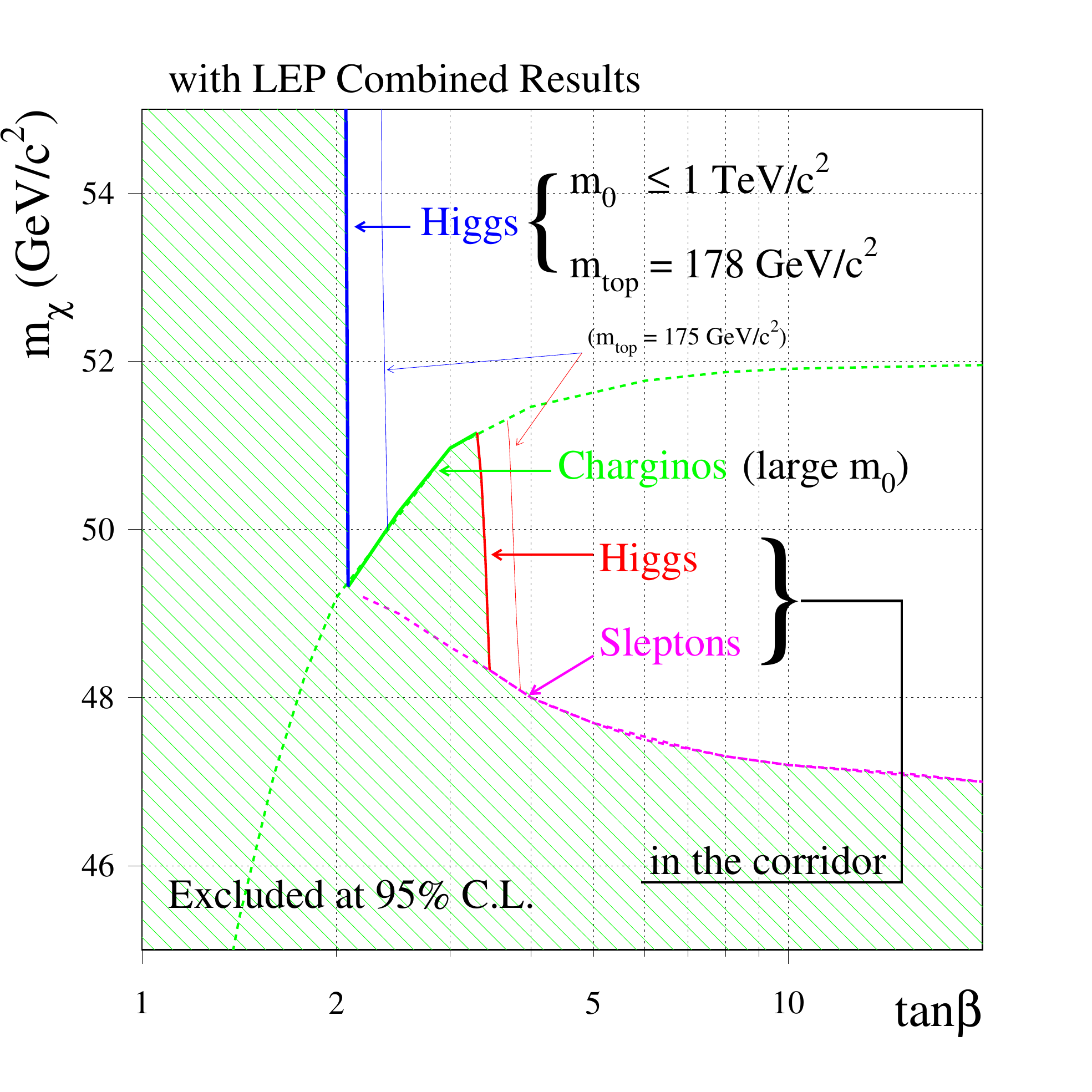}
\caption{\label{LSP}
Lower mass limit for the lightest neutralino as a function of
$\tan\beta$, inferred in the conventional scenario from searches at
LEP for charginos, sleptons, and neutral Higgs
bosons~\cite{leplsp}. The dashed contour is the limit obtained for
large $m_0$.}
\end{figure}

Direct searches for the lightest neutralino had been performed at
lower energy $e^+e^-$ colliders, PEP and PETRA, in the reaction
$e^+e^-\to\gamma\tilde\chi_1^0\tilde\chi_1^0$, where the photon from
initial state radiation is used to tag the production of an invisible
final state.  At LEP, at or above the $Z$ resonance, the irreducible
background from $e^+e^-\to\gamma\nu\bar\nu$ is too large to obtain
competitive results.  Furthermore, production via $s$-channel $Z$
exchange may simply vanish, e.g., if the LSP is a pure photino, while
production via $t$-channel selectron exchange can be made negligible
if selectrons are sufficiently heavy. Indirect limits on the mass of
the LSP can however be obtained within constrained models. With
gaugino mass unification, $m_{\tilde\chi_1^0}$ is typically half the
chargino mass. As a result, the chargino mass limit translates into a
$\tilde\chi_1^0$ mass lower limit of 52~GeV for heavy sfermions and
large $\tan\beta$. If sfermion mass unification is used in addition, a
limit of 47~GeV is obtained at large $\tan\beta$, independent of
$m_0$. This limit is set by searches for sleptons in the corridor. For
low values of $\tan\beta$, constraints from the Higgs boson searches
can be used, as was shown in Sec.~\ref{subsec:LEPHiggs} for benchmark
scenarios. A complete scan over $m_0$, $m_{1/2}$, $\mu$ and
$\tan\beta$ was performed and, for each parameter set, the maximum $h$
mass predicted was compared to the experimental limit, and the
constraints from chargino and slepton searches were included.  The
translation of the scan result in terms of excluded domain in the
$(\tan\beta,m_{\tilde\chi_1^0})$ plane is shown in Fig.~\ref{LSP},
from which a $\tilde\chi_1^0$ mass lower limit of 47~GeV is
derived~\cite{leplsp}. Within the more constrained mSUGRA scenario,
wherein $\mu$ is calculated from the other parameters, this limit
becomes 50~GeV~\cite{lepsugra}.
 
{\bf Squarks:} On general grounds, the mass reach for strongly
interacting particles is expected to be substantially higher at the
Tevatron than at LEP. For some specific configurations, however, the
searches at the Tevatron become inefficient, in which cases the
results obtained at LEP remain of interest.  This is particularly
relevant for third generation squarks which may be substantially
lighter than the other squarks, as motivated in
Sec.~\ref{sec:supersymmetry}. The lighter third generation mass
eigenstates are simply denoted stop and sbottom, $\tilde t$ and
$\tilde b$, in the following.

In the mass range accessible at LEP, and given the chargino mass limit
which effectively forbids $\tilde t\to b\tilde\chi^+$, the stop is
expected to decay into a charm quark and a neutralino, $\tilde t\to
c\tilde\chi_1^0$, as long as
$m_{\tilde{t}}<m_W+m_b+m_{\tilde\chi^0_1}$~\cite{Hikasa:1987db}.
Because this decay is a flavor-changing loop process, the stop
lifetime can be large enough to compete with the hadronization time,
and the simulation programs were adjusted to take this feature into
account. The final state from stop pair production exhibits an
acoplanar jet topology, for which no signal was observed above
standard model backgrounds. As already explained for staus, the amount
of mixing between the weak eigenstates can be such that the stop does
not couple to the $Z$ boson. In this worst case scenario, stop mass
lower limits ranging from 96 to 99~GeV were obtained, depending on the
$\tilde\chi_1^0$ mass, as long as
$m_{\tilde{t}}-m_{\tilde\chi^0_1}-m_c > 5$~GeV~\cite{lepstop}. For
smaller $\tilde{t} - \tilde\chi^0_1$ mass differences, long-lived
$R$-hadrons may be produced in the stop hadronization process. The
production of such $R$-hadrons and their interaction in the detector
material were taken into account in a dedicated search, from which a
stop mass lower limit of 63~GeV was derived, valid for any
$m_{\tilde{t}}-m_{\tilde\chi^0_1}$~\cite{Heister:2002hp}. For specific
parameter choices, and in spite of the slepton mass limits, it can be
that the $\tilde{t}\to b\ell\tilde\nu$ decay is kinematically allowed,
in which case it is dominant. From a search for events exhibiting
jets, leptons and missing energy, a stop mass lower limit of 96~GeV
was obtained, valid for sneutrino masses smaller than
86~GeV~\cite{lepstop}.

The case of a light sbottom is much simpler, as the tree-level
$\tilde{b}\to b\tilde\chi_1^0$ decay mode is dominant. From searches
for acoplanar $b$-flavored jets, a mass lower limit of about 95~GeV
was obtained in the worst case scenario where the sbottom does not
couple to the $Z$~\cite{lepstop}.

\subsubsection{Searches at the Tevatron}
\label{subsubsec:TevSUSY}

The program most widely used for the calculation of SUSY particle
production cross sections at the Tevatron is {\sc
prospino}~\cite{Beenakker:1996ch}, which provides next-to-leading
order accuracy. The results reported below were generally obtained
with the CTEQ6.1M PDF set~\cite{Pumplin:2002vw,Stump:2003yu}. Various
codes were used to calculate the low energy SUSY spectrum from initial
parameters at the grand unification scale: {\sc
suspect}~\cite{Djouadi:2002ze}, {\sc softsusy}~\cite{Allanach:2001kg},
and {\sc isajet}~\cite{Paige:2003mg}.  This may introduce slight
inconsistencies when comparing results in different channels or from
different experiments in terms of parameters at the high scale.  The
production of SUSY particles was in general simulated with {\sc
pythia}~\cite{Sjostrand:2006za}, with decays modeled with {\sc
sdecay}~\cite{Muhlleitner:2003vg} or with {\sc isasugra} as
implemented in {\sc pythia}. Typically, SM backgrounds were simulated
with {\sc alpgen}~\cite{Mangano:2002ea} for the production of $W$ and
$Z$ bosons in association with jets, or with {\sc pythia} otherwise.

As already mentioned in Sec.~\ref{subsec:strategies}, the channels
most relevant for SUSY particle searches at hadron colliders are the
production of squarks and gluinos on the one hand, of electroweak
gauginos on the other.  For squarks and gluino, the search is
conducted in events exhibiting a jets+\met\
topology~\cite{Hinchliffe:1982iz,Kane:1982hw,Ellis:1984yw}, while for
electroweak gauginos, it is conducted in the trilepton final
state~\cite{Baer:1986vf,Nath:1987sw,Barbieri:1991vk}.

{\bf Generic squarks and gluinos:} Depending on the squark and gluino
mass hierarchy, different pair production processes via the strong
interaction are expected to contribute in $p\bar p$ collisions at the
Tevatron: $\tilde{q}\tilde{\bar{q}}$ and, to a lesser extent,
$\tilde{q}\tilde{q}$, if $m_{\tilde{q}} \ll m_{\tilde{g}}$; $\tilde
g\tilde g$ if $m_{\tilde{g}} \ll m_{\tilde{q}}$; and all of these
processes, as well as $\tilde{q}\tilde{g}$, if the squark and gluino
masses are similar.  If $m_{\tilde{q}}< m_{\tilde{g}}$, squarks are
expected to decay directly into a quark and a gaugino, $\tilde{q}\to
q\tilde\chi$, where $\tilde\chi$ is dominantly $\tilde\chi^0_1$ for
$\tilde{q}_R$, and $\tilde\chi^\pm$ or $\tilde\chi^0_2$ for
$\tilde{q}_L$.  If $m_{\tilde{g}}< m_{\tilde{q}}$, gluinos are
expected to decay via virtual squark exchange into a quark, an
antiquark, and a gaugino, $\tilde{g}\to q\bar{q}\tilde\chi$, where
$\tilde\chi$ is typically $\tilde\chi^\pm$ or
$\tilde\chi^0_2$~\cite{Barnett:1987kn,Baer:1989hr}. The heavier
gauginos further decay into a fermion-antifermion pair and an LSP,
$\tilde\chi^0_1$, so that there is always some missing \et\ in the
final state. More detailed predictions can be made only within a
specific model such as mSUGRA.

The aforementioned production processes have been searched for by CDF
and D\O\ in topologies involving at least two jets, four jets and
three jets, all with large $\met$. Initial and final state radiation
of soft jets can increase further those jet multiplicities. A first
class of background to squark and gluino production arises from
processes with intrinsic $\met$, such as ($W\to\ell\nu$)+jets, where
the lepton escapes detection, or ($Z\to\nu\nu$)+jets, which is
irreducible. Monte Carlo simulations were used to estimate those
backgrounds, after calibration on events where leptons from
$W\to\ell\nu$ or $Z\to\ell\ell$ are detected. Another class of
background is due to multijet production by strong
interaction. Although there is no intrinsic \met\ in such events, fake
\met\ can arise from jet energy mismeasurements (and also real \met\
from semileptonic decays of heavy flavor hadrons). In such events, the
\met\ distribution decreases quasi-exponentially, and the direction of
the \met\ tends to be close to that of a mismeasured jet.  Requiring
sufficiently large \met\ and applying topological selection criteria
allows this background to remain under control. While D\O\ applied
criteria tight enough to reduce this background to a negligible level,
CDF estimated its remaining contribution based on simulations
calibrated on control samples.

\begin{figure}
\includegraphics[width=8.5cm]{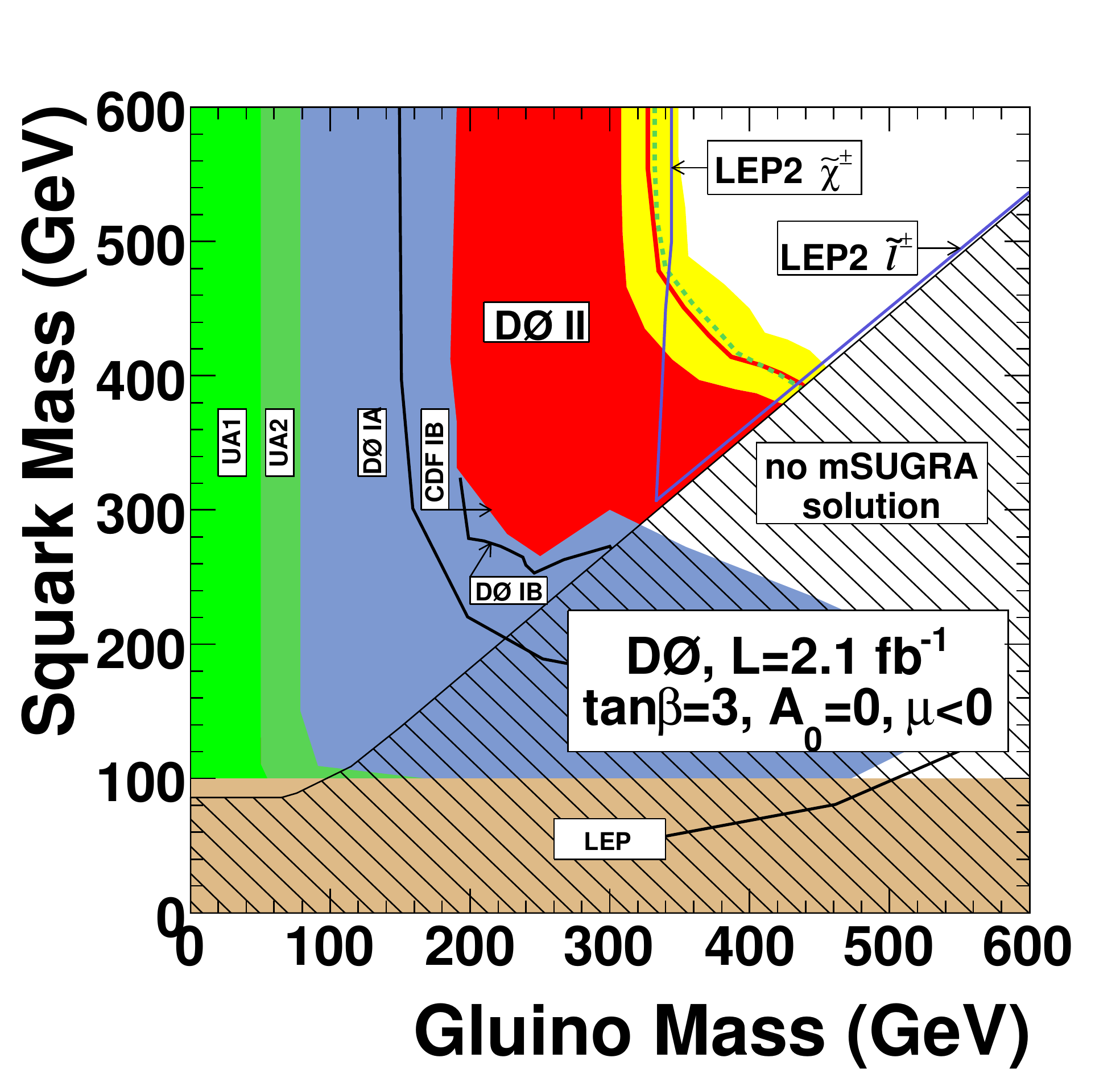}
\caption{\label{sqgl}
Region in the ($m_{\tilde{g}},m_{\tilde{q}}$) plane excluded by
D\O~\cite{:2007ww} and by earlier experiments. The red curve
corresponds to the nominal scale and PDF choices. The yellow band
represents the uncertainty associated with these choices. The blue
curves represent the indirect limits inferred from the LEP chargino
and slepton searches.}
\end{figure}

No excesses of events were observed over SM backgrounds, which was
translated into exclusion domains in the plane of squark and gluino
masses.  To this end, a specific SUSY model had to be chosen, so that
the masses and decay modes of all the gauginos involved in the decay
chains could be determined. The model used by both CDF and D\O\ was
mSUGRA, with $A_0=0$, $\mu<0$, and $\tan\beta=5$~(CDF) or 3~(D\O). The
production of all squark species was considered, except for the third
generation (CDF) or for stops (D\O), and the squark mass quoted was
the average of the masses of the squarks considered. Finally, the
large theoretical uncertainties associated to the choices of PDFs and
of the factorization and renormalization scales had to be taken into
account when turning cross section upper limits into exclusion domains
in terms of masses. Based on an integrated luminosity of 2.1~\invfb,
D\O\ excluded the domain shown in Fig.~\ref{sqgl}, from which lower
limits of 379 and 308~GeV were derived for the squark and gluino
masses, respectively, as well as a lower limit of 390~GeV if
$m_{\tilde{q}}=m_{\tilde{g}}$~\cite{:2007ww}. Similar results were
obtained by the CDF collaboration~\cite{Aaltonen:2008rv}.

\begin{figure}
\includegraphics[width=8.5cm]{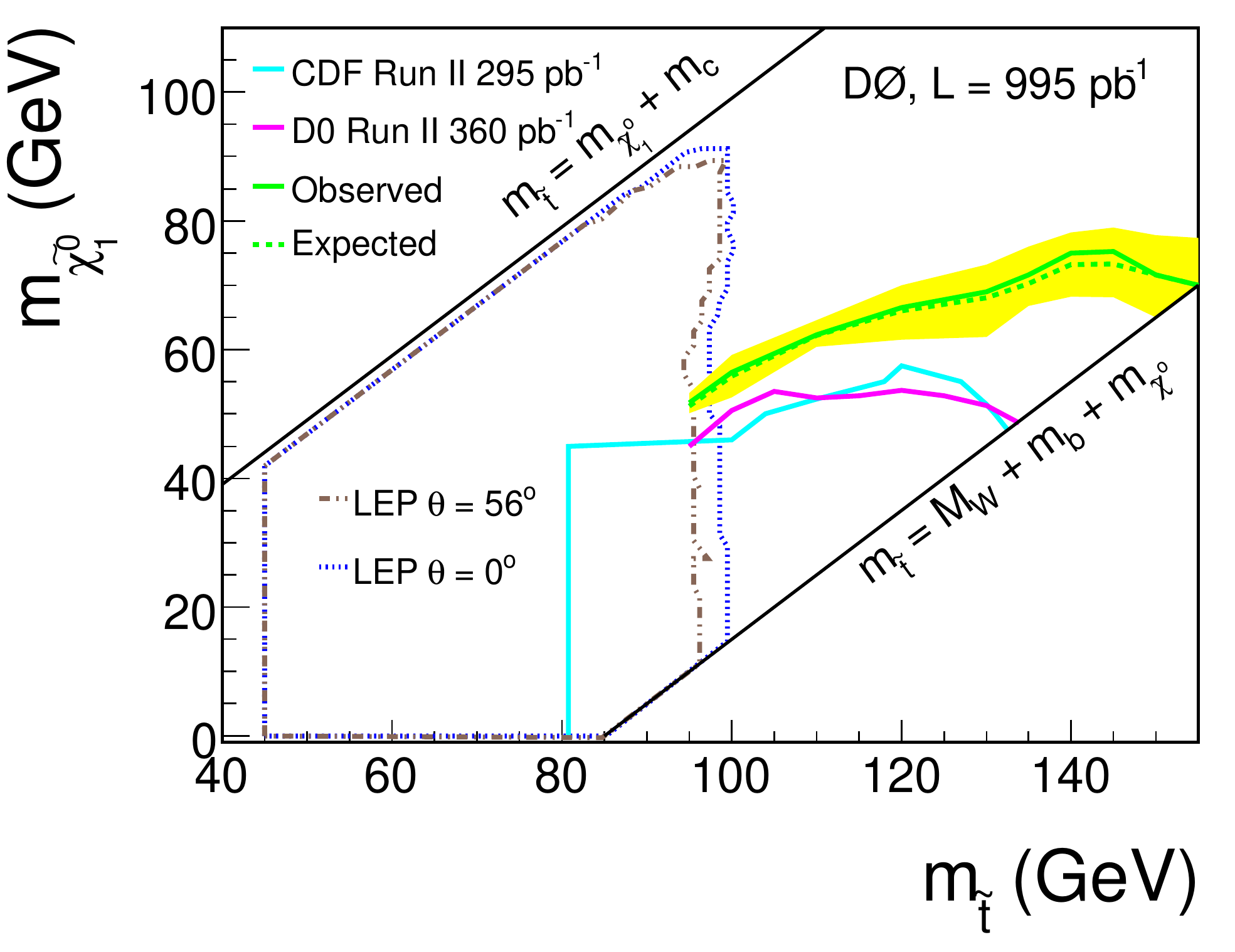}
\caption{\label{stop}
Region in the ($m_{\tilde{t}},m_{\tilde\chi^0_1}$) plane excluded by
D\O~\cite{Abazov:2008rc} and by earlier experiments. The solid curve
corresponds to the nominal scale and PDF choices. The yellow band
represents the uncertainty associated with these choices.}
\end{figure}

{\bf Third generation squarks:} As already mentioned, a stop NLSP
decays into a charm quark and a neutralino as long as
$m_{\tilde{t}}<m_W+m_b+m_{\tilde\chi^0_1}$.  The final state from stop
pair production therefore consists in acoplanar charm jets and
\met. Because only one of the squark species is now produced, the
cross section is smaller than for generic squarks, and the mass reach
is therefore lower. As a consequence, the jets are softer, and there
is also less \met. The corresponding loss of sensitivity was
attenuated by making use of heavy-flavor tagging, which resulted in
the exclusion domain shown in Fig.~\ref{stop}, obtained by
D\O~\cite{Abazov:2008rc} from an analysis of 1~\invfb\ of data.  It
can be seen that a stop mass of 150~GeV is excluded for
$m_{\tilde\chi^0_1}=65$~GeV.  In spite of the larger mass reach at the
Tevatron, the LEP results remain the most constraining for $\tilde{t}
- \tilde\chi_1^0$ mass differences smaller than $\simeq 40$~GeV.
Similar searches were performed for a sbottom NLSP decaying into
$b\tilde\chi_1^0$~\cite{Abazov:2006fe,Aaltonen:2007sw}, with better
sensitivity due to a more efficient heavy-flavor tagging for $b$ than
for $c$ quarks. A mass lower limit of 222~GeV was obtained by D\O\ for
$m_{\tilde\chi^0_1}<60$~GeV, based on 310~\invpb\ of data.

\begin{figure}
\includegraphics[width=8.5cm]{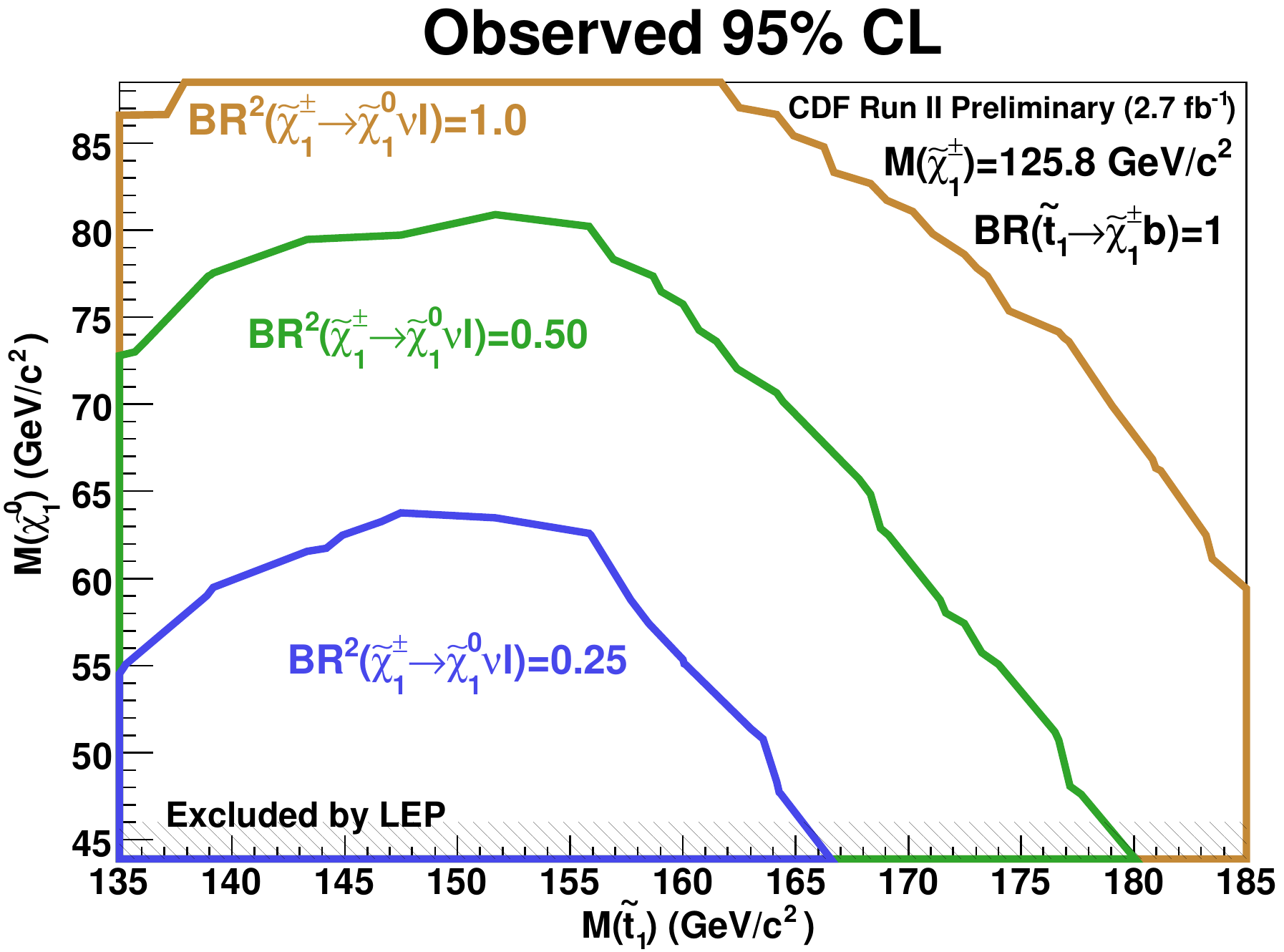}
\caption{\label{stoptocha}
Regions in the ($m_{\tilde{t}},m_{\tilde\chi^0_1}$) plane excluded by
CDF~\cite{CDFstop} for $m_{\tilde\chi^\pm}=125.8$~GeV and for various
values of the branching fraction for the
$\tilde\chi^\pm\to\ell\nu\tilde\chi_1^0$ decay.}
\end{figure}

Other mass hierarchies were considered, where the stop or sbottom is
not the NLSP. Three-body stop decays, $\tilde{t}\to b\ell\tilde\nu$,
are dominant if kinematically allowed and when
$\tilde{t}\to\tilde\chi^+ b$ is not, which is possible for some model
parameter choices in spite of the mass limits on charged sleptons
available from LEP. The final states investigated by D\O\ comprised
two muons or a muon and an electron, with $b$ jets and \met. Based on
an analysis of 400~\invpb\ of data, the largest stop mass excluded is
186~GeV, for $m_{\tilde\nu}=71$~GeV~\cite{:2007im}. If the chargino
is lighter than the stop, the $\tilde{t}\to b\tilde\chi^+$ decay is
dominant. A search was performed by CDF in the two lepton, two $b$
jets and \met\ final state, with a sensitivity depending on the
branching fraction of the chargino leptonic decay,
$\tilde\chi^\pm\to\ell\nu\tilde\chi_1^0$, which is enhanced for light
sleptons. An example of an excluded domain in the
$(m_{\tilde{t}},m_{\tilde\chi_1^0})$ plane is shown in
Fig.~\ref{stoptocha}~\cite{CDFstop}, based on 2.7~\invfb\ of data. In
both of those searches, the background from top quark pair production
was a major challenge. Yet another mass hierarchy was considered by
CDF, namely that where the sbottom is the only squark lighter than the
gluino. In such a configuration, the $\tilde{g}\to b\tilde{b}$ decay
is dominant, and gluino pair production then leads to a final state of
four $b$ jets and \met. This search was performed in a data sample of
2.5~\invfb, and lead to excluded sbottom masses as large as 325~GeV
for gluino and LSP masses of 340 and 60~GeV,
respectively~\cite{Aaltonen:2009fm}.

{\bf Charginos and neutralinos:} The associated production of
charginos and neutralinos, $p\bar p\to\tilde\chi^\pm\tilde\chi_2^0$,
is an electroweak process mediated by $s$-channel $W$ and $t$-channel
squark exchanges.  Leptonic decays,
$\tilde\chi^\pm\to\ell^\pm\nu\tilde\chi^0_1$ and
$\tilde\chi^0_2\to\ell^+\ell^-\tilde\chi^0_1$, are mediated by $W$ and
$Z$ exchange, respectively, and by slepton exchange. If sleptons are
light, leptonic decays can be sufficiently enhanced for searches in
final states consisting of three leptons and \met\ to become sensitive
in spite of production cross sections of a fraction of a picobarn. An
additional challenge is the rather small energy carried by the final
state leptons in the chargino and neutralino mass domain to which the
searches at the Tevatron are currently sensitive.

In both the CDF~\cite{Aaltonen:2008pv} and D\O~\cite{Abazov:2009zi}
analyses, only two leptons were required to be positively identified
as electrons or muons.\footnote{The D\O\ analysis also considered
final states with a muon and one or two $\tau$ leptons identified.}
Allowing the third lepton to be detected as an isolated charged
particle track provided sensitivity to final states including a $\tau$
lepton that decays into hadrons.  In the CDF analysis, the trilepton
final state was split into topologies with different signal to
background ratios, the purest being when the three leptons are
positively identified as electrons or muons with tight criteria. In
the D\O\ approach, different selections were optimized according to
the amount of energy available to the lepton candidates. The ultimate
background for these trilepton searches is associated $WZ$ production.
 
\begin{figure}
\includegraphics[width=8.5cm]{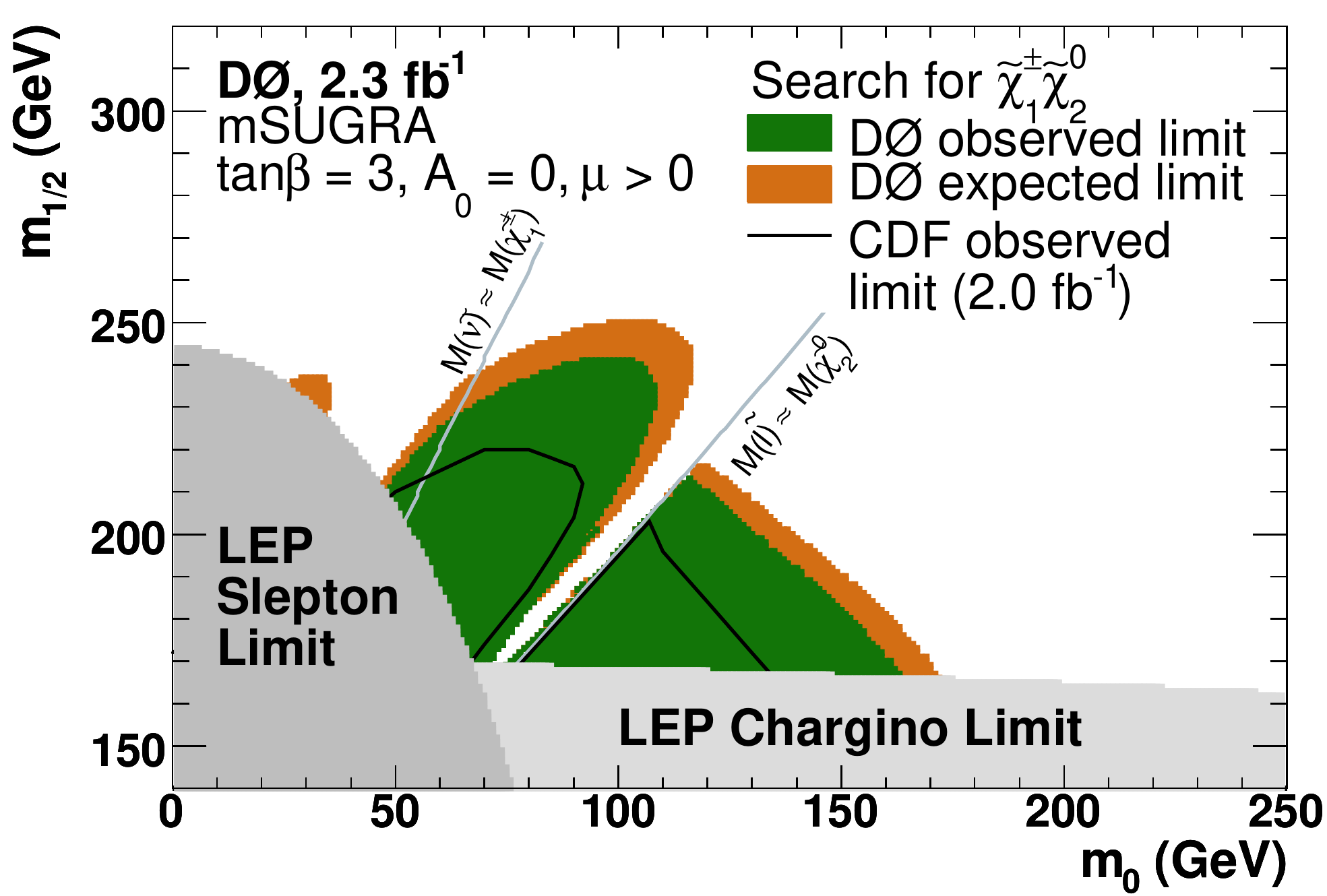}
\caption{\label{tril}
Regions in the ($m_0,m_{1/2}$) plane excluded 
by the D\O\ search for trileptons~\cite{Abazov:2009zi}.}
\end{figure}

The D\O\ search, based on an integrated luminosity of 2.3~\invfb,
excludes regions in the mSUGRA parameter space as shown in
Fig.~\ref{tril} for $A_0=0$, $\tan\beta=3$ and $\mu>0$. It can be seen
that the domain excluded at LEP is substantially extended by these
trilepton searches. The interruption in the exclusion domain is due to
configurations where the small $\tilde\chi_2^0 - \tilde\ell$ mass
difference results in one of the final state leptons carrying too
little energy, thus preventing efficient detection. Requiring only two
leptons to be identified, but with same charge sign in order to reduce
the otherwise overwhelming SM backgrounds, should provide sensitivity
in that region, as was shown by D\O\ in an analysis based on a smaller
data sample~\cite{Abazov:2005ku}. The same-sign dilepton signature was
also considered in an earlier CDF analysis~\cite{:2007mu}.
 
\subsection{Searches in non-canonical scenarios}
\label{subsec:noncanon}

\subsubsection{$R$-parity violation}
\label{subsubsec:RPV}

Searches for SUSY with $R$-parity violation were performed at LEP, the
Tevatron and HERA. Both $R$-parity conserving pair production of SUSY
particles and $R$-parity violating resonant single SUSY particle
production were considered.  The produced particles were subsequently
subject to either direct or indirect (via a cascade to the LSP)
$R$-parity violating decays.  Unless otherwise specified, a single
$R$-parity violating coupling is assumed to be non-vanishing in the
following, large enough for the lifetime of the LSP to be safely
assumed to be negligible.

{\bf Searches at LEP:} Extensive searches for pair production were
performed at LEP, involving all possible $R$-parity violating
couplings. The possible final states are numerous, ranging from four
leptons and missing energy for $\tilde\chi_1^0$ pair production, with
decays mediated by a $\lambda$-type coupling, e.g., $\tilde\chi_1^0\to
e\mu\nu$, to ten hadronic jets and no missing energy for chargino pair
production, with $\tilde\chi^\pm\to q\bar q'\tilde\chi_1^0$ followed
by a $\tilde\chi_1^0$ decay into three quarks via a $\lambda''$-type
coupling, e.g., $\tilde\chi_1^0\to udd$. The results of these searches
are at least as constraining as in the canonical
scenario~\cite{leprpv,Heister:2002jc,Abdallah:2003xc,%
Achard:2001ek,Abbiendi:2003rn}.

The production of a sneutrino resonance via a $\lambda_{1j1}$ coupling
was also investigated. No signal was observed, and mass lower limits
almost up to the center-of-mass energy were set for sufficiently large
values of the $R$-parity violating coupling
involved~\cite{Barate:2000en,Heister:2002kq,Abdallah:2002wt,%
Acciarri:1997wc,Abbiendi:1999wm}.

{\bf Searches at HERA:} As explained in Sec.~\ref{subsec:strategies},
the HERA $ep$ collider is most effective in the searches for
$R$-parity violating resonant single squark production via a
$\lambda'$-type coupling. Direct and indirect squark decays were
investigated, and the search results were combined to lead to squark
mass lower limits up to
275~GeV~\cite{Aktas:2004tm,Aktas:2004ij,:2006je}, within mild model
assumptions, for a $\lambda'$ coupling of 0.3, i.e., with
electromagnetic strength.

{\bf Searches at the Tevatron:} A fully general search for all
$R$-parity violating couplings is not possible at the Tevatron, as it
was at LEP. For instance, $\lambda''$ couplings lead to multijet final
states with no or little missing energy, which cannot be distinguished
from standard multijet production. Searches have therefore been
designed for specific choices of couplings leading to distinct
signatures.

Gaugino pair production followed by indirect decays has been
extensively studied by both CDF~\cite{Abulencia:2007mp} and
D\O~\cite{Abazov:2006nw} in the case of a $\lambda$-type coupling. The
final state is expected to contain four charged leptons, with flavors
depending on the indices in the $\lambda_{ijk}$ coupling, and \met\
due to two neutrinos. For $m_0=1$~TeV, $\tan\beta=5$, and $\mu>0$, the
chargino mass lower limits obtained by D\O\ from an analysis of
360~\invpb\ of data are 231, 229, and 166~GeV for the $\lambda_{121}$,
$\lambda_{122}$, and $\lambda_{133}$ couplings, respectively, with
reduced sensitivity in the last case due to the occurrence of $\tau$
leptons in the final state.

Stop pair production, with $\tilde t\to b\tau$ via a $\lambda'_{333}$
coupling has been searched by CDF~\cite{Brigliadori:2008vf} in the
topology where one $\tau$ lepton decays into an electron or a muon,
and the other into hadrons. From an analysis of 322~\invpb\ of data, a
stop mass lower limit of 151~GeV was derived.

Resonant smuon or sneutrino production could be mediated by a
$\lambda'_{211}$ coupling. With indirect decays, the final state would
exhibit at least one muon and two jets. This topology was investigated
by D\O~\cite{Abazov:2006ii}, and an excluded domain was set in the
$(m_{\tilde\mu},\lambda'_{211})$ plane, leading to a smuon mass lower
limit of 363~GeV for $\lambda'_{211}=0.1$, and for $A_0=0$,
$\tan\beta=5$, and $\mu<0$.

Resonant sneutrino production mediated by a $\lambda'_{i11}$ coupling
was also investigated by CDF and
D\O~\cite{Abulencia:2005nf,Acosta:2005ij,Abulencia:2006xm,Abazov:2007zz},
now assuming that the sneutrino decays directly via a $\lambda$-type
coupling. The final states considered were $ee$, $e\mu$, $\mu\mu$, and
$\tau\tau$. The sneutrino mass limits obtained depend on the product
of the two couplings involved.

\subsubsection{Gauge-mediated SUSY breaking}
\label{subsubsec:GMSB}

As already explained, the LSP in GMSB is a very light gravitino, and
the phenomenology depends essentially on the nature of the NLSP, a
neutralino or a stau, possibly almost mass degenerate with $\tilde
e_R$ and $\tilde\mu_R$, and on its lifetime.

\begin{figure}
\includegraphics[width=8.5cm]{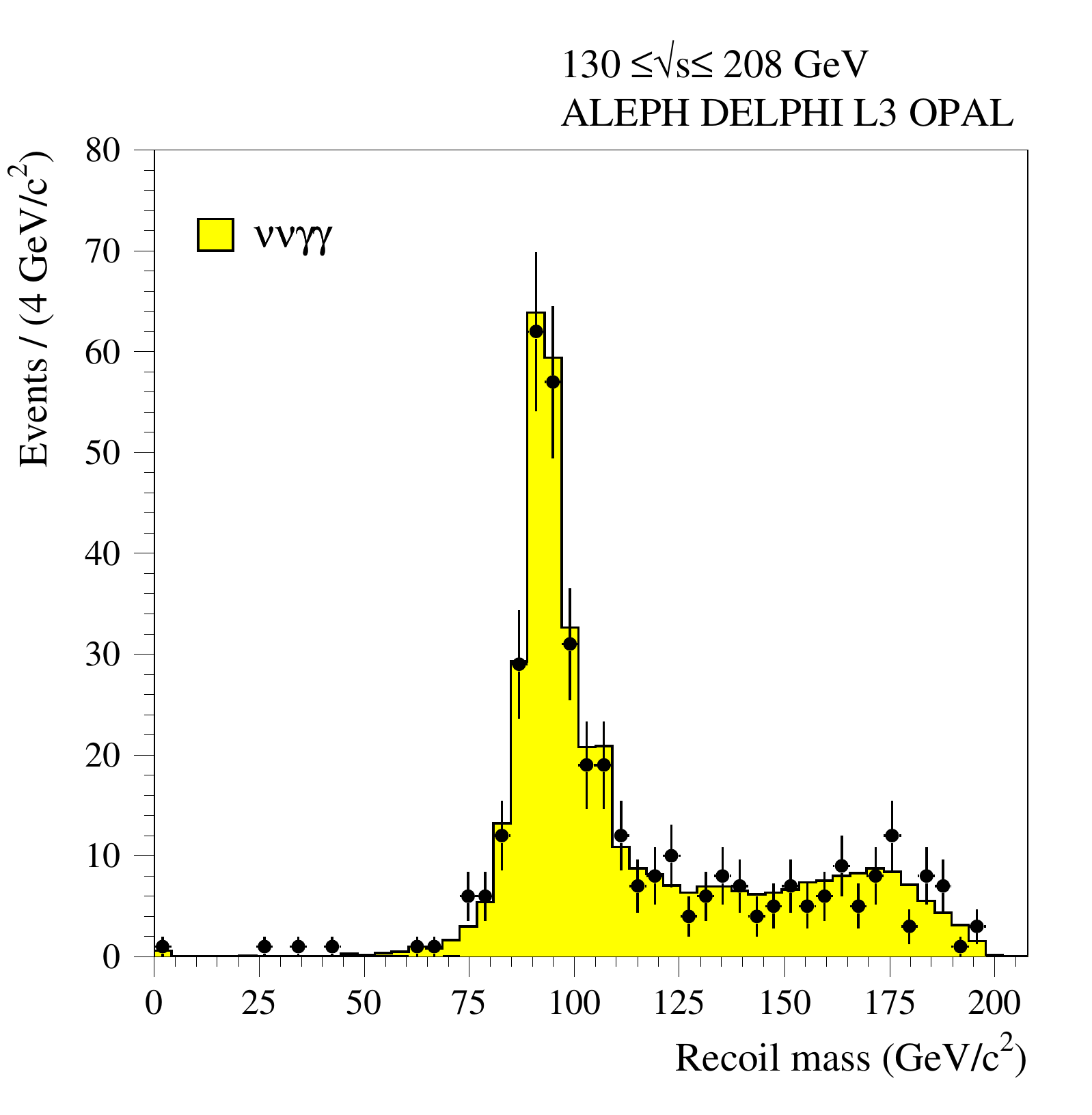}
\caption{\label{gaga}
Mass of the invisible system recoiling against pairs of photons at 
LEP~\cite{lepgaga}.}
\end{figure}

{\bf Neutralino NLSP:} In the mass range of current interest, a
neutralino NLSP decays into a photon and a gravitino,
$\tilde\chi_1^0\to\gamma\tilde{G}$. Pair production of such a
neutralino at LEP would therefore lead, assuming prompt decays, to a
final state of two acoplanar photons and missing energy. As can be
seen in Fig.~\ref{gaga}, no excess was observed above the SM
background from $e^+e^-\to (Z^{(\ast)}\to\nu\bar\nu)\gamma\gamma$. In
GMSB, $\tilde\chi_1^0$ has a large Bino component, so that pair
production in $e^+e^-$ interactions proceeds via selectron $t$-channel
exchange. An excluded domain in the $(m_{\tilde
e_R},m_{\tilde\chi_1^0})$ plane was therefore derived~\cite{lepgaga},
ruling out the GMSB interpretation (in terms of selectron pair
production) of an anomalous $ee\gamma\gamma+$~\met\ event that had
been observed by CDF~\cite{Abe:1998ui} during Run I of the Tevatron.

Searches were also performed at LEP for photons not pointing toward
the interaction point, which could arise from non-prompt decays of a
neutralino NLSP. For even longer lifetimes, the phenomenology becomes
identical to that of the canonical scenario. The results of these
various searches for a neutralino NLSP were combined with those in
various topologies expected to arise from heavier SUSY particle
production to lead to a robust neutralino mass lower limit of 54~GeV
within the minimal GMSB
framework~\cite{Heister:2002vh,Abbiendi:2005gc}.

Searches for acoplanar photons with large \met\ were performed at the
Tevatron by both CDF~\cite{cdfgagashort} and D\O~\cite{:2007is}.  This
topology is expected to arise whenever SUSY particles are pair
produced, which subsequently decay to a neutralino NLSP with
negligible lifetime.  No excess of events was observed over the
backgrounds due to photon misidentification or from fake \met, all
determined from data.  These results were interpreted within the
``Snowmass slope SPS 8'' benchmark GMSB model~\cite{Allanach:2002nj}
where the only free parameter is the effective SUSY breaking scale
$\Lambda$. The other parameters were fixed as follows: $N_5=1$
messenger, a messenger mass of $2\Lambda$, $\tan\beta=15$, and
$\mu>0$.  Neutralino NLSP masses smaller than 138~GeV are excluded by
the CDF analysis, based on 2~\invfb\ of data.

\begin{figure}
\includegraphics[width=8.5cm]{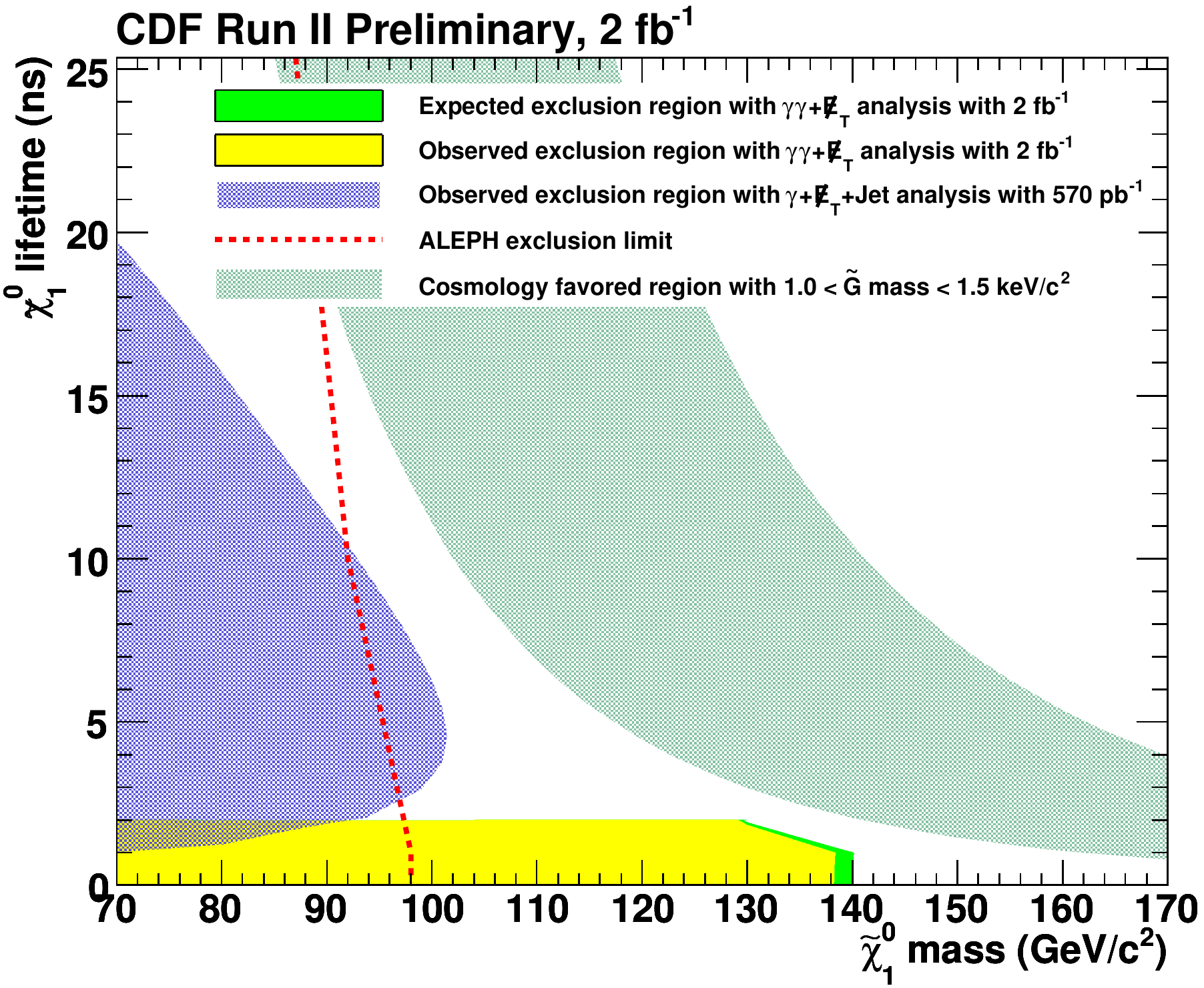}
\caption{\label{CDFgaga}
Domain excluded by CDF in the plane of neutralino-NLSP mass and
lifetime~\cite{cdfgagashort,Abulencia:2007ut}.}
\end{figure}

The possibility of non-prompt neutralino NLSP decays was also
investigated by CDF~\cite{Abulencia:2007ut}, making use of the timing
information of their calorimeter. No signal of delayed photons was
observed in a data sample of 570~\invpb, from which an excluded domain
in the plane of the mass and lifetime of the NLSP was inferred, as
shown in Fig.~\ref{CDFgaga} together with the result of
Ref.~\cite{cdfgagashort}.

\begin{figure}
\includegraphics[width=7.5cm]{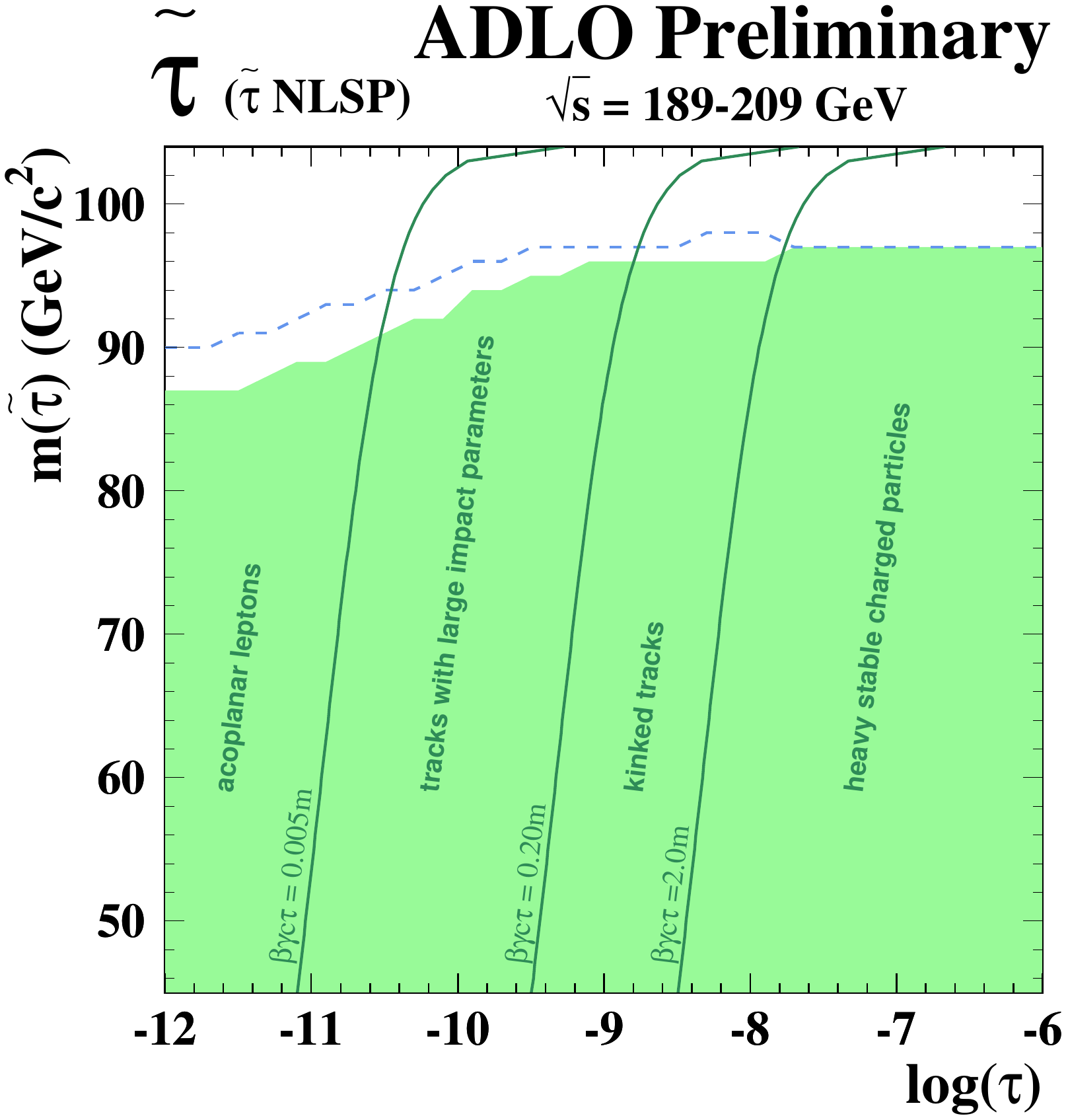}                 
\caption{\label{LEPstau} Domain excluded at LEP~\cite{lepgmsb} in the
$(\log_{10} \tau, m_{\tilde{\tau}})$ plane, where $\tau$ is the
lifetime in seconds and $m_{\tilde{\tau}}$ is the mass of a stau NLSP
in GMSB. The green shaded region is the excluded domain, and the
dashed blue curve is the expected boundary of the exclusion region.}
\end{figure}

{\bf Stau NLSP:} For prompt $\tilde\tau\to\tau\tilde{G}$ decays, the
final state arising from stau pair production at LEP is the same as in
the canonical scenario with a very light $\tilde\chi_1^0$. For very
long lifetimes, the searches for long lived charginos already reported
apply. Searches for in-flight decays along charged particle tracks
were designed to address intermediate lifetimes. The combination of
all these searches allowed a stau NLSP mass lower limit to be set from
87 to 97~GeV, depending on the stau lifetime, as shown in
Fig.~\ref{LEPstau}~\cite{lepgmsb}.

\subsubsection{Other non-canonical scenarios}
\label{subsubsec:misc}

A number of searches were performed at LEP and at the Tevatron in
other non-canonical scenarios.

{\bf Stable charged particles:} In anomaly mediated SUSY breaking, the
LSP is wino-like, and the $\tilde\chi^\pm - \tilde\chi_1^0$ mass
difference is therefore small.  As a result, stable charginos are not
unlikely. The searches designed at LEP in the canonical scenario for
large $M_2$ values apply here equally well.  At the Tevatron, a search
was performed by D\O\ for pairs of charged massive stable particles
that could result from chargino pair production. Such particles would
behave like slow moving muons that could be detected as delayed
signals in the muon system. No significant excess of delayed muons was
observed in 1.1~\invfb\ of data, and a mass lower limit of 206~GeV was
set on long-lived wino-like charginos~\cite{Abazov:2008qu}.

A search for stable stops was performed by CDF in 1~\invfb\ of data,
using a high \pt\ muon trigger and their time-of-flight
detector. Stable stops hadronize to form $R$-hadrons which behave like
slow muons. A model for the interactions of those $R$-hadrons with the
detector material was constructed, within which a stop mass lower
limit of 249~GeV was derived~\cite{Aaltonen:2009ke}.

{\bf Stable or long-lived gluinos:} Models have been built where the
gluino could be the LSP and therefore stable, if $R$-parity is
conserved~\cite{Raby:1997bpa,Baer:1998pg,Raby:1998xr}.  Alternatively,
gluinos may decay, but with long lifetimes.  This occurs, for example,
in models with ``split SUSY,'' unnatural models in which all squarks
and sleptons are very heavy, but the gauginos remain at the
electroweak
scale~\cite{ArkaniHamed:2004fb,Giudice:2004tc,Giudice:2004tcerratum}.
Since gluino decays are mediated by squark exchange, the gluino
becomes long-lived.

Although gluinos cannot be produced directly in $e^+e^-$ interactions,
they could be produced via gluon splitting, e.g., $e^+e^-\to
q\bar{q}g^\ast\to q\bar{q}\tilde{g}\tilde{g}$, and hadronize into
metastable ``$R$-hadrons.'' The QCD predictions for four-jet events
would therefore be modified. Gluinos could also be produced in the
decay of heavier squarks.  Dedicated analyses were performed at
LEP~\cite{Heister:2003hc,Abdallah:2002qi}, leading to a mass lower
limit of 27~GeV for a stable gluino.

A search for long-lived gluinos was also performed by D\O\ with
410~\invpb\ of data~\cite{Abazov:2007ht}.  After hadronization into an
$R$-hadron, a long-lived gluino could come to rest in the calorimeter
and decay later on, during a bunch crossing different from the one
during which it was created~\cite{Arvanitaki:2005nq}. The main decay
mode expected is $\tilde{g}\to g\tilde\chi^0_1$, leading to an
hadronic shower originating from within the calorimeter and not
pointing toward the $p\bar p$ interaction region, in an otherwise
empty event.  No excess of this anomalous topology was observed over
the background due to cosmic muons or to the beam halo. The gluino
mass lower limits derived depend on the lifetime $\tau_{\tilde{g}}$,
the branching fraction $\cal B$ for the decay mode considered, the
$\tilde\chi^0_1$ mass, and the cross section $\sigma_R$ for the
conversion of a neutral $R$-hadron into a charged one in the
calorimeter. As an example, a mass limit of 270~GeV was obtained for
$\tau_{\tilde{g}}<3$~hours, ${\cal B} = 1$,
$m_{\tilde\chi_1^0}=50$~GeV, and $\sigma_R=3$~mb.

\section{Summary}
\label{sec:Summary}

Supersymmetry is one of the most promising ideas for extending the
standard model.  When realized at the weak scale, many SUSY models
provide natural, and even elegant, solutions to the most pressing
problems in particle physics today by stabilizing the gauge hierarchy,
providing dark matter candidates, and accommodating force unification,
both with and without gravity.  In addition, the general framework of
weak-scale SUSY is flexible enough to encompass a wide variety of new
phenomena, including extended Higgs sectors, missing energy,
long-lived and metastable particles, and a host of other signatures of
new physics.  Searches for SUSY are therefore also searches for other
forms of new physics which, even if less profoundly motivated, are, of
course, also important to pursue.

In this review, we have comprehensively summarized the state of the
art in searches for SUSY at the high energy frontier.  Although this
is a continuously evolving subject, this review provides a snapshot of
the field at a particularly important time, when final results from
LEP and HERA are in hand, the Tevatron experiments have reported deep
probes of many supersymmetric models with several $\fb^{-1}$ of data,
and the LHC will soon begin operation.

This review has summarized searches for both supersymmetric Higgs
bosons and standard model superpartners.  In the Higgs sector, SUSY
requires a light neutral Higgs boson. This Higgs boson could be
standard-model like, but it could also have non-standard couplings.
In addition, it is accompanied by other Higgs bosons, both neutral or
charged. The most stringent constraints on a SM-like Higgs boson
currently come from LEP, with a mass lower limit of 114.4~GeV that
applies in the MSSM at low $\tan\beta$. Furthermore, the LEP
experiments set a lower limit of 93~GeV on the lightest neutral Higgs
boson of the MSSM, independent of $\tan\beta$. The MSSM parameter
space has now been further restricted by the Tevatron experiments.
For example, $\tan\beta$ values larger than 40 are excluded for
$m_A=140$~GeV. For charged Higgs bosons, LEP excludes masses below
78.6~GeV, and the Tevatron experiments have extended this mass limit
to $\sim 150$~GeV for very large values of $\tan\beta$.

For superpartners, the bounds are, of course, model-dependent, but the
main results may be summarized as follows.  The searches at LEP have
constrained the masses of all SUSY particles, except for the gluino
and the LSP, to be larger than approximately 100~GeV in most SUSY
scenarios. Furthermore, an indirect lower limit on the mass of a
neutralino LSP has been set at 47~GeV in the MSSM with gaugino and
sfermion mass unification.  The higher center-of-mass energy at the
Tevatron has allowed tighter mass limits to be obtained for strongly
interacting SUSY particles: 379 and 308~GeV for squarks and gluinos,
respectively, within the mSUGRA framework at low $\tan\beta$. In that
same model, domains beyond the LEP reach were also probed by searches
for associated chargino-neutralino production.

In the near future, the first indication for SUSY at high energy
colliders could be the observation of a light neutral Higgs boson at
the Tevatron.  Of course, such a discovery is not proof of SUSY ---
only the discovery of superpartners would unambiguously establish SUSY
as being realized in nature.  Once collisions begin at the LHC and the
detectors are sufficiently understood, it will not take more than
$\sim 1$~\invfb\ to discover squarks and gluinos with masses less than
$\sim 1.5$~TeV~\cite{Aad:2009wy,Ball:2007zza}.  A new
era will then begin during which the whole SUSY spectrum will have to
be deciphered, and the properties of the SUSY model established. Many
more \invfb\ will be needed for that purpose, and to unravel the
spectrum of SUSY Higgs bosons.

\section*{Acknowledgments} 

The work of JLF was supported in part by NSF grants PHY--0239817 and
PHY--0653656, NASA grant NNG05GG44G, and the Alfred P.~Sloan
Foundation. JFG is supported by the CNRS/IN2P3 (France).  The work of
JN was supported by DOE grant DE--FG02--91ER40664.



\providecommand{\href}[2]{#2}\begingroup\raggedright\endgroup

\end{document}